\newif\ifjcp
\jcptrue  

\ifjcp
    \documentclass[twocolumn,aip,jcp,reprint,floatfix]{revtex4-2}
\else
    \documentclass[journal=jacsat,manuscript=article]{achemso}
    \setkeys{acs}{articletitle = true}
    \newcommand{\onlinecite}[1]{\hspace{-1 ex} \nocite{#1}\citenum{#1}}
    \SectionNumbersOn
\fi

\usepackage{float}
\usepackage{xcolor}
\usepackage{graphicx}
\usepackage{dcolumn}
\usepackage{bm}
\usepackage{amsmath}
\usepackage{braket}
\usepackage{dsfont} 
\usepackage{hyperref}

\DeclareFixedFont{\ttb}{T1}{txtt}{bx}{n}{9} 
\DeclareFixedFont{\ttm}{T1}{txtt}{m}{n}{9}  

\usepackage{color}
\definecolor{deepblue}{rgb}{0,0,0.5}
\definecolor{deepred}{rgb}{0.6,0,0}
\definecolor{deepgreen}{rgb}{0,0.5,0}

\usepackage{listings}

\newcommand\pythonstyle{\lstset{
language=Python,
basicstyle=\ttm,
morekeywords={self},              
keywordstyle=\ttb\color{deepblue},
emph={MyClass,__init__},          
emphstyle=\ttb\color{deepred},    
stringstyle=\color{deepgreen},
frame=tb,                         
showstringspaces=false
}}

\lstnewenvironment{python}[1][]
{
\pythonstyle
\lstset{#1}
}
{}

\newcommand\pythonexternal[2][]{{
\pythonstyle
\small
\lstinputlisting[#1]{#2}}}

\newcommand{\cl}[1]{{\mathbf{#1}}}
\newcommand{\ci}[1]{{#1}}

\newcommand{\cumulant}{{\tilde{K}}}
\newcommand{\approxcumulant}{{K}}

\newcommand{\toadd}[1]{{#1}}
\newcommand{\toremove}[1]{\ignorespaces}

\newcommand{\setinfo}{%
    \title{On the effective reconstruction of expectation values from {\em ab initio} quantum embedding}%
    \author{Max Nusspickel}%
    \affiliation{Department of Physics, King's College London, Strand, London WC2R 2LS, U.K.}%
    \author{Basil Ibrahim}%
    \affiliation{Department of Physics, King's College London, Strand, London WC2R 2LS, U.K.}%
    \author{George H. Booth}%
    \email{george.booth@kcl.ac.uk}%
    \affiliation{Department of Physics, King's College London, Strand, London WC2R 2LS, U.K.}%
}

\ifjcp
\else
\fi

\ifjcp\else
    \setinfo
\fi
\begin{document}
\ifjcp
    \setinfo
\fi
\newcommand*{\cmtmax}[1]{\textcolor{blue}{Max: #1}}
\newcommand*{\ghb}[1]{\textcolor{cyan}{George: #1}}
\newcommand*{\basil}[1]{\textcolor{violet}{Basil: #1}}
\newcommand{\ubar}[1]{\text{\b{$#1$}}}

\begin{abstract}
    Quantum embedding is an appealing route to fragment a large interacting quantum system into several smaller auxiliary `cluster' problems to exploit the locality of the correlated physics. In this work we critically review approaches to recombine these fragmented solutions in order to compute non-local expectation values, including the total energy. Starting from the democratic partitioning of expectation values used in density matrix embedding theory, we motivate and develop a number of alternative approaches, numerically demonstrating their efficiency and improved accuracy as a function of increasing cluster size for both energetics and non-local two-body observables in molecular and solid state systems. These approaches consider the $N$-representability of the resulting expectation values via an implicit global wave~function across the clusters, as well as the importance of including contributions to expectation values spanning multiple fragments simultaneously, thereby alleviating the fundamental locality approximation of the embedding. We clearly demonstrate the value of these introduced functionals for reliable extraction of observables and robust and systematic convergence as the cluster size increases, allowing for significantly smaller clusters to be used for a desired accuracy compared to traditional approaches in {\em ab initio} wave~function quantum embedding.
\end{abstract}

\ifjcp
    \maketitle
\fi



\section{Introduction}

Quantum chemical methods to describe explicit correlations in an {\em ab initio} many-electron system can be highly accurate, though their applicability is often stymied by a steep computational scaling with respect to system size which (despite significant recent progress) limits their use for extended systems \cite{Booth2013,C2CP23927B,Zhang2019,Gruber2018,McClain2017,doi:10.1021/acs.jctc.1c00985}. To combat this, the locality of this correlated physics is increasingly exploited, enabling a reduction in scaling to be competitive compared to mean-field or density functional approaches, whilst remaining free from empiricism \cite{doi:10.1063/1.5126216, doi:10.1021/cr200093j}. The field of `local correlation' methods in quantum chemistry generally build these locality constraints in the particle-hole excitation picture of the system, localizing each of these spaces separately \cite{doi:10.1142/9789812776815_0003,Usvyat2018,Schafer2021}. While highly related, `quantum embedding' approaches from condensed matter physics are also increasingly coming to the fore as an alternative paradigm and applied to quantum chemical and {\em ab initio} systems \cite{Sun2016}. 

A loose (and necessarily imperfect) characterization of a key difference in these approaches could be that quantum embedding does not build this locality from a particle-hole picture---rather, a fully local set of `atomic-orbital-like' degrees of freedom are chosen initially (which will in general have neither fully occupied or unoccupied mean-field character), which we will call the `fragment' space, though is also often called the `impurity' space for historical reasons in traditional quantum embedding literature. A larger space is then constructed by augmenting these fragment orbitals with additional orbitals (often called `bath' orbitals). These are designed to reproduce the quantum fluctuations, entanglement and/or hybridization between the fragment and the rest of the system, as characterized by some tractable (generally mean-field) level of theory which can be performed on the full system. These individual local quantum problems of the fragment and bath orbitals define a `cluster', which is then solved to provide the correlated properties of the original fragment space, potentially with a subsequent self-consistency then applied to update the original mean-field/low-level theory on the full system.


The general algorithm in most quantum embeddings is therefore summarized as 
a) fragment the system, b) for each fragment, construct a bath space describing the coupling to the wider system, c) solve an interacting problem in the cluster space of each fragment via a `high-level' correlated method, d) extract properties of the system, e) optionally, a self-consistency is performed to embed the correlated effects from the cluster model back into the low-level full system method to update the coupling between fragment and environment. There are a large number of choices and variations within this general framework, including (but not limited to) how the bath space is defined (including the choice of `low-level' theory) \cite{Fertitta2018,PhysRevB.102.165107,PhysRevB.101.045126}, how the interacting cluster Hamiltonian is constructed and solved \cite{PhysRevB.64.165114,PhysRevB.100.115134,PhysRevB.92.155132,PhysRevB.104.245114}, and the choice of self-consistent requirements \cite{doi:10.1063/1.5108818,PhysRevB.102.085123,doi:10.1021/acs.jctc.1c01061}. Furthermore, the fundamental quantum variables by which these quantities over the different spaces are characterized can vary, with dynamical mean-field theory and its variants working in a Green's function (dynamical) formalism \cite{Metzner1989,Georges1996,Kotliar2006,KAROLAK201011,doi:10.1021/acs.jctc.9b00934,doi:10.1021/acs.jctc.8b00927,Zhu2021}, while density matrix embedding theory (DMET) and its variants work in a wave~function (static) formalism \cite{Knizia2012,Knizia2013,Bulik2014,Wouters2016,Welborn2016,Cui2020,Pham2020,Lau2021,PhysRevB.104.035121,doi:10.1021/acs.jctc.8b01009,https://doi.org/10.1002/cpa.21984,doi:10.1126/science.abm2295,PhysRevResearch.2.043259,doi:10.1021/acs.jpclett.2c01915,https://doi.org/10.48550/arxiv.2209.03202}, although these two approaches are not fundamentally distinct, and can also be rigorously connected via a common framework \cite{PhysRevB.99.115129,Fertitta2019,Sriluckshmy2021}. 

Much recent progress has been made in these various quantum embeddings and their application to {\em ab initio} systems, including the use of quantum computation as a high-level solver \cite{PhysRevB.105.125117,https://doi.org/10.1002/qua.26975,D2SC01492K,DMET_QC,PhysRevResearch.3.033230}.
The key point in all of these embedding approaches however, is that the scaling with respect to the full system size is defined by only the scaling of the low-level (often mean-field) method, given the local nature of the auxiliary cluster problems. Furthermore, the choice of high-level solver and arbitrary atomic-orbital-like fragmentation allows for spaces which capture strong (albeit still local) correlation effects, beyond the traditional constraints of the particle-hole picture of most local quantum chemistry.

In this work we focus on a critical aspect of quantum embedding which we believe has received less attention, but which has substantial ramifications for its accuracy and applicability. This concerns how non-local properties (including the total energy) of the full system can be reconstructed from the independent cluster solutions of each fragment. We will assess the effect of the inherent locality approximation of quantum embedding on the convergence of different functionals of these non-local expectation values, and motivate and demonstrate new approaches which substantially accelerate convergence with respect to fragment and bath size in the embedding. While this is quite a technical work, the outcome is that general design principles by which different functionals can be devised become clear, including the $N$-representability of these estimates.

Here, we focus on wave function-based quantum embedding (we believe that the ability and approach for constructing appropriate functionals in a Green's function perspective is clearer). 
We start from the density matrix embedding theory as the parent wave~function approach to quantum embedding \cite{Knizia2012,Wouters2016}. 
We always consider fragments consisting of a single atom only and, where we seek to systematically improve expectation values, enlarge the cluster spaces by adding additional interacting bath orbitals.
We believe that this is an efficient and `black-box' approach, and avoids the ambiguities and non-monotonic improvements in the alternative of defining a systematic expansion in the fragment sizes for {\em ab initio} systems, which also can suffer from reducing the symmetry of the problem \cite{PhysRevX.12.011046}. 

The expansion of the bath space is defined from the approximate interacting density matrix (or instantaneous hybridization) between the fragment and environment at a simple approximate second-order perturbation theory~(MP2), and is controlled by a single cutoff parameter as detailed in Ref.~\onlinecite{PhysRevX.12.011046}. This provides cluster-specific bath natural orbitals~(BNOs) as a controllable, reliable and well-defined expansion of the bath space (and hence overall cluster space) of a given fragment. Furthermore, in this work we will neglect considerations of self-consistency of the original mean-field (beyond where self-consistency is required for meaningful extraction and comparison of expectation values, e.g. to control total electron number). More extensive self-consistency to qualitatively change the original full-system mean-field will be considered in future work \cite{PhysRevB.102.085123,doi:10.1063/1.5108818}, but is unlikely to change the conclusions of this work, especially as convergence with cluster size (either via expansion of the fragment or, as in this work, an interacting bath) obviates the effect of self-consistency.


We start the paper recapitulating the original DMET `democratic partitioning' for expectation values, which can be computed via fragment-projection of the reduced density matrices of each fragment \cite{Wouters2016}. We then describe an improved approach for the total energy based around a cumulant functional for the two-body effects. We move on to an approach based around direct projection of wave function amplitudes, rather than density matrices, formally satisfying $N$-representability conditions (not satisfied in the aforementioned density-matrix approaches) and substantially improving expectation values. Finally, practical approaches and further approximations to these will be described to define an efficient protocol for arbitrary expectation values and high-level wave function descriptions. These different approaches are benchmarked for energetics and other non-local properties (spin correlation functions) on both molecular systems via the W4-11 test set \cite{KARTON2011165}, and periodic systems, finding an efficient approach for reconstruction of non-local observables in quantum embedding of general fragmented systems.

\subsection{Summary of findings}

\begin{figure}
    \centering
    \includegraphics[width=1\linewidth]{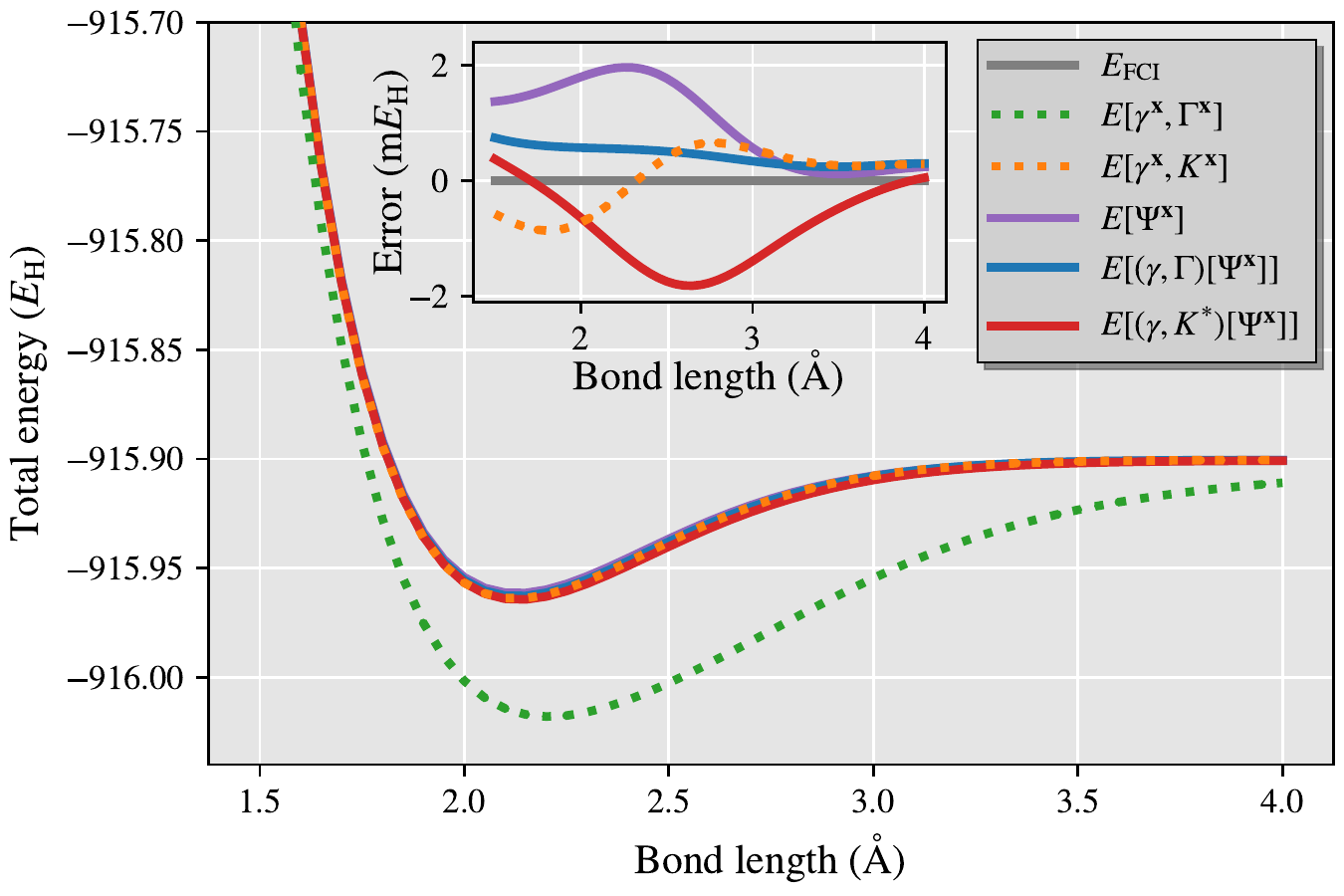}
    \caption{Binding energy of the Chlorine dimer in a STO-6G basis compared to FCI, with $E[\gamma^{\cl{x}},\Gamma^{\cl{x}}]$ corresponding to the democratically partitioned energy expression of traditional DMET \cite{Wouters2016}. Different energies correspond to different total energy functionals, constructed from the same solutions to the two DMET embedding problems (with fragment spaces of the (L{\"o}wdin-orthogonalized) core and valence orbitals of each Chlorine atom). The two cluster spaces are defined by 10 orbitals (9 fragment and 1 bath), with the full space comprising 18 orbitals. The motivation and definition of these different energy functionals from the embedded wave functions are given in the rest of this work, with more details on this system (and others like it) discussed in section~\ref{sec:example}.}
    \label{fig:cl2_intro}
\end{figure}

In Figure~\ref{fig:cl2_intro}, we show results from the different total energy functionals described in this work, as a representative illustration of the significant difference that this choice can make in a simple system (Chlorine dimer, minimal basis, two atomic fragments). For all the different total energy functionals shown in Fig.~\ref{fig:cl2_intro}, the same identical fragment and DMET cluster space is used, and the same (exact) solver and wave function are found from each cluster \cite{Knizia2012, Knizia2013}. The only difference is the choice of energy functional to reconstruct the total energy from the two cluster solutions, with $E[\gamma^{\mathbf{x}},\Gamma^{\mathbf{x}}]$ representing the traditional `democratically partitioned' DMET energy used primarily in the literature to date \cite{Wouters2016}.

As a summary regarding the reconstruction of expectation values from cluster solutions in this work, we find the main conclusions to be the following:
\begin{enumerate}
    \item It is advantageous to separate factorizable products of lower-rank contributions to expectation values where possible. In this way, we can construct e.g. the two-electron energy from the one-body density matrix and two-body cumulant, rather than the two-body density matrix directly. We demonstrate that this improves estimators via the inclusion of implicit cross-cluster contributions to the expectation values. Many of the improvements in this work arise from implicitly building in contributions to expectation values (or the wave~function itself) from products of different cluster contributions, coupling the individual cluster solutions in a non-local fashion, and minimizing the local approximations inherent to the embedding framework.
    \item $N$-representability of estimators can be used as an effective guiding principle, i.e. that they can be derived from a valid `global' wave function from the combined cluster wave function solutions. We show how this can be achieved exactly. Defining this `global' wave function ensures conservation of many quantum numbers e.g. total electron number, which removes the necessity for costly (and sometimes ill-defined) self-consistency conditions, and can allow for variational estimators \cite{https://doi.org/10.1002/cpa.21984}. Beyond this, enabling convergence with respect to bath size of each fragment can render self-consistency entirely unnecessary.
    \item It is possible to construct estimators from this well-defined global wave function by casting the expectation value as a functional of the wave function amplitudes of each cluster, rather than combining density matrices or observables directly. Each factor of the wave function amplitudes in the expectation value has their {\em occupied} indices (symmetrically) projected onto the fragment spaces of each cluster to avoid double-counting, and a sum over all cluster solutions for each factor of the wave~function parameters is introduced.
    \item Along with dramatically improved estimates, this wave~function approach furthermore avoids the requirement for optimizations of chemical potentials, as the global electron number is strictly conserved, and the condition of the union of the fragment spaces is only that they span the occupied, rather than the full orbital space of the original system in order to converge to exact results (e.g. as the (interacting) bath is expanded to completeness for each cluster), vastly reducing the burden of large fragment spaces spanning virtual orbitals for calculations in realistic basis sets.
    \item If the approach above results in an intractable scaling with respect to number of fragments in the system, then a principled approximation can be made, which we show can fortuitously lead to effective cancellation of errors and an even faster convergence of the ground state energy to the exact global expectation values.
\end{enumerate}

We motivate and evidence these conclusions in the following sections, resulting in our final recommendation for an approach to one- and two-body expectation values from DMET and related wave~function quantum embeddings in {\em ab initio} systems, while retaining at most an $\mathcal{O}(N^3)$ scaling with system size in the evaluation of these observables. We detail the practical implementation of these schemes for both an exact high-level solver, and an (arbitrary order) coupled-cluster framework. 
All results can be reproduced from our recently released {\tt Vayesta} codebase for quantum embedding \footnote{Embedding code and documentation can be found at https://github.com/BoothGroup/Vayesta}, which interfaces with the {\tt PySCF} code \cite{Sun2020,Sun2018}, with scripts to generate many of the results of this work (including the input required for Fig.~\ref{fig:cl2_intro}) found in the Supplementary Information (SI).

\section{Global expectation values from cluster density matrices}\label{sec:expect_from_dm}

\subsection{Democratic partitioning of density matrices} \label{sec:dem_part}


Expectation values in DMET derived from operators which span more than one fragment are calculated from `democratically partitioned' reduced density-matrices (RDMs) \cite{Wouters2016}. These can be written as
%
%
%
\begin{align}
    \gamma_{pq} &= \sum_\cl{x}^{N_\mathrm{frag}}
    ( \hat{P}^\cl{x} \gamma^\cl{x} )_{pq} \label{eq:demo_dm1} \\
    \Gamma_{pqrs} &= \sum_\cl{x}^{N_\mathrm{frag}}
    ( \hat{P}^\cl{x} \Gamma^\cl{x} )_{pqrs} \label{eq:demo_dm2}
    ,
\end{align}
where $\gamma$ and $\Gamma$ are one- and two-body reduced density matrices formed from the high-level solution of each cluster problem, as
\begin{alignat}{2}
    &\gamma_{pq}^{\cl{x}} &&= \braket{\Psi^{\cl{x}} | \hat{c}_p^\dagger \hat{c}_q | \Psi^{\cl{x}} } , \label{eq:proj1dm} \\
    &\Gamma_{pqrs}^{\cl{x}} &&= \braket{\Psi^{\cl{x}} | \hat{c}_p^\dagger \hat{c}_r^\dagger \hat{c}_s \hat{c}_q | \Psi^{\cl{x}}} , \label{eq:proj2dm}
\end{alignat}
where the high-level cluster wave function,  $|\Psi^{\cl{x}} \rangle$, includes the contribution from the doubly occupied environmental orbitals of each cluster. $\hat{P}^\cl{x}$ is an operator, which introduces a projection onto the fragment subspace of the cluster~$\cl{x}$ in a symmetrically averaged fashion, e.g.
%

\begin{align} \label{eq:projector_dm}
( \hat{P}^\cl{x} \gamma^\cl{x} )_{pq} &=
\frac{1}{2} \sum_t^{N_\mathrm{mo}} ( P_{pt}^\cl{x} \gamma^\cl{x}_{tq} + P_{qt}^\cl{x} \gamma^\cl{x}_{pt} ),\\
( \hat{P}^\cl{x} \Gamma^\cl{x} )_{pqrs} &=
\frac{1}{4} \sum_t^{N_\mathrm{mo}}  ( P_{pt}^\cl{x} \Gamma^\cl{x}_{tqrs} + P_{qt}^\cl{x} \Gamma^\cl{x}_{ptrs} \\
                    &\qquad \qquad+ P_{rt}^\cl{x} \Gamma^\cl{x}_{pqts} + P_{st}^\cl{x} \Gamma^\cl{x}_{pqrt} ) \nonumber.
\end{align}
In these, $\cl{x}$ labels both the $N_\mathrm{frag}$ fragments and clusters, for which there is an unambiguous one-to-one correspondence, and $P_{pq}^\cl{x}$ is the fragment projection matrix acting over the whole molecular orbital~(MO) space, as
\begin{equation}\label{eq:projector}
    P^{\cl{x}}_{pq} = \sum_{x}^{N_\mathrm{cl}^\cl{x}}
    \left( \mathbf{C}^T \mathbf{S} \mathbf{C}_\mathrm{f}^\cl{x} \right)_{px}
    \left( \mathbf{C}^T \mathbf{S} \mathbf{C}_\mathrm{f}^{\cl{x}} \right)_{qx}
    ,
\end{equation}
with $\mathbf{C}$ representing the MO coefficients, $\mathbf{S}$ the atomic orbital~(AO) overlap matrix,
and $\mathbf{C}_\mathrm{f}^\cl{x}$ the coefficients of the fragment orbitals in cluster~$\cl{x}$.
Note that a post mean-field cluster solver will only modify the 1-RDM in the cluster--cluster part,
whereas the 2-RDM also acquires contributions in off-diagonal cluster--environment parts.
As a result, the projection in Eq.~\eqref{eq:demo_dm1} can be perform\toadd{ed} in the respective cluster spaces,
whereas in Eq.~\eqref{eq:demo_dm2} it has to be performed in the full system space, in order to take
these changes in the off-diagonal parts into account.

The projection onto the fragment space of each cluster is required since the bath spaces overlap significantly between different clusters, and we must project out any double counting arising from contributions from overlapping bath spaces. In contrast, the partitioning of the system into fragment spaces is considered to be a disjoint fragmentation, where the set of all fragment orbitals are an orthonormal set with no overlapping fragment spaces in different clusters.
From the democratically partitioned density-matrices, the total energy can be calculated as
\begin{equation}\label{eq:e_tot}
    E_\mathrm{tot} = E_\mathrm{nuc} +
    \sum_{pq}^{N_\mathrm{mo}} h_{pq} \gamma_{qp}  
    + \frac{1}{2} \sum_{pqrs}^{N_\mathrm{mo}} (pq|rs) \Gamma_{pqrs}
    .
\end{equation}
Note that in practice one can avoid forming the full system density-matrices to calculate the total energy
and instead calculate energy contributions directly from the individual cluster density-matrices over purely the cluster degrees of freedom.
The contribution from the doubly occupied environmental orbitals can be integrated out, by forming the Coulomb- and exchange potential of the
unentangled occupied orbitals of each DMET cluster via an effective one-electron potential \cite{}. This leads to the more common expression for the DMET energy \cite{Wouters2016}, equivalent to Eq.~\eqref{eq:e_tot} via construction of democratically partitioned density matrices of Eq.~\eqref{eq:projector_dm}, as
\begin{equation} \label{eq:dempart_energy}
\begin{split}
E[\gamma^{\mathbf{x}},\Gamma^{\mathbf{x}}] = E_\mathrm{nuc} + &\sum_{\cl{x}}^{N_\mathrm{frag}} \left[ 
\sum_{pq}^{N_\mathrm{cl}^\cl{x}} \tilde{h}^{\cl{x}}_{pq} (\hat{P}^\cl{x} \gamma^\cl{x} )_{qp}  \right. \\
&+ \left. \frac{1}{2} \sum_{pqrs}^{N_\mathrm{cl}^\cl{x}} (pq|rs) (\hat{P}^\cl{x} \Gamma^\cl{x} )_{pqrs} \right] ,
\end{split}
\end{equation}
where $\gamma^\cl{x}$ and $\Gamma^\cl{x}$ refer to the cluster reduced density matrices, with the projector purely acting in this cluster space, and $N_\mathrm{cl}^{\cl{x}}$ denoting the number of orbitals in the cluster $\cl{x}$. The energetic effect of these states (static Coulomb and exchange contributions) is then included via the construction of the $\tilde{h}^{\cl{x}}$, which includes the potential from these unentangled states to the one-body hamiltonian as $\frac{1}{2} \sum_{mn} (pq||mn) \gamma^{\cl{x}}_{\mathrm{core},mn}$ where $\gamma^{\cl{x}}_{\mathrm{core}}$ is the density matrix from these core states \cite{}.
We can exploit fragments that are (by symmetry) in identical chemical environments by only computing the cluster solutions and energy contributions of symmetry-unique fragments. In the rest of this work, the expression $E[\gamma^{\mathbf{x}},\Gamma^{\mathbf{x}}]$ will denote this democratically partitioned energy functional, shorthand for $E[\{\gamma^{\mathbf{x}}\},\{\Gamma^{\mathbf{x}}\}]$, denoting that the energy is computed from the set of individually constructed cluster one- and two-body RDMs.

This energy expression is exact when two conditions are satisfied. First, it requires that the fragmentation of the full system is complete, i.e. that the union of the fragment spaces spans all degrees of freedom of the system.
This condition ensures that the trace of the sum of the different fragment projectors is exactly equal to the total number of orbitals in the system, or alternatively, that
\begin{equation}
    \sum_{\cl{x}}^{N_{\textrm{frag}}} {\hat P}^{\cl{x}} = \hat{\mathds{1}} . \label{eq:comp_proj}
\end{equation}
While it is a relatively mild condition to ensure that the combined fragment spaces span the generally localizable occupied space, ensuring that they span the (generally much larger) high-energy virtual space is harder to achieve and leads to much larger fragment spaces. This has required DMET simulations in realistic basis sets to augment fragment spaces with projected atomic orbitals~(PAOs)\cite{PULAY1983151} to ensure this condition is fulfilled for reasonable results \cite{Cui2020}.
The second criteria which must be fulfilled, is that the individual cluster density matrices must be exact, which in general will require the clusters of each fragment themselves to be enlarged to completeness, either by increasing the size of the fragment or (interacting) bath space.
This ensures that $|\Psi^\cl{x} \rangle \rightarrow |\Psi \rangle$ and the projected density matrices of the clusters (Eq.~\ref{eq:projector_dm}) are equivalent to the projections of the exact density matrix over the whole system. Combined with the completeness of the projector (Eq.~\ref{eq:comp_proj}), this will lead to the exact energy from Eq.~\eqref{eq:dempart_energy}. 

Away from this exact limit, there are a number of drawbacks to this approach to compute properties from the DMET solutions. Firstly, the reconstructed full-system density matrices (Eqs.~\eqref{eq:demo_dm1} and \eqref{eq:demo_dm2}) from the projected cluster solutions are {\em not} $N$-representable, meaning that they cannot be derived from a valid wave function. This can be seen as the democratically partitioned 1-RDM of Eq.~\eqref{eq:demo_dm1}, 
\begin{equation}\label{eq:part_rdm1}
    \gamma_{pq} = \sum_\cl{x}^{N_\mathrm{frag}} \hat{P}^\cl{x} \braket{\Psi^\cl{x} | c_p^\dagger c_q |\Psi^\cl{x}  } ,
\end{equation}
cannot be rewritten as a simple expectation value~$\braket{\tilde{\Psi} | c_p^\dagger c_q | \tilde{\Psi}}$ of some wave~function~$\ket{\tilde{\Psi}}$. As a specific consequence, this can result in eigenvalues (occupation numbers) becoming negative or greater than two (in a restricted basis), violating the Pauli principle, and removing any variational guiding principle in the method \cite{https://doi.org/10.1002/cpa.21984}. Furthermore, conserved quantities and good quantum numbers in the individual cluster solutions such as electron number ($N$), spin and its $z$-projection ($S_z$, $S^2$) and other symmetries are not maintained in these composite full system descriptions.

To mitigate some of these effects, a global chemical potential (or potentially a fragment-specific chemical potential) is almost universally optimized in DMET to ensure that at least an exact, integer number of electrons is recovered in these democratically partitioned density matrices \cite{Wouters2016,doi:10.1063/1.4891861}. This can move electrons between the fragment and bath of each correlated cluster solution, such that the known global electron number is maintained as a constraint. While this corrects one known global quantum number in the density matrices, it does not correct others, and does not in general restore $N$-representability of the full system RDMs. Furthermore, this necessitates costly additional self-consistent loops over the high-level calculations. Furthermore, this requirement of a chemical potential optimization to a known total electron number further underlines the importance of the constraint of Eq.~\eqref{eq:comp_proj}, ensuring that the full (occupied and virtual) space is spanned by the fragments, as all can be partially occupied in the correlated state and the democratically partitioned density matrices must trace to the correct electron number. Numerical demonstration of the breaking of these $N-$representability constraints in the democratically partitioned RDMs will be given in Sec.~\ref{sec:example} (with and without a global chemical potential optimization), also showing the deleterious effect on properties and computed energetics of the system that result. 

\subsection{Democratic partitioning of cumulants} \label{sec:part_cumulant}

%

In the following, we propose a simple alternative for the construction of democratically partitioned two-body density-matrices
from DMET clusters, from which expectation values such as the energy can be calculated.
Instead of partitioning the 2-RDM directly as in Eq.~\eqref{eq:demo_dm2}, we partition the two-body cumulant,~$\cumulant$,
defined (in a restricted basis)\footnote{We note that the two-body cumulant would often be denoted by $\lambda$, but define by $\cumulant$ to avoid potential confusions with the lambda-amplitudes of coupled-cluster theory used later in the text.} via
\begin{equation}\label{eq:cumulant}
    \Gamma_{pqrs} = \gamma_{pq} \gamma_{rs} - \frac{1}{2} \gamma_{pr}\gamma_{sq} + \cumulant_{pqrs}.
\end{equation}
The non-cumulant (disconnected) contributions to the 2-RDM can then be reconstructed from the democratically-partitioned one-body
density-matrix, given by Eq.~\eqref{eq:demo_dm1}, such that Eq.~\eqref{eq:demo_dm2} is replaced by 
%
\begin{equation}\label{eq:demo_cumulant}
    \Gamma_{pqrs} = \gamma_{pq} \gamma_{rs} - \frac{1}{2} \gamma_{pr}\gamma_{sq}
    + \sum_\cl{x}^{N_\mathrm{frag}} ( \hat{P}^\cl{x} \cumulant^\cl{x} )_{pqrs}
    .
\end{equation}
The difference between Eq.~\eqref{eq:demo_dm2} and \eqref{eq:demo_cumulant} lies purely in the non-cumulant contribution to the two-body
density matrix, which can be written as the product of democratically partitioned 1-RDMs. In the standard DMET partitioning of Eq.~\eqref{eq:demo_dm2} these are taken from a single embedding problem at a time,
whereas the partitioning of Eq.~\eqref{eq:demo_cumulant} contains `cross-cluster' contributions, which can be seen by inserting Eq.~\eqref{eq:demo_dm1} into the first term of Eq.~\eqref{eq:demo_cumulant}:
\begin{equation}\label{eq:non-cumulant-cross-cluster}
    \gamma_{pq} \gamma_{rs} =
    \sum_\cl{x}^{N_\mathrm{frag}}
    \sum_\cl{y}^{N_\mathrm{frag}}
    ( \hat{P}^\cl{x} \gamma^\cl{x} )_{pq}
    ( \hat{P}^\cl{y} \gamma^\cl{y} )_{rs}
    .
\end{equation}
In this way, the non-local (correlated) one-body physics of two distinct clusters, $\cl{x} \neq \cl{y}$,
contribute to global two body expectation values; the same  is not possible in Eq.~\eqref{eq:demo_dm2}.
This is expected to be important in cases where the orbitals $p$ and $q$ are far from the orbitals~$r$ and $s$, and are not spanned together in any single DMET cluster. In this case, the conventional DMET partitioning of Eq.~\eqref{eq:demo_dm2} will not account for the relaxation of the external Coulomb- and exchange potential of a fragment due to the (potentially correlation-induced) density changes within the other orbital set. In contrast, this will be implicitly included in the partitioning according to Eq.~\eqref{eq:demo_cumulant}. This is the key physics where we expect a partitioning of cumulants to be superior for two-body physics to the traditional democratic partitioning approach of Sec.~\ref{sec:dem_part} for non-local expectation values.
We will denote any total energies resulting from the democratically partitioned cumulant approach described in this section as $E[\gamma^{\cl{x}},K^{\cl{x}}]$ in the rest of this work.

\begin{figure}
    \centering
    \includegraphics[width=1\linewidth]{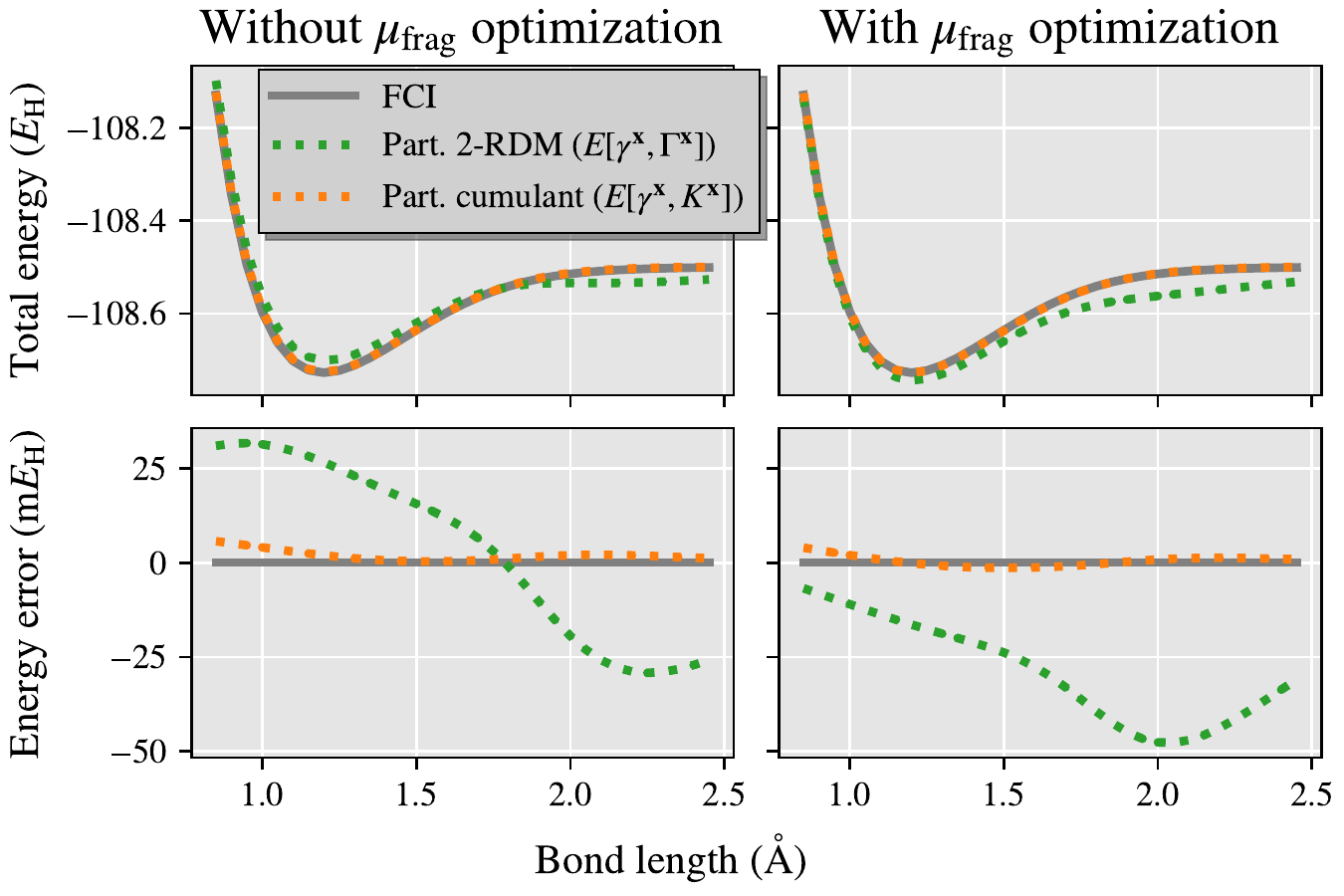}
    \caption{Dissociation curve of N$_2$ in the STO-6G basis,
    calculated from two atomic DMET embedding problems. The same DMET-FCI cluster solution is used in both the partitioned density matrix (Sec.~\ref{sec:dem_part}) and partitioned cumulant (Sec.~\ref{sec:part_cumulant}) energy expressions, $E[\gamma^{\cl{x}},\Gamma^{\cl{x}}]$ and $E[\gamma^{\cl{x}}, K^{\cl{x}}]$ respectively, defined as given by Eqs.~\eqref{eq:dempart_energy} and \eqref{eq:e_dmet_new}.
    Plots of the right have optimized a global chemical potential to ensure that the democratically partitioned 1RDM traces to the correct number of electrons, while the plots on the left omit this chemical potential optimization.}
    \label{fig:demo_cumulant_n2}
\end{figure}

We present a simple example showing the difference between the partitioned density matrices and the partitioned cumulant approach in Fig.~\ref{fig:demo_cumulant_n2}, for a representative system of N$_2$ in a minimal basis set. The fragment space consists of the five symmetrically (L{\"o}wdin) orthogonalized atomic orbitals of a single atom (the 1s, 2s, 2p$_x$, 2p$_y$, 2p$_z$ spaces) with the DMET bath consisting of an additional three bath orbitals consistent with the bond order of the dimer in a minimal basis. The DMET cluster in this example therefore contains 10 electrons in 8 orbitals, which is compared to the full system of 14 electrons and 10 orbitals, and is solved with exact diagonalization (FCI). It is found that regardless of whether the fragment chemical potential is optimized or not, the democratically partitioned cumulant results in a substantial improvement in the energy, with the partitioned cumulant being almost unaffected by this chemical potential optimization. This is further consistent with the Cl$_2$ dimer shown in Fig.~\ref{fig:cl2_intro} and analyzed in more detail in Sec.~\ref{sec:example}, while the improvement in non-energetic properties (the spin--spin correlation function) from the cumulant partitioning will be shown in Sec.~\ref{sec:spin_corr_fns}.

A minor technical detail to mention is that we use a slightly different partitioning in practice than that of Eq.~\eqref{eq:cumulant} throughout this work. The correlated cluster 1-RDM can be decomposed as $\gamma = \gamma^0 + \Delta \gamma$, where $\gamma^0$ is the reference mean-field density-matrix. We then work with a slightly modified cumulant definition, $\approxcumulant$, as
\begin{equation}\label{eq:approx_cumulant_1}
\begin{split}
    \Gamma_{pqrs} &=
    \gamma_{pq}^0 \gamma_{rs}^0
    + \gamma_{pq}^0 \Delta\gamma_{rs}
    + \Delta\gamma_{pq} \gamma_{rs}^0 \\
    &- \frac{1}{2} ( \gamma_{pr}^0\gamma_{sq}^0
    + \gamma_{pr}^0\Delta\gamma_{sq}
    + \Delta\gamma_{pr} \gamma_{sq}^0 )
    + \approxcumulant_{pqrs} .
\end{split}
\end{equation}
The approximate cumulant $\approxcumulant$ of Eq.~\eqref{eq:approx_cumulant_1} therefore contains the
$(\Delta \gamma)^2$-terms which are not present in the true cumulant $\cumulant$, with the relation between the two given as
\begin{equation}\label{eq:approx_cumulant}
    \approxcumulant_{pqrs} = \cumulant_{pqrs}
    + \Delta\gamma_{pq} \Delta\gamma_{rs}
    - \frac{1}{2} \Delta\gamma_{pr} \Delta\gamma_{sq}
    .
\end{equation}
The final expression for the 2-RDM is then
\begin{equation}\label{eq:demo_approx_cumulant}
\begin{split}
    \Gamma_{pqrs} &=
    \gamma_{pq}^0 \gamma_{rs}^0
    + \gamma_{pq}^0 \Delta\gamma_{rs}
    + \Delta\gamma_{pq} \gamma_{rs}^0 \\
    &- \frac{1}{2} ( \gamma_{pr}^0\gamma_{sq}^0
    + \gamma_{pr}^0\Delta\gamma_{sq}
    + \Delta\gamma_{pr}\gamma_{sq}^0 )\\
    &+ \sum_\cl{x}^{N_\mathrm{frag}} ( \hat{P}^\cl{x} \approxcumulant^\cl{x} )_{pqrs}
    ,
\end{split}
\end{equation}
where (in contrast to the democratically partitioned 2-RDM of Eq.~\ref{eq:demo_dm2}),
the projection of the cumulants can be performed in the cluster space only, since the correlated cluster solver will only lead to a non-zero cumulant in this space.
This allows the total energy functional in terms of $\gamma^\cl{x}$ and $\approxcumulant^\cl{x}$ to be written as
\begin{equation}\label{eq:e_dmet_new}
\begin{split}
E[\gamma^{\mathbf{x}},\approxcumulant^{\mathbf{x}}] &= E_\mathrm{HF}[\gamma^0] + \sum_{pq} F_{pq}[\gamma^0] \Delta\gamma_{qp} \\
&+  \frac{1}{2} \sum_{\cl{x}}^{N_\mathrm{frag}} \sum_{pqrs}^{N_\mathrm{cl}^\cl{x}} (pq|rs) (\hat{P}^\cl{x} {\approxcumulant}^\cl{x} )_{pqrs}
,
\end{split}
\end{equation}
where $E_\mathrm{HF}[\gamma^0]$ and $F[\gamma^0]$ are the Hartree--Fock energy and Fock matrix corresponding to the reference mean-field density-matrix $\gamma^0$, and $\Delta \gamma$ is the correlated part of the democratically-partitioned 1-RDM, formed from the cluster density matrices, $\gamma^{\cl{x}}$, as shown in Eq.~\eqref{eq:demo_dm1}.
Note that Eq.~\eqref{eq:e_dmet_new}, in contrast to the conventional DMET energy functional~\eqref{eq:dempart_energy} which it replaces, does not involve a cluster-specific effective core-Hamiltonian.

We find that democratically partitioning the approximate cumulant instead of the true cumulant gives almost identical results
for all systems tested in this paper.
However, in the case of the true cumulant, the Fock matrix of the correlated 1-RDM, $F[\gamma^0 + \Delta \gamma]$, would need to be constructed to calculate the energy, a step that scales as $\mathcal{O}(N^4)$
with respect to the system size~$N$.
%
%
%
For this reason, we use the democratically-partitioned approximate cumulant and calculate the total energy according to Eq.~\eqref{eq:e_dmet_new}, referred to as $E[\gamma^{\cl{x}}, K^{\cl{x}}]$ in the results of this paper.


\section{Global expectation values from cluster wave functions} \label{sec:partwfn}

While the democratically-partitioned cumulant described in the previous section can dramatically improve two-body properties (as will be further shown later), it still suffers from the same fundamental problem discussed in Sec.~\ref{sec:dem_part}. This is that the resulting density matrices are not generally $N$-representable, i.e. they do not correspond to a physical fermionic wave~function over the system. In this section, we go a step further and present an alternative paradigm of directly combining the cluster wave functions rather than RDMs to overcome this representability issue. This was first proposed in a correlation energy functional in Ref.~\onlinecite{PhysRevX.12.011046} (described in Sec.~\ref{sec:lin_energy_func}), but is now generalized and expanded.

The basic idea is to consider an implicit full-system wave~function reconstructed from the cluster wave~functions themselves, as
    \begin{equation}\label{eq:part_wf_linear}
        \ket{\Psi} = \sum_\cl{x}^{N_\mathrm{frag}} \hat{P}^\cl{x} \ket{\Psi^\cl{x}} .
    \end{equation}
From this global state, expectation values can be defined which resolves $N$-representability issues and results in improved estimators. For instance, the 1-RDM can be written as
\begin{equation}\label{eq:part_wf_rdm1}
    \gamma^{(\text{part. WF})}_{pq} = \sum_\cl{xy}^{N_\mathrm{frag}} \braket{\hat{P}^\cl{x} \Psi^\cl{x} | \hat{c}_p^\dagger \hat{c}_q   |\hat{P}^\cl{y} \Psi^\cl{y}  }
\end{equation}
As long as all projected cluster wave~functions~$\hat{P}^\cl{x} \ket{\Psi^\cl{x}}$ are physical, fermionic wave~function in their own right,
then the linear superposition of Eq.~\eqref{eq:part_wf_linear} is a physical, fermionic wave function as well.
As a result, the 1-RDM of Eq.~\eqref{eq:part_wf_rdm1} is trivially $N$-representable,
as it is simply $\braket{\Psi | c_p^\dagger c_q | \Psi}$ with the ``global'' wave~function~$\ket{\Psi}$ given by Eq.~\eqref{eq:part_wf_linear}.
Higher order density-matrices can be constructed in the same way and are thus also $N$-representable. Furthermore, the full system energy computed from RDMs constructed in this fashion will be variational if the cluster solvers are themselves variational methods.
%
It is not difficult to rationalize that expectation values would also be improved from the formulation as sketched for the 1-RDM in Eq.~\eqref{eq:part_wf_rdm1} compared to the democratically partitioned RDMs. This is because a large number of `cross-cluster' contributions to these expectation values are included, where wave~function amplitudes from different clusters are combined beyond their mean-field contributions found in Eq.~\eqref{eq:part_rdm1}.

\subsection{Partitioning of general wave functions}\label{sec:part_linear_wf}
    
So far we have not specified what we mean by the projection of a cluster wave~function as $\hat{P}^\cl{x} \ket{\Psi^\cl{x}}$ in practice.
Since the DMET bath orbitals guarantee that the mean-field reference determinant $\ket{\Phi}$ is represented exactly in each cluster (ignoring the unentangled environment orbitals) \cite{Knizia2012}, it is convenient to define $\hat{P}^\cl{x}$ in terms of its action on the correlated part of the wave~function only, i.e. our full system wave function can be written as
\begin{equation}\label{eq:part_wf_linear_ph}
    \ket{\Psi} = \ket{\Phi} + \sum_\cl{x}^{N_\mathrm{frag}} \hat{P}^\cl{x} \ket{\Delta \Psi^\cl{x}} ,
\end{equation}
where $\ket{\Delta \Psi^\cl{x}} = \ket{\Psi^\cl{x}} - \ket{\Phi}$ and intermediate normalization is assumed.

We then choose to represent the correlated cluster wave~function part $\ket{\Delta \Psi}$ in the basis of particle--hole excitations around the reference determinant in a linear wave~function ansatz
\begin{equation}\label{eq:ci_wf}
    \begin{split}
    \ket{\Delta \Psi} &= (\hat{C}_1 + \hat{C}_2 + \dots ) \ket{\Phi} \\
    &= \sum_{i} \sum_{a} c_i^a \ket{\Phi_i^a}
    + \sum_{ij} \sum_{ab} c_{ij}^{ab} \ket{\Phi_{ij}^{ab}} + \dots
    ,
    \end{split}
    \end{equation}
    %
where we use $i,j,\dots$ ($a,b,\dots$) to represent general occupied (virtual) orbitals in this work.
We choose the fragment projection of this wave function to be defined by its symmetric action on the occupied coefficient indices, for example
\begin{align}\label{eq:part_c1}
    &( \hat{P}^\cl{x} C_1^\cl{x} )_i^a = \sum_{k} P^\cl{x}_{ik} (C^\cl{x}_1)_k^a , \\
    \label{eq:part_c2}
    &( \hat{P}^\cl{x} C_2^\cl{x} )_{ij}^{ab}
    = \frac{1}{2} \sum_{k} \left[ P^\cl{x}_{ik} (C^\cl{x}_2)_{kj}^{ab} + P^\cl{x}_{jk} (C^\cl{x}_2)_{ik}^{ab}\right] ,
\end{align}
in the case of the single and double excitation coefficients in the cluster, $C^{\cl{x}}_1$ and $C^{\cl{x}}_2$, with generalization to higher-order excitation levels straightforward. \toadd{Indices in these expressions correspond to the cluster canonicalized orbitals, with a} critical consequence of the DMET (and BNO) bath construction being that the $n$-fold excitations of the cluster space are entirely spanned by the $n$-fold excitations of the full system due to the coincidence of their reference states. The fragment projection operators acting in the occupied-only space of the cluster can be formulated as
\begin{equation}
    P^{\cl{x}}_{ik} = \sum_{x \in \cl{x}}
    \left[ (\mathbf{C}^{\cl{x}})^T \mathbf{S} \mathbf{C}_\mathrm{f}^\cl{x} \right]_{ix}
    \left[ (\mathbf{C}^{\cl{x}})^T \mathbf{S} \mathbf{C}_\mathrm{f}^{\cl{x}} \right]_{kx} \label{eq:clust_proj_op}
\end{equation}
where $\mathbf{C}_\mathrm{f}^{\cl{x}}$ represents the AO coefficients of the fragment orbitals~($x$) of fragment~$\cl{x}$, and $\mathbf{C}^{\cl{x}}$ represents the coefficients of the occupied cluster orbitals (in contrast to the projector defined in the full-system MO space of Eq.~\ref{eq:projector})\toadd{, with $i$, $k$ representing occupied orbitals of cluster~$\cl{x}$}. We note that this projection operator is not diagonal in the occupied orbital basis of the cluster.
Furthermore, we have crucially chosen to only apply the projection to the \textit{occupied} dimensions of the wave~function coefficients in Eqs.~(\ref{eq:part_c1}, \ref{eq:part_c2}).
This is an important difference compared to the projection of density matrices according to Eq.~\eqref{eq:projector_dm},
where the projector always needs to act on general indices, which enumerate both occupied and virtual orbitals.
The significance of this is that the initial fragmentation of the system now only needs to ensure that the entire occupied space is spanned by the union of all fragments, in order for the projector to be complete. If this is satisfied, the only remaining approximation results from the deviation of each cluster wave~function from exactness (which can be systematically resolved via increasing the cluster/bath space). 

This is a significant advantage, ensuring that the expectation values become exact as the bath space of each cluster is increased, without requiring the fragment spaces to span the virtual space of the system. The requirement for completeness of the fragment projectors is now relaxed from Eq.~\eqref{eq:comp_proj}, and can be written as
\begin{equation}
    \mathrm{Tr} \left[ \sum_{\cl{x}}^{N_{\textrm{frag}}} {\hat P}_{\cl{x}} \gamma^0 {\hat P}_{\cl{x}} \right] = N_e , \label{eq:comp_wfn_proj}
\end{equation}
where $N_e$ is the number of electrons.
This quality is especially impactful in larger basis sets required for quantitative accuracy, as fragment sizes are now {\em independent} of the size of this basis, with the required virtual space captured instead by the bath expansion.
A fragmentation spanning the occupied space is for example guaranteed by the choice of intrinsic atomic orbitals~(IAOs) for the fragment spaces\cite{Knizia2013IAO}, used in the main results of this work unless otherwise specified. We note that neither the representation of the wave~function amplitudes in a particle--hole basis, nor the choice of projection of just the occupied space in itself precludes the solver from capturing strong correlation, since we do not assume a truncation of the excitation levels which are represented in the wave~function. The ability to capture strong correlation effects (for an exact high-level solver) is determined by the suitability of the cluster space, and is invariant to the choice of representation of the resulting wave function. It could however be possible that a more efficient projection operator could be formulated for e.g. strongly correlated lattice models, where the requirement of the fragmentation spanning the virtual space is not difficult to fulfil. However, for {\em ab initio} quantum embeddings, we do not anticipate that this would be beneficial.

    
\subsection{Partitioning of exponential wave function forms}\label{sec:part_exp_wf}

While all wave~functions can be written in the linear form of Eq.~\eqref{eq:ci_wf}, we now show that there are benefits in casting the wave~function to an exponential parameterization for partitioning in this space. This form (common to coupled-cluster methods\cite{RevModPhys.79.291}) can be written as
\begin{equation}\label{eq:exp_wf_ansatz}
\begin{split}
    \ket{\Delta\Psi} &= ( \mathrm{e}^{\hat{T}_1 + \hat{T}_2 + \dots} - 1 ) \ket{\Phi} = \sum_{i}^{N_\mathrm{occ}} \sum_{a}^{N_\mathrm{vir}} t_i^a \ket{\Phi_i^a} \\
    &+ \sum_{ij}^{N_\mathrm{occ}} \sum_{ab}^{N_\mathrm{vir}}
    \left( t_{ij}^{ab} + t_i^a t_j^b \right) \ket{\Phi_{ij}^{ab}} + \dots
\end{split}
\end{equation}
The amplitudes can be converted between the linear and exponential forms more generally \cite{Lehtola2017} via the relations
\begin{align}
    \hat{C}_1 &= \hat{T}_1, \label{eq:c1_t1} \\
    \hat{C}_2 &= \hat{T}_2 + \frac{(\hat{T}_1)^2}{2}, \label{eq:c2_t2} \\
    \hat{C}_3 &= \hat{T}_3 + \frac{(\hat{T}_2 \hat{T}_1 + \hat{T}_1 \hat{T}_2)}{2} + \frac{(\hat{T}_1)^3}{3!}, \label{eq:c3_t3} \\
    & \dots . \nonumber
\end{align}
For ground state wave~functions, the norm of the $T$-amplitudes of the exponential ansatz generally
decays more quickly with respect to the excitation rank than the $C$-amplitudes of the linear ansatz.
The exponential wave~function ansatz thus allows for an alternative way to partition the wave~function
by projecting the $T$-amplitudes, rather than $C$-amplitudes, which by analogy with Eqs.~(\ref{eq:part_c1}, \ref{eq:part_c2}) can be written as
\begin{align}\label{eq:part_t1}
    &( \hat{P}^\cl{x} T_1^\cl{x} )_i^a = \sum_{k}^{N_\mathrm{occ}^\cl{x}} P^\cl{x}_{ik} (T^\cl{x}_1)_k^a , \\
    \label{eq:part_t2}
    &( \hat{P}^\cl{x} T_2^\cl{x} )_{ij}^{ab}
    = \frac{1}{2} \sum_{k}^{N_\mathrm{occ}^\cl{x}}  \left[ P^\cl{x}_{ik} (T^\cl{x}_2)_{kj}^{ab} + P^\cl{x}_{jk} (T^\cl{x}_2)_{ik}^{ab}\right]
    ,
\end{align}
%

To illustrate the difference between the partitioning of the global wave~function via its $C$- and $T$-amplitudes,
we can compare the wave~function ans{\"a}tze truncated at the double-excitation level. For the partitioning of the linear $C$-amplitude representation of the wave~function (Sec.~\ref{sec:part_linear_wf}), we obtain
\begin{equation}\label{eq:global_linear_wf}
\begin{split}
    \ket{\Delta \Psi} =
    &\sum_\cl{x}^{N_\mathrm{frag}} \sum_{i}^{N_\mathrm{occ}^\cl{x}}  \sum_{a}^{N_\mathrm{vir}^\cl{x}}  ( \hat{P}^\cl{x} C^\cl{x}_1 )_i^a \ket{\Phi_i^a} \\
    + &\sum_\cl{x}^{N_\mathrm{frag}}  \sum_{ij}^{N_\mathrm{occ}^\cl{x}}  \sum_{ab}^{N_\mathrm{vir}^\cl{x}}  ( \hat{P}^\cl{x} C^\cl{x}_2 )_{ij}^{ab} \ket{\Phi_{ij}^{ab}} 
\end{split}
\end{equation}
while for the exponential form of the wave~function, we achieve 
\begin{equation}\label{eq:global_exp_wf}
\begin{split}
    \ket{\Delta \Psi} =
    &\sum_\cl{x}^{N_\mathrm{frag}}  \sum_{i}^{N_\mathrm{occ}^\cl{x}} \sum_{a}^{N_\mathrm{vir}^\cl{x}} ( \hat{P}^\cl{x} T^\cl{x}_1 )_i^a \ket{\Phi_i^a} \\
    + &\sum_\cl{x}^{N_\mathrm{frag}}  \sum_{ij}^{N_\mathrm{occ}^\cl{x}} \sum_{ab}^{N_\mathrm{occ}^\cl{x}} ( \hat{P}^\cl{x} T^\cl{x}_2 )_{ij}^{ab} \ket{\Phi_{ij}^{ab}} \\
    + &\sum_{\cl{x}\cl{y}}^{N_\mathrm{frag}} 
    \sum_{i}^{N_\mathrm{occ}^\cl{x}}  \sum_{a}^{N_\mathrm{vir}^\cl{x}}
    \sum_{j}^{N_\mathrm{occ}^\cl{y}} \sum_{b}^{N_\mathrm{vir}^\cl{y}}
    ( \hat{P}^\cl{x} T^\cl{x}_1 )_{i}^{a} ( \hat{P}^\cl{y} T^\cl{y}_1 )_{j}^{b} \ket{\Phi_{ij}^{ab}} .
\end{split}
\end{equation}
While the first two terms in Eqs.~(\ref{eq:global_linear_wf}, \ref{eq:global_exp_wf}) are equivalent and
involve only a single summation over clusters, the last term of Eq.~\eqref{eq:global_exp_wf}
(representing disconnected double excitations) has a double loop over pairs of clusters.
%
%
As a result, a single excitation in cluster~$\cl{x}$ and a different single excitation in cluster~$\cl{y}$
can contribute together to form a double excitation of the global wave~function, even if the two clusters are far apart
and do not overlap.
Similarly, for wave~functions truncated at the triple excitation level, two independent clusters~(e.g. in $T_2^\cl{x} T_1^\cl{y}$),
or three independent clusters~(e.g. in $T_1^\cl{x} T_1^\cl{y} T_1^\cl{z}$) can combine to contribute disconnected triple excitations
to the global wave~function.
The same is not possible for the linear~wave~function partitioning, for which both connected and disconnected contributions
must always come from a single cluster only.

This partitioning of the global wave~function at the level of the $T$-amplitudes therefore results in non-local `cross-cluster' information being built into the solution from the quantum embedding -- this time on the level of the reconstructed global wave~function itself, rather than similar cross-cluster information being built into the 2-RDM (see Sec.~\ref{sec:part_cumulant}) or other expectation values (see Eq.~\ref{eq:part_wf_rdm1}). The beneficial impact of these cross-cluster contributions is a key tenet of this work, and serves to mitigate the impact of the local approximations inherent to quantum embedding by not treating the cluster solutions as contributing independently to final expectation values.

Finally, we should stress that while this framework of partitioning the global $T$-amplitudes of the wave~function naturally fits with the use of coupled-cluster as a high-level solver, we are not inherently constrained to this. If we avoid truncating the rank of the $T$-operator, then any wave~function can be cast into the exponential form of Eq.~\eqref{eq:exp_wf_ansatz} for the high-level solution of the cluster problems, and benefit in this way. 
%
In particular, a linear wave~function obtained from a FCI cluster solver can be converted to the exponential representation via Eqs.~(\ref{eq:c1_t1}--\ref{eq:c3_t3}), projected, and finally recombined into a global exponential wave~function to benefit from these cross-cluster contributions. Nevertheless, the partitioning of the global $T$-amplitudes of the wave~function also particularly suited to coupled-cluster high-level solvers, which are increasingly being used as a compact and accurate approach which enable access to larger fragment and bath spaces, and have recently been shown to be accurate in a variety of embedding contexts \cite{Shee2019, Zhu2019,doi:10.1021/acs.jctc.1c00712,PhysRevB.103.155158,PhysRevX.12.011046}.

%
%
%

\subsection{Expectation values from linear functionals} \label{sec:lin_energy_func}

In this section, we consider the use of this implicit `global' wave~function to compute expectation values of operators which commute with the Hamiltonian, and which can therefore be calculated from functionals which are linear in the wave~function.
A trivial example of this is the total energy, as its associated operator is the Hamiltonian itself.
Projecting the time-independent Schr{\"o}dinger equation from the left with the Hartree--Fock determinant~$\langle \Phi|$ and assuming intermediate normalization~$\braket{\Phi|\Psi}=1$, it follows that
\begin{equation}\label{eq:e_proj_hf}
    E[\Psi] = \braket{\Phi | \hat{H} | \Psi} ,
\end{equation}
which is linear in $\Psi$. This is the traditional energy functional of e.g. coupled-cluster theory, but holds for any state.

Since the Hamiltonian operator of an electronic structure problem involves up to two-body
interactions, only the double excitations contribute to the correlation energy
\begin{equation}\label{eq:linear_energy_func}
    E_\mathrm{corr} = E - E_\mathrm{HF} = \sum_{ij}^{N_\mathrm{occ}} \sum_{ab}^{N_\mathrm{vir}} c_{ij}^{ab} \left[2(ia|jb) - (ib|ja) \right]
    ,
\end{equation}
where the single excitations do not contribute for a Hartree--Fock reference state, due to Brillouin's theorem (though a single-body contribution can easily be included if using a non-Hartree--Fock reference).
Projecting the $C_2$-amplitudes in Eq.~\eqref{eq:linear_energy_func} according to Eq.~\eqref{eq:part_c2} enables us to partition the correlation energy into cluster contributions
as
\begin{equation}\label{eq:linear_energy_func_2}
\begin{split}
    E[\Psi^{\cl{x}}] &= E_\mathrm{HF} + \sum_\cl{x}^{N_\mathrm{frag}} E_\mathrm{corr}^\cl{x}\\
    &= E_\mathrm{HF} + \sum_{\cl{x}}^{N_\mathrm{frag}}
    \sum_{ij}^{N_\mathrm{occ}^\cl{x}} \sum_{ab}^{N_\mathrm{vir}^\cl{x}}  (\hat{P}^\cl{x} C^\cl{x}_2)_{ij}^{ab} \left[2(ia|jb) - (ib|ja) \right]
    .
\end{split}
\end{equation}

These contributions can be formulated from the individual cluster solutions, where the orbital indices run over the occupied or virtual states of each cluster, and where the projection operator of Eq.~\eqref{eq:clust_proj_op} can be constructed in this cluster orbital space, to evaluate the energy efficiently in a cost linear in the number of clusters.
This is the energy functional introduced and used in Ref.~\citenum{PhysRevX.12.011046}, and will be denoted by $E[\Psi^{\cl{x}}]$ throughout this work, to indicate that it is constructed directly from the cluster wave~function amplitudes {\em without} requiring intermediate cluster density matrices.
The use of this functional for energies is common to quantum chemical local correlation methods, for example within the PNO \cite{Riplinger2013} or cluster-in-molecules approaches \cite{doi:10.1021/jp100782u}. However, the `fragments' in these methods are then generally defined in terms of (localized) purely occupied orbitals, instead of the general `atomic-orbital-like' fragment spaces which are used in the context of quantum embedding and which in this work require the use of explicit (non-diagonal) projection operators as described.

This functional can also be used within a fragment projection of the $T$-amplitudes of an exponential representation of the high-level cluster wave~function (see Sec.~\ref{sec:part_exp_wf}). In this case, inter-cluster contributions arise from $(\hat{P}^{\cl{x}}T_1^{\cl{x}} )( \hat{P}^{\cl{y}}T_1^{\cl{y}})$ terms. This physics is neglected in the $C$-amplitude projection of above in the case of no overlap between clusters $\cl{x}$ and $\cl{y}$. 
However, in the examples in this work, the projection of the $C$- and $T$-amplitudes gave almost indistinguishable results---unsurprising given the small contribution of the overall $T_1^2$ contribution to the energy and the fact that the local portion of this is still included.
For this reason, and to reduce the computational cost of this energy functional and keep consistency with Ref.~\onlinecite{PhysRevX.12.011046}, all results in this work denoted $E[\Psi^{\cl{x}}]$ will correspond to the projection of the $C$-amplitudes of the cluster state (even for coupled-cluster high-level solvers) as shown in Eq.~\eqref{eq:linear_energy_func_2}, though the projection of $T$-amplitudes may be preferred in the future.

\subsection{Partitioned wave~function density matrices} \label{sec:proj_wfn_dms}

General expectation values, with operators that do not commute with the Hamiltonian,
require a quadratic functional of the wave~function, of
the form~$\braket{\Psi | \hat{O} |\Psi}$.
In the case of one- and two-body expectation values, $O_1$ and $O_2$, these
can also be expressed in terms of the one- and two-body reduced density-matrices
(which in turn depend quadratically on the wave~function) as
\begin{alignat}{2}
    &O_1[\Psi] &&= \braket{\Psi | \hat{O}_1 | \Psi} = 
    \mathrm{Tr} \: [\gamma_{pq} \braket{p | \hat{O}_1 | q} ] , \\
    &O_2[\Psi] &&= \braket{\Psi | \hat{O}_2 | \Psi} = 
    \mathrm{Tr} \: [\Gamma_{pqrs} \braket{pr | \hat{O}_2 | qs}] .
\end{alignat}
The one- and two-body RDMs are thus important quantities to be able to compute for expectation values beyond the energy.
Of course, with knowledge of the density matrices one can also calculate the total energy via Eq.~\eqref{eq:e_tot}. For most quantum chemical methods, where a globally stationary state is found, the result will be identical to the projected energy of Eq.~\eqref{eq:e_proj_hf}.
However, within the quantum embedding framework the wave function is partitioned into fragment contributions which are then calculated in an approximate fashion and we do not expect this equivalency to hold. Therefore, while the individual cluster wave functions are invariant to this choice of energy functional, the reconstructed full system total energy may not be.
Indeed, there is likely benefit in calculating the total energy as a symmetric (quadratic) expectation value (or equivalently from the RDMs), since the error in the energy will reduce quadratically with the error in the wave~function, rather than linearly as in the energy functional of Eq.~\eqref{eq:e_proj_hf}. 

In sections~\ref{sec:part_linear_wf} and \ref{sec:part_exp_wf} we showed how global wave~functions can be defined in terms of contributions of projected cluster wave~functions.
In principle, these global wave~functions then fully and uniquely determine their corresponding full system reduced density-matrices.
For example, the 1-RDM of a partitioned linear wave function
can be written as
\begin{eqnarray} \label{eq:FCI_1rdm}
    \gamma_{pq} & = & \braket{\Psi | \hat{c}_p^\dagger \hat{c}_q |\Psi}
    = \braket{\Phi +\Delta\Psi | \hat{c}_p^\dagger \hat{c}_q | \Phi + \Delta \Psi}\\
    & = & \gamma_{pq}^0
    + \sum_\cl{x}^{N_\mathrm{frag}} \sum_{\lambda \in \cl{x}} (\hat{P}^\cl{x} C^\cl{x})_\lambda
    \braket{\Phi | \hat{c}_p^\dagger \hat{c}_q | \Phi_\lambda } \nonumber \\
    & + & \sum_\cl{x}^{N_\mathrm{frag}} \sum_{\lambda \in \cl{x}} (\hat{P}^\cl{x} C^\cl{x})_\lambda \braket{\Phi_\lambda | \hat{c}_p^\dagger \hat{c}_q | \Phi} \nonumber \\
    & + & \sum_{\cl{x},\cl{y}}^{N_\mathrm{frag}} \sum_{\lambda \in \cl{x}} \sum_{\mu \in \cl{y}}
    (\hat{P}^\cl{x} C^\cl{x})_\lambda
    (\hat{P}^\cl{y} C^\cl{y})_\mu
    \braket{\Phi_\lambda | \hat{c}_p^\dagger \hat{c}_q | \Phi_\mu} ,
\end{eqnarray}
where $\lambda$ and $\mu$ enumerate the excitations (to all included orders) within the respective cluster in the particle-hole basis.
Note that in contrast to the democratically partitioned 1-RDM of Eq.~\eqref{eq:part_rdm1},
the 1-RDM of a partitioned wave~function contains a summation over cluster pairs,
thus incorporating simultaneous excitations within two different clusters, as indicated above and in Eq.~\eqref{eq:part_wf_rdm1}.

For a partitioned exponential wave~function the global state itself already contains
nested summations over projected cluster wave function $T$-amplitude contributions (see Eq.~\eqref{eq:global_exp_wf}). The resulting 1-RDM will thus have even higher order contributions than the $C$-amplitude projection (going up to three simultaneous cluster contributions for a high-level wave~function represented up to $T_2$, and up to five simultaneous clusters for the 2-RDM). Both approaches result in explicitly \mbox{$N$-representable} RDMs, but again the $T$-amplitude projection can be argued to have a larger number of physical non-local `cross-cluster' contributions and therefore represents a better approximation. In Sec.~\ref{sec:practice} we will detail the technical aspects of how all of these cross-cluster contributions can be efficiently computed while retaining no more than DFT scaling in the full system size (or number of clusters). Total energies computed from these $T$-amplitude projected RDMs as described in this section will be denoted $E[(\gamma, \Gamma)[\Psi^{\cl{x}}]]$ throughout this work, to signify that these are density matrices derived from the global partitioned wave~function.

\section{A simple example: Chlorine dimer} \label{sec:example}

Before describing specific implementational details in Sec.~\ref{sec:practice}, we consider a simple example to corroborate the claims in the work so far, before a more extensive set of test systems is considered in Sec.~\ref{sec:results}. We consider the binding of Cl$_2$ in a minimal basis. The choice of a minimal basis is intentional, since this means that we can easily compare to exact FCI results, and that the same (L{\"o}wdin orthogonalized) atomic orbital space can be used for the fragments in all approaches. For larger basis sets, the democratically partitioned density matrix and cumulant approaches require all orbitals to be spanned by the fragment spaces. However the projected wave~function approaches require only (at least) the occupied space to be assigned to fragments for the projector to be complete. Therefore for more realistic basis sets, the fragment spaces would generally be chosen according to different criteria, with the fragment spaces of the former needing to grow with basis size. Using a minimal basis therefore avoids this issue and allows the different methods to be compared on the same footing. The DMET cluster space in this example therefore consists of 18 electrons in 9 fragment orbitals plus a single DMET bath orbital (due to the minimal basis) in each of the two symmetrically equivalent clusters, while the full space consists of 34 electrons in 18~orbitals.

\begin{figure}[!htb]
\centering
    \includegraphics[width=1\linewidth]{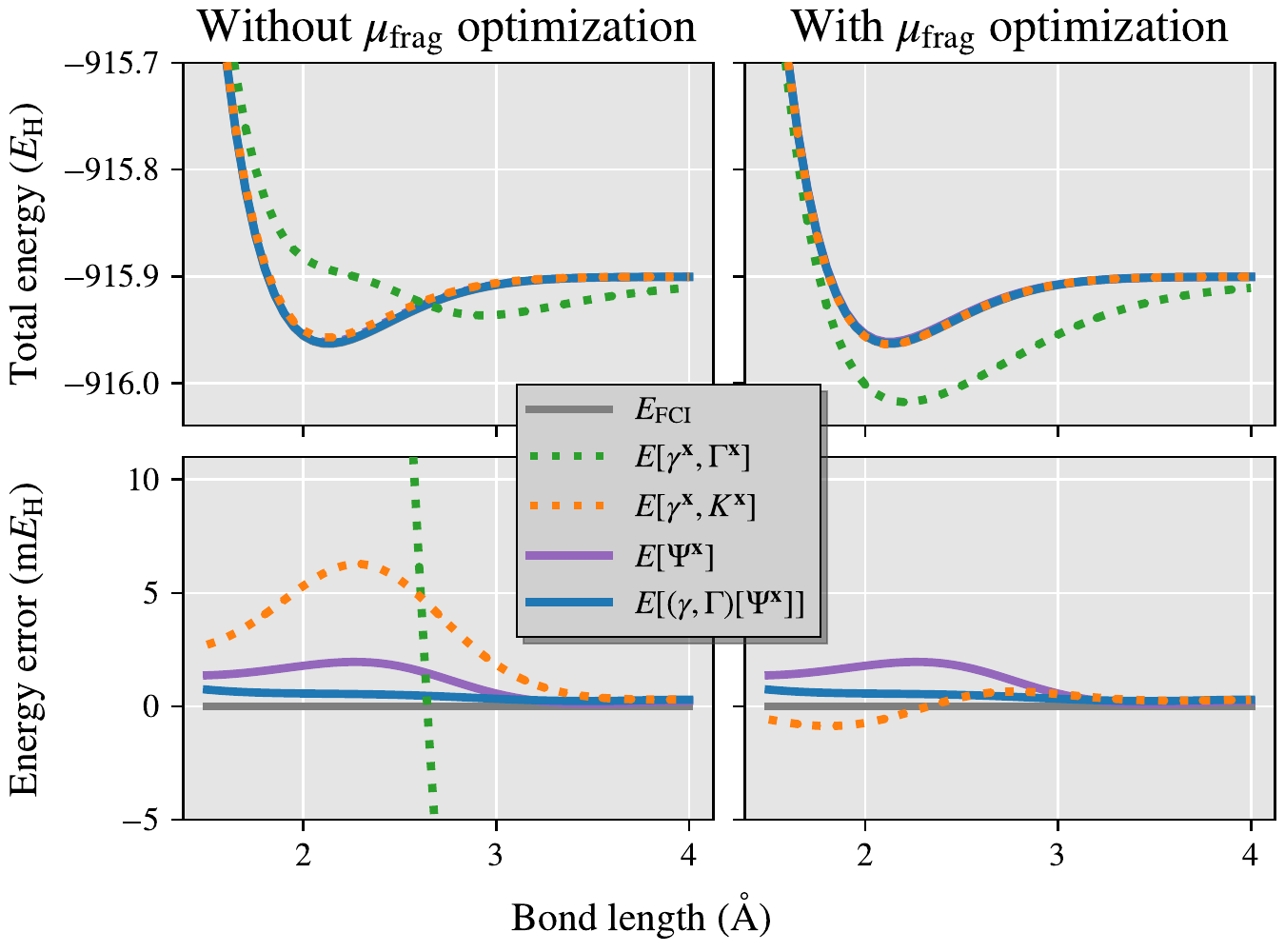}
    \caption{Comparison of different proposed DMET energy expressions for the binding of Cl$_2$ in a STO-6G basis. Each cluster consists of the atomic orbitals of one atom, with a single DMET bath orbital on the other, with all clusters solved exactly. Right column results include a fragment chemical potential optimization to constrain the total number of electrons, while this is omitted in the left column.}
    \label{fig:cl2_energy}
\end{figure}

The interacting cluster Hamiltonian is solved exactly via FCI \cite{KNOWLES198975}, and we compare the four different energy expressions outlined to far in Fig.~\ref{fig:cl2_energy}: the democratically-partitioned RDM energy common to DMET approaches to date ($E[\gamma^{\cl{x}}, \Gamma^{\cl{x}}]$ from Sec.~\ref{sec:dem_part}), the democratically partitioned cumulant energy ($E[\gamma^{\cl{x}}, K^{\cl{x}}]$ from Sec.~\ref{sec:part_cumulant}), the linear energy functional ($E[\Psi^{\cl{x}}]$ from Sec.~\ref{sec:lin_energy_func}) and the energy from the RDMs of the $T$-amplitude projected global wave~function ($E[(\gamma, \Gamma)[\Psi^{\cl{x}}]]$ from Sec.~\ref{sec:proj_wfn_dms}). We include calculations with and without a global chemical potential optimization to ensure that the right number of electrons are conserved for the full system 1-RDM, but no further optimization of a correlation potential is considered here. When used (right column of Fig.~\ref{fig:cl2_energy}), this fragment chemical potential, $\mu_{\mathrm{frag}}$, is added to the one-electron cluster Hamiltonian within the fragment space only, and optimized such that the density-matrix traces to the correct number of electrons \cite{Wouters2016} (these $\mu_{\mathrm{frag}}$-optimized results are the same as shown in Fig.~\ref{fig:cl2_intro}). All results are subject to identical locality approximation of DMET, whereby the full system is represented only by the two wave functions in the cluster subspaces, allowing for a faithful comparison of the quality of the reconstructed expectation values. An example Python script, showing how these calculations can be performed using \texttt{PySCF} and \texttt{Vayesta} can be found in
the SI.

We find in Fig.~\ref{fig:cl2_energy} that the democratically partitioned RDM energy is changed significantly by the optimization of the fragment chemical potential, but in neither case are the results reasonable, with unphysical results without $\mu_{\mathrm{frag}}$-optimization, and substantial overbinding of the dimer when this is included. These results are also found in many other systems (barring hydrogen dimers and chains, where we find the democratically partitioned density matrix energy accurate, as has been noted elsewhere \cite{Knizia2013,Wouters2016,Fertitta2019,PhysRevB.102.085123}). The democratically partitioned cumulant approach is also changed by the $\mu_{\mathrm{frag}}$-optimization, but is much less sensitive to this, with both results already significantly improved over the democratically partitioned RDM energy. The chemical potential optimization further drops the non-parallelity error of the democratically partitioned cumulant approach from $\sim 6\,\mathrm{m}E_\mathrm{H}$ to just over $1\,\mathrm{m}E_\mathrm{H}$ in this system.

We also show the energies derived from partitioned wave~functions over the system, either calculating the energy from a linear or quadratic functional (via the RDMs) of the wave function probability amplitudes. It is shown that these are also very accurate, which we can ascribe to two properties of these functionals. The first is that they are intrinsically $N$-representable, ensuring that they fulfil the physical constraints of being derivable from a wave function. As a consequence, they necessarily correspond to the correct, integer number of electrons, and therefore obviates the necessity of chemical potential optimization which has no effect (therefore also substantially reducing the cost of the calculations). The second rationalization for the particularly good performance of the partitioned wave~function RDM energy is due to the introduction of contributions from products of wave~function amplitudes from different clusters, coupling the cluster solutions together. Furthermore, for the $E[(\gamma,\Gamma)[\Psi^{\cl{x}}]]$ energy, these contributions result in an error in the energy which is rigorously quadratic rather than linear in the wave function error, and introduces a variationality (for a variational method) into the results which results in the smallest non-parallelity error of 0.5m$E_H$.


\begin{figure}[!htb]
\centering
    \includegraphics[width=1\linewidth]{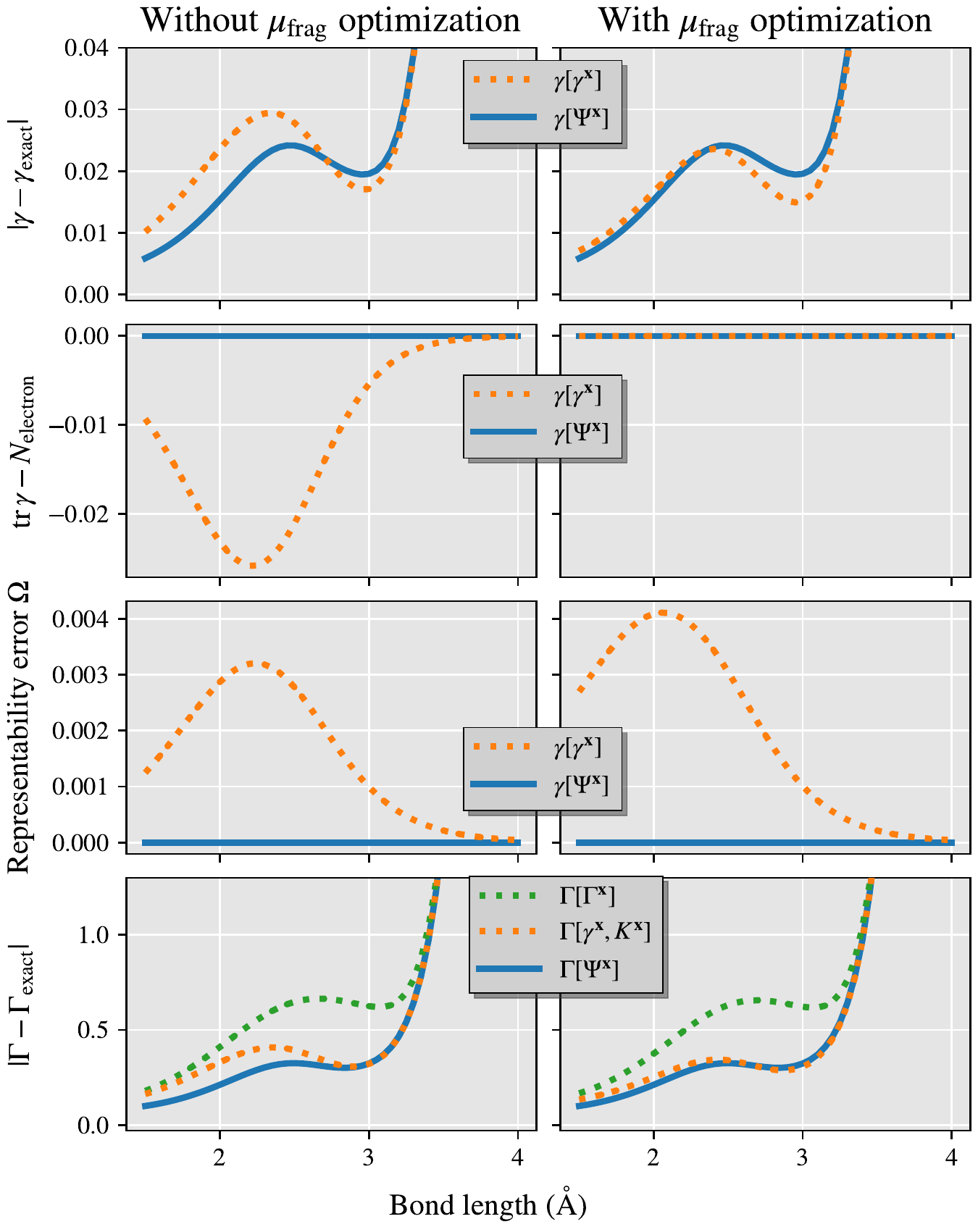}
    \caption{Analysis of the 1- and 2-RDMs from direct democratic partitioning (dotted lines) compared to RDMs constructed from partitioning of the wave function (solid line), for the same Cl$_2$ system as Fig.~\ref{fig:cl2_energy}. The results are shown without (left) and with (right) chemical potential optimization over the DMET clusters. Top to bottom rows show the errors in the norm, trace and $N$-representability conditions (Eq.~\ref{eq:Nrep_error}) of the 1-RDM for the different approaches, with the bottom row showing the norm error in the 2-RDM including the democratically partitioned cumulant approach.}
    \label{fig:cl2_dms}
\end{figure}

In order to analyze the different approaches in more detail, in Fig.~\ref{fig:cl2_dms} we compare the 1- and 2-RDMs resulting from the democratically partitioned RDMs ($\gamma[\gamma^{\cl{x}}]$ for the 1-RDM, and $\Gamma[\Gamma^{\cl{x}}]$ or $\Gamma[\gamma^{\cl{x}},K^{\cl{x}}]$ for the 2-RDM from the direct democratic partitioning and the cumulant partitioning respectively). We also compare to the RDMs constructed from the partitioned wave~function expansion ($\gamma[\Psi^{\cl{x}}]$ and $\Gamma[\Psi^{\cl{x}}]$). We consider the norm of the errors compared to the exact full system 1- and 2-RDMs, as well as the electron number error from the trace of the 1-RDMs. We also introduce a measure of the $N$-representability error of the 1-RDM, $\Omega$, which sums the deviation from allowable occupations of the system orbitals as 
\begin{equation}
    \Omega = \sum_{i: n_i < 0} |n_i| + \sum_{i: n_i > 2} |n_i - 2| , \label{eq:Nrep_error}
\end{equation}
where~$n_i$ are the eigenvalues of~$\gamma$.

While the overall norm error of the 1-RDMs seem relatively similar between the approaches (especially after chemical potential optimization), it can be seen that there is still a significant electron number error (without chemical potential optimization), and $N$-representability error in this quantity. While the chemical potential optimization fixes the electron number error, it is actually found to {\em increase} the $N$-representability error. In contrast, the partitioned wave function approach ensures that the electron numbers are exactly fulfilled even without a chemical potential, and that the $N$-representability error is strictly eliminated. For the 2-RDM errors, it is clear that significant improvements are found by going to a democratically partitioned cumulant compared to the 2-RDM directly, further indicating that it is the non-cumulant error in the 2-RDM which is contributing to the significant energy errors in the standard DMET energy functional.

\section{From concepts to practice} \label{sec:practice}

While the previous sections describe the principle and numerical advantages behind the reconstruction of a global wave function, in practice we will often want this to be implicit as the reconstruction of the global correlated wave function will in general be prohibitive in cost. Instead, we want to directly compute the RDMs or expectation values of interest from this state in low-polynomial time with respect to system size, and without combining the wave function amplitudes into an explicit global state. This allows for the cluster wave~functions to remain distributed in their fragmented cluster representations. We focus in this section on the efficient computation of the 1- and 2-RDMs, from which all two-body properties can be computed, with the principle of an implicit global wave function demonstrated in Sec.~\ref{sec:proj_wfn_dms} underlying this construction. 

We restrict ourselves here to a projection of the $T$-amplitudes of the global state, represented in an exponential form. This is consistent with the approach in the previous section, will maximize the number of non-local combinations of cluster solutions, and is a natural choice for coupled-cluster high-level solvers (although as explained in Sec.~\ref{sec:part_exp_wf}, other solvers can be cast in this form). Furthermore, we focus on an efficient implementation of the cluster solver truncated at the $T_2$-amplitudes, i.e. CCSD (or MP2), which is a cost-effective high-level cluster solver for {\em ab initio} systems \cite{Shee2019, Zhu2019,doi:10.1021/acs.jctc.1c00712,PhysRevB.103.155158,PhysRevX.12.011046}, yet is sufficiently computationally inexpensive for us to rigorously demonstrate the convergence of these functionals with respect to bath size in Sec.~\ref{sec:results}. Extensions to other high-level cluster solvers (e.g. FCI) can proceed via Eq.~\ref{eq:FCI_1rdm}, and will be described in future work. \toadd{We stress here that this approach to compute expectation values from a global wave~function with alternative (i.e. strongly correlated) solvers would not lead to an increase in scaling with respect to full system size compared to the algorithm shown. Indeed, the fact that coupled-cluster RDMs involve high-degree polynomials of the cluster solution variables (i.e. $T_n$ amplitudes) results in the formal scaling with system size being generally larger than solvers such as FCI (where the RDMs are just quadratic in the cluster $C_n$ variables).}

Explicit reconstruction of the global CCSD $T$-amplitudes (which we call the `global $T_2$ algorithm') would require a memory overhead scaling as $\mathcal{O}(N^4)$ and computational scaling of $\mathcal{O}(N^5)$ and $\mathcal{O}(N^6)$ for the 1- and 2-RDMs respectively (where $N$ is a measure of full system size) via the usual CCSD equations \cite{doi:10.1063/1.460915}. We reduce this scaling down to $\mathcal{O}(N^2)$ cost in memory and $\mathcal{O}(N^3)$ time for the 1-RDM construction via direct use of the cluster amplitudes and construction of appropriate intermediates without introducing any additional approximations. This is described in Sec.~\ref{sec:1rdm_part_wf}, which we denote the `distributed $T_2$ algorithm', and is crucial to ensure that this step is a sub-leading scaling compared to the initial mean-field calculation and for applicability to large systems. For the construction of the 2-RDM and properties derived from it, the requirement of a similar quadratic scaling with system size necessitate the introduction of a further approximation, which is described in Sec.~\ref{sec:2rdm_part_wf}, with rigorous validation of this further approximation demonstrated in Sec.~\ref{sec:results}.

%
%
%
%


\subsection{One-RDM from cluster wave functions} \label{sec:1rdm_part_wf}

In order to construct density-matrices at the CCSD level, both wave~function $T$-amplitudes and the Lagrange multipliers (or $\Lambda$-amplitudes) are required. These are optimized within each cluster after the $T$-amplitudes are found, giving rise to a set of $\Lambda^{\cl{x}}$ amplitudes for each cluster, $\cl{x}$. The principle of reconstruction of a `global' set of $\Lambda$-amplitudes follows symmetric projection of the occupied indices of each cluster $\Lambda^{\cl{x}}$-amplitudes onto the fragment space (defined analogously to Eqs.~\ref{eq:part_t1} and \ref{eq:part_t2}), and a summation over the cluster solutions, as
\begin{align}
    & \lambda_{a}^{i} = \sum_\cl{x}^{N_\mathrm{frag}} ( \hat{P}^\cl{x} \Lambda_1^\cl{x} )_{a}^{i} \label{eq:part_l1}, \\
    & \lambda_{ab}^{ij} = \sum_\cl{x}^{N_\mathrm{frag}} ( \hat{P}^\cl{x} \Lambda_2^\cl{x} )_{ab}^{ij} \label{eq:part_l2}
    .
\end{align}

All expectation values in coupled cluster (including the 1-RDM considered here) involve sums of expressions which are polynomials in the $T$-amplitudes, and up to linear in the $\Lambda$-amplitudes. The central idea of the efficient `distributed' approach is to avoid the explicit reconstruction of the global $T$- and $\Lambda$-amplitudes entirely, and instead iterate over tuples of cluster solutions, directly summing in their corresponding contribution to the 1-RDM. The number of clusters which needs to be looped over in order to include all cross-cluster contributions to the desired expectation value (or 1-RDM in this instance) depends on the maximum order of the polynomial in the expression, i.e. the total number of $T$- or $\Lambda$-amplitudes which are contracted together.

For the 1-RDM, this maximum order is formally three, corresponding to contractions of the type $\Lambda_1 T_1^2$. This corresponds to the maximum rank of simultaneous `cross-cluster' information in the construction of this quantity (i.e. triplets of cluster solutions are required). However, we can reduce this by first explicitly constructing the global $T_1$- and $\Lambda_1$-amplitudes from the cluster solutions according to Eqs.~(\ref{eq:part_t1}, \ref{eq:part_l1}), since this will only require a scaling of $\mathcal{O}(N^2)$ and $\mathcal{O}(N^3)$ in memory and time respectively, which is an acceptable overhead compared with the initial mean-field calculation. This is in contrast to the analogous explicit construction of the global $T_2$-amplitudes, which we have already argued is prohibitive. The fragment projection for $T_1$ and $\Lambda_1$ is therefore performed in each cluster, and then projected back into the full system MO space and summed over all clusters. Furthermore, we store these global $T_1$- and $\Lambda_1$-amplitudes on every memory partition of a distributed memory parallel calculation. The key result is that we now only need to loop over cluster tuples up to the maximum rank of distributed (i.e. $T_2$ or $\Lambda_2$) amplitudes in the 1-RDM expression, which is only quadratic, resulting from an $L_2 T_2$ contraction. This requires only an $\mathcal{O}(N^2)$ double summation over pairs of cluster amplitudes (which can be further reduced asymptotically down to $\mathcal{O}(N)$ as described later). If certain contractions require a $T_1$ to be contracted with a cluster distributed $T_2$, then the global $T_1$ amplitudes can be projected into the required cluster space of the $T_2$.

We demonstrate this distributed amplitude algorithm for the example of the CCSD contribution to the occupied--occupied block of the 1-RDM, given as
%
\begin{equation}\label{eq:dm1_oo}
\begin{split}
    \Delta\gamma_{ij} &= 
    \Delta\gamma_{ij}^{(\Lambda_1,T_1)} + \Delta\gamma_{ij}^{(\Lambda_2,T_2)} \\
    &= \sum_{a}^{N_\mathrm{vir}} \lambda_a^i t_j^a
    + \sum_k^{N_\mathrm{occ}}  \sum_{ab}^{N_\mathrm{vir}}  \lambda_{ab}^{ik} \left( 2 t_{jk}^{ab} - t_{jk}^{ba}\right) 
    .
\end{split}
\end{equation}
The first term of this contribution can readily be calculated from the explicitly combined global singles amplitudes
with $\mathcal{O}(N^3)$ scaling.
%
%
For the second term, we can calculate the contribution arising from a specific cluster pair ($\cl{x}, \cl{y}$), by using the $\Lambda_2$- and $T_2$-amplitudes corresponding to these clusters, as
\begin{equation}\label{eq:dm1_oo_l2t2}
\begin{split}
    \Delta\gamma^\cl{xy}_{i_\ci{x} j_\ci{y}} = 
    &\sum_{k_\ci{x}}^{N_\mathrm{occ}^{\cl{x}}}
    \sum_{a_\ci{x} b_\ci{x}}^{N_\mathrm{vir}^{\cl{x}}}
    \sum_{k_\ci{y}}^{N_\mathrm{occ}^{\cl{y}}}
    \sum_{a_\ci{y} b_\ci{y}}^{N_\mathrm{vir}^{\cl{y}}}
    S_{a_\ci{x} a_\ci{y}}^\cl{xy}
    S_{b_\ci{x} b_\ci{y}}^\cl{xy}
    S_{k_\ci{x} k_\ci{y}}^\cl{xy} \\
    & [\hat{P}^\cl{x} \Lambda^\cl{x}_2]_{a_\ci{x} b_\ci{x}}^{i_\ci{x} k_\ci{x}}
    \left( 2 [\hat{P}^\cl{y} T^\cl{y}_2]_{j_\ci{y} k_\ci{y}}^{a_\ci{y} b_\ci{y}}
    - [\hat{P}^\cl{y} T^\cl{y}_2]_{j_\ci{y} k_\ci{y}}^{b_\ci{y} a_\ci{y}}\right)
    ,
\end{split}
\end{equation}
where we use the notation~$i_\ci{x}$ ($a_\ci{x}$) to indicate an occupied (virtual) orbital of cluster~$\cl{x}$,
and $S^\cl{xy}$ to represents the overlap matrix between the occupied (or virtual, as indicated by the subscript indices) orbitals of cluster~$\cl{x}$ with cluster~$\cl{y}$, given by
%
\begin{equation}\label{eq:overlap_xy}
    S_{p_\ci{x} p_\ci{y}}^\cl{xy} = \sum_{\alpha}^{N_\mathrm{ao}} C_{\alpha p_\ci{x}}^\cl{x}
    \left[ \sum_\beta^{N_\mathrm{ao}} S_{\alpha\beta} C_{\beta p_\ci{y}}^\cl{y}
    \right]
    .
\end{equation}
These $S^\cl{xy}$ matrices can be precomputed in $\mathcal{O}(N^3)$ for all cluster pairs, where the columns of $C^\cl{x}$ are the relevant orbitals of cluster~$\cl{x}$ in the AO representation, and $S$ is the AO~overlap matrix.
%
%
We then sum the 1-DM contributions of Eq.~\eqref{eq:dm1_oo_l2t2} over all cluster pairs, according to
%
%
\begin{equation}\label{eq:dm1_oo_sum_xy}
\begin{split}
    \Delta\gamma_{ij}^{(\Lambda_2,T_2)}
    = \sum_\cl{x}^{N_\mathrm{frag}} \sum_{i_\ci{x}}^{N_\mathrm{occ}^\cl{x}} R_{i i_\ci{x}}^\cl{x}
    \left[ \sum_\cl{y}^{N_\mathrm{frag}}  \sum_{j_\ci{y}}^{N_\mathrm{occ}^\cl{y}} R_{j j_\ci{y}}^\cl{y}
    \Delta\gamma^{\cl{xy}}_{i_\ci{x} j_\ci{y}} \right]
    ,
\end{split}
\end{equation}
where $R$ represents the overlap between cluster MOs and MOs of the full system, i.e.
%
%
\begin{equation}\label{eq:overlap_x_mf}
    R_{p p_\ci{x}}^{\cl{x}} = \sum_{\alpha}^{N_\mathrm{ao}} C_{\alpha p} \left[ \sum_\beta^{N_\mathrm{ao}} S_{\alpha\beta} C_{\beta p_\ci{x}}^\cl{x}
    \right]
    .
\end{equation}

To analyse the computational scaling of the distributed amplitude algorithm, we note that the number of
clusters, AOs, and full system MOs grows linearly with the system size, whereas the number of cluster orbitals per cluster remains constant.
As a result, the calculation of all cluster pair contributions~\eqref{eq:dm1_oo_l2t2} scales as $\mathcal{O}(N^2)$, while Eqs.~(\ref{eq:overlap_xy}--\ref{eq:overlap_x_mf})
can be evaluated in $\mathcal{O}(N^3)$ time, if intermediates are formed as indicated by the brackets.
An analogous loop over cluster pairs can also be used to calculate the virtual--virtual and mixed occupied-virtual blocks of the 1-RDM,
resulting in an overall~$\mathcal{O}(N^3)$ scaling algorithm.
%
%

\begin{figure}
    \centering
    \includegraphics[width=1\linewidth]{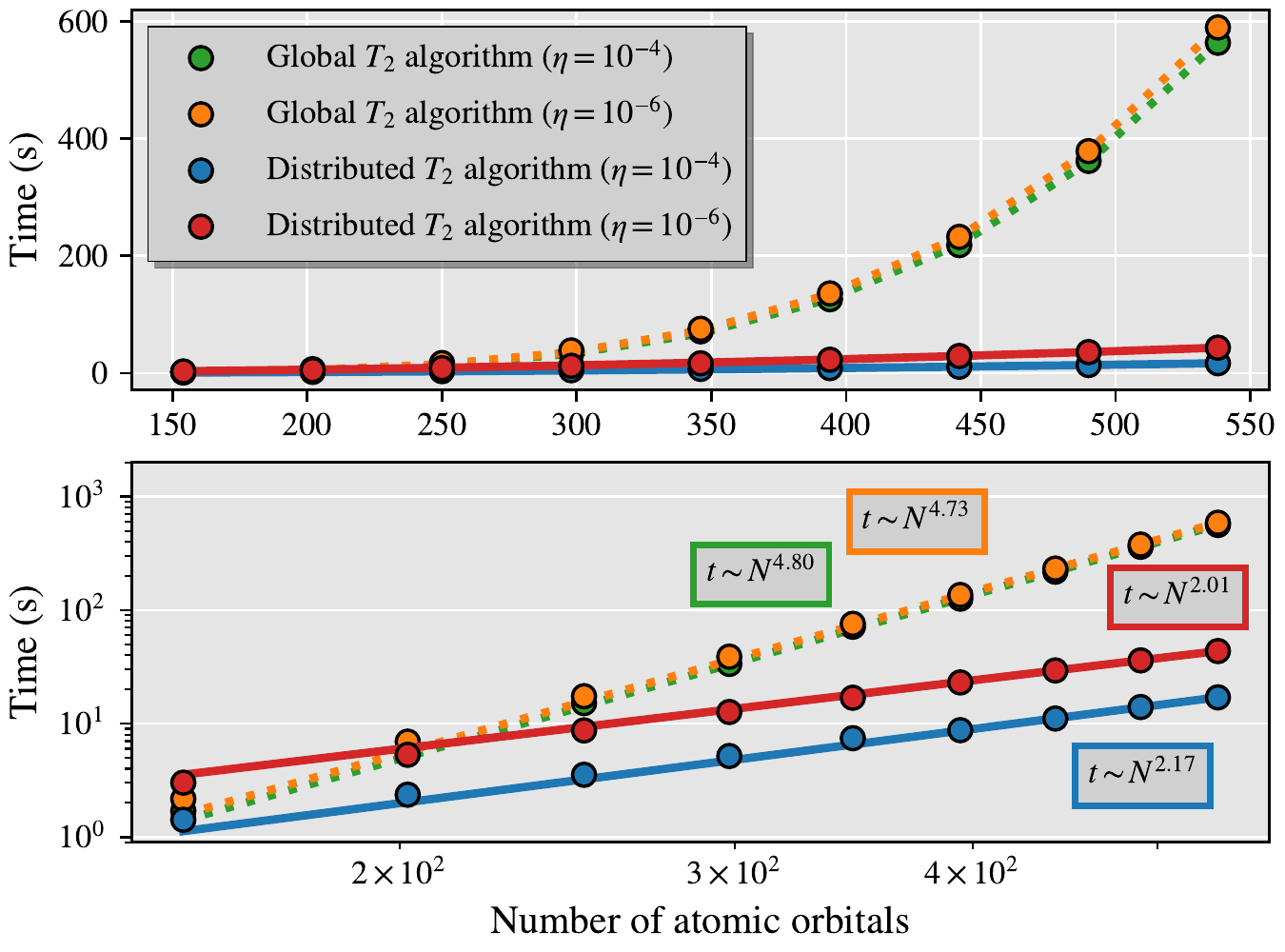}
    \caption{Timings for the construction of the full system \mbox{1-RDM} from the partitioned wave function for two different bath truncation thresholds, $\eta = 10^{-4}$ and $\eta = 10^{-6}$, for the `global' and `distributed' $T_2$ algorithms.
    Increasing the bath size has little
    effect on the global~$T_2$ algorithm, whereas it leads to an increase in the prefactor (but not the exponent) of the distributed~$T_2$ algorithm.
    }
    \label{fig:dm1_scaling}
\end{figure}

In Fig.~\ref{fig:dm1_scaling} we compare the scaling of the global amplitude and distributed amplitude algorithms, for a series of alkanes between hexane (C$_6$H$_{14}$) and C$_{22}$H$_{46}$ in a cc-pVDZ basis and atomic IAO fragmentation. To extend beyond a minimal DMET bath size, we augment the bath expansion with `bath natural orbitals' (BNO), which can be systematically enlarged in order to describe the beyond-mean-field coupling of the fragment to the environment at the level of approximate MP2 theory \cite{PhysRevX.12.011046}. This bath construction is similar to pair natural orbitals (PNO)---though traditional PNOs are applied only for the virtual coupling of a localized electron pair \cite{Edmiston1966,Meyer1973,Riplinger2013}. The completeness of this bath space is controlled by a threshold $\eta$, where the bath becomes complete as $\eta \rightarrow 0$, and is limited to just the traditional DMET bath as $\eta$ becomes large. More details on the construction and motivation of this bath expansion can be found in Ref.~\onlinecite{PhysRevX.12.011046}.

From the timing data for the 1-RDM construction in Fig.~\ref{fig:dm1_scaling} for two different bath space sizes, we fit the polynomial $t(N) = a N^b$ to the last six data points of each curve to determine the overall computational scaling. 
For the global amplitude algorithm, this is close to the expected scaling with $b\sim5$,
whereas for the distributed amplitude algorithm
we find an exponent close to 2 indicating a quadratically-scaling algorithm, instead of the expected exponent of 3. This indicates the small prefactor of the efficient $\mathcal{O}(N^3)$ operations of Eqs.~(\ref{eq:overlap_xy}--\ref{eq:overlap_x_mf}), whereas the evaluation of Eq.~\eqref{eq:dm1_oo_l2t2}, although scaling as $\mathcal{O}(N^2)$, has a significant larger prefactor and dominates the computational costs for the tested system sizes. 

Despite this favorable scaling, we employ a number of additional techniques, to further improve the efficiency of this algorithm. These include a singular value decomposition of each cluster overlap space, in order to find the most compact domain for describing the inter-cluster physics. In the large system limit, this also naturally leads to a $\mathcal{O}(N)$ scaling in the 1-RDM construction, due to the fact that each cluster will only have appreciable overlap with $\mathcal{O}(1)$ other clusters. In addition, we consider efficiency gains that can be made from $\mathbf{k}-$point sampled periodic systems, effective distributed memory parallelism, and a compressed representation of the projected amplitudes. These technical improvements in the 1-RDM construction are detailed in Appendix A.

\subsection{`In-cluster' approximation to the two-RDM} \label{sec:2rdm_part_wf}

In principle, an analogous approach to above can be used to recover any expectation value from the implicit partitioned wave function, with the scaling determined by the maximum number of products of $T_n$ or $\Lambda_n$ in any of the terms, where $n>1$. For the 2-RDM, this increases to three (from two for the 1-RDM), which would necessitate looping over triplets of cluster solutions (noting that a similar SVD approach to screen the overlap would still asymptotically reduce to linear scaling). Note that including the $T_1$ and $\Lambda_1$ contributions, the 2-RDM would include simultaneous contributions of up to five cluster amplitudes, but by explicitly reconstructing these $T_1$ and $L_1$ amplitudes globally, only looping over three clusters~($\cl{x}, \cl{y}, \cl{z}$) would be required. However, including the projections back to the full space and noting the $\mathcal{O}(N^4)$ storage of the global 2-RDM, these scalings are likely prohibitive if we aim for the quantum embedding to maintain a scaling with full system size which is no more than mean-field theory.

To return to (at most) a $\mathcal{O}(N^3)$ scaling approach, an additional approximation is thus required.
As a first step, we split the 2-RDM into products of 1-RDMs and an (approximate) cumulant, $\approxcumulant$, defined according to Eq.~\eqref{eq:approx_cumulant_1}.
The non-cumulant part can be treated without further approximations, using the efficient calculation of the 1-RDM
presented in the previous section. This ensures that all the cross-cluster contributions in the non-cumulant part are exactly included from an implicit valid wave function.
The cumulant part, on the other hand, will be treated in an `in-cluster approximation', meaning that
all products of wave~function amplitudes are taken within a single cluster at a time, rather than explicitly including triplets of clusters.
%
%
%

In order to illustrate this, let us consider a contribution to the cumulant, $\delta\approxcumulant$,
which involves a $(\Lambda_2 T_2^2)$ triple product of double (de)excitation amplitudes.
If evaluated exactly according to the implicit partitioned wave function, this would require contributions from all triplets~$(\cl{x}, \cl{y}, \cl{z})$ of clusters, as
%
%
\begin{equation}\label{eq:cumulant_xyz}
\begin{split}
    \delta \approxcumulant^{\cl{xyz}}_{i_\ci{y} a_\ci{z} j_\ci{y} b_\ci{z}} =
    &\sum_{k_\ci{x} l_\ci{x}}
    \sum_{c_\ci{x} d_\ci{x}}
    \sum_{c_\ci{y} d_\ci{y}}
    \sum_{k_\ci{z} k_\ci{z}}
    S^\cl{xy}_{c_\ci{x} c_\ci{y}}
    S^\cl{xy}_{d_\ci{x} d_\ci{y}}
    S^\cl{xz}_{k_\ci{x} k_\ci{z}}
    S^\cl{xz}_{l_\ci{x} l_\ci{z}}\\
    & [\hat{P}^\cl{x} \Lambda^\cl{x}_2]_{c_\ci{x} d_\ci{x}}^{k_\ci{x} l_\ci{x}}
    [\hat{P}^\cl{y} T^\cl{y}_2]_{i_\ci{y} j_\ci{y}}^{c_\ci{y}d_\ci{y}} [\hat{P}^\cl{z} T^\cl{z}_2]_{k_\ci{z} l_\ci{z}}^{a_\ci{z} b_\ci{z}}
    .
\end{split}
\end{equation}
In the in-cluster approximation, this contribution is replaced by
\begin{equation}\label{eq:cumulant_incluster}
    \delta\approxcumulant^{\cl{x}}_{i_\ci{x} a_\ci{x} j_\ci{x} b_\ci{x}} =
    \sum_{k_\ci{x} l_\ci{x}}
    \sum_{c_\ci{x} d_\ci{x}}
    [\hat{P}^\cl{x} \Lambda^\cl{x}_2]_{c_\ci{x} d_\ci{x}}^{k_\ci{x} l_\ci{x}} [T^\cl{x}_2]_{i_\ci{x} j_\ci{x}}^{c_\ci{x} d_\ci{x}} [T^\cl{x}_2]_{k_\ci{x} l_\ci{x}}^{a_\ci{x} b_\ci{x}}
    ,
\end{equation}
which only contains amplitudes from cluster~$\cl{x}$.
Note that Eq.~\eqref{eq:cumulant_incluster} only contains a single fragment-space projector, which is applied to the $\Lambda$-amplitude,
compared to Eq.~\eqref{eq:cumulant_xyz}, where every amplitude is projected.
In general, the number of projectors in a term has to agree to the number of nested loops over clusters the given term is summed over. This ensures preservation of a core principle of the embedding, that we guarantee an exact result, free from double-counting, as the bath space increases to completeness, as long as the sum of the fragment spaces spans the entire occupied space (Eq.~\ref{eq:comp_wfn_proj}).
Instead of projecting the $\Lambda$-amplitude, we could have also chosen to project the first or second $T$-amplitude in Eq.~\eqref{eq:cumulant_incluster}, resulting in a different in-cluster approximation, which nevertheless becomes exact in the full-bath limit.
However, projecting the $\Lambda$-amplitudes yields a simple recipe,
since all contributions to the 2-RDM cumulant are at most linear in $\Lambda_1$ or $\Lambda_2$. Some limited experimentation with other options for the projected amplitudes did not seem to significantly change the results.
The only term this prescription does not work for is the $\Lambda^0$ terms of coupled-cluster expressions, which results in a MP2-like contribution
\begin{equation}
    \delta \approxcumulant_{iajb} = t_{ij}^{ab} + t_i^a t_j^b - \frac{1}{2}\left[ t_{ij}^{ba} + t_i^b t_j^a \right]
    ,
\end{equation}
in which case we use projected $T$-amplitudes, according to
\begin{equation}
\begin{split}
    \delta \approxcumulant_{i_\ci{x} a_\ci{x} j_\ci{x} b_\ci{x}}^\cl{x} &=
    [\hat{P}^\cl{x} T^\cl{x}_2 ]_{i_\ci{x} j_\ci{x}}^{a_\ci{x} b_\ci{x}}
    + [\hat{P}^\cl{x} T_1^\cl{x} ]_{i_\ci{x}}^{a_\ci{x}} [ T_1^\cl{x} ]_{j_\ci{x}}^{b_\ci{x}}\\
    &- \frac{1}{2} \left(
    [\hat{P}^\cl{x} T^\cl{x}_2 ]_{i_\ci{x} j_\ci{x}}^{b_\ci{x} a_\ci{x}}
    + [\hat{P}^\cl{x} T_1^\cl{x}]_{i_\ci{x}}^{b_\ci{x}} [ T_1^\cl{x} ]_{j_\ci{x}}^{a_\ci{x}}
    \right)
    .
\end{split}
\end{equation}
We note here that the `global' $T_1$- and $\Lambda_1$-amplitudes as defined in the last section are {\em not} used in this `in-cluster' cumulant approximation, ensuring that only a single cluster summation is used for all terms. Relaxation of this constraint will be investigated in the future.

This `in-cluster' approximation to the two-body cumulant reduces its computation to sums over single cluster wave function contributions, avoiding the computational effort associated with the \toadd{fully non-local approach (as used for the 1-RDM in Sec.~\ref{sec:1rdm_part_wf}). The computational effort to compute the two-body cumulant contributions is now negligible compared to the 1-RDM construction, which requires contributions from all pairs of clusters.} However, we do not want to build the whole full-system cumulant (which would require a prohibitive $\mathcal{O}(N^4)$ scaling in memory), but rather calculate expectation values directly from this sum over the independent cluster wave functions. We can therefore \toadd{directly} construct the energy functional with this `in-cluster' approximated cumulant, which we denote $E[(\gamma, K^*)[\Psi^{\cl{x}}])$ in this work (the asterisk denoting the `in-cluster' approximation to the two-body cumulant), as
\begin{equation}\label{eq:e_incluster}
\begin{split}
E[(\gamma,\approxcumulant^*)[\Psi^\cl{x}]] &= E_\mathrm{HF}[\gamma^0] + \sum_{pq} F_{pq}[\gamma^0] \Delta\gamma_{qp}[\Psi^\cl{x}] \\
&+  \frac{1}{2} \sum_{\cl{x}}^{N_\mathrm{frag}} \sum_{pqrs}^{N_\mathrm{cl}^\cl{x}} (pq|rs) {\approxcumulant^*}[\hat{P}^\cl{x} \Psi^\cl{x}] )_{pqrs}
,
\end{split}
\end{equation}
where ${\approxcumulant^*}[\hat{P}^\cl{x} \Psi^\cl{x}]$ indicates that each term in the expression for the (approximate) cumulant of Eq.~\eqref{eq:approx_cumulant_1} is computed from products of contributions from the {\em same} cluster solution (the `in-cluster approximation'). This allows the two-body contributions from each cluster to be computed entirely independently, as opposed to $\Delta \gamma [\Psi^{\cl{x}}]$ which is assumed to contain all relevant products of different cluster solutions (see Sec.~\ref{sec:1rdm_part_wf}).
Other static two-body expectation values can be derived using this `in-cluster' two-body cumulant analogously. In Sec.~\ref{sec:spin_corr_fns} and Appendix B, we go beyond energetics to consider the two-point instantaneous spin correlation function, which can be computed in this same framework via reconstruction of a global 1-RDM, with cluster-local two-body contributions via the `in-cluster' approximated cumulant.

While this approach bears similarities to the democratically partitioned cumulant for two-body properties of Sec.~\ref{sec:part_cumulant} (specifically that the disconnected part of the 2-RDM is treated separately to include cross-cluster contributions), it has significant advantages. In particular, the 1-RDM part is treated in a fully $N$-representable way, with contributions from all triplets of different cluster wave functions, as well as the fact that the projections acting on the occupied spaces mean that fragments only need to span the complete occupied space to ensure convergence to `in-method' exactness as the bath spaces of each fragment are enlarged. We consider this a critical aspect to allow for practical extension of the embedding to realistic basis sets.

Although this `in-cluster' approximation is justified primarily by numerical expediency, it is still a controllable approximation in the same fashion as the embedding approximation itself, via an expansion of the bath space, guaranteeing systematic improvability of the approximation to exactness. However, it is also clear that the `in-cluster' approximation to the two-body cumulant does result in a loss of strict $N$-representability of the resulting 2-RDM. This is because it is no longer derivable from a full system wave function, since there is no single set of `global' $T$- and $\Lambda$-amplitudes used for all expressions. This can be seen in the first set of results in Fig.~\ref{fig:cl2_intro} which presents all energy functionals used in this work. While the error in the energy functional which builds the RDMs from the explicit full system partitioned wave function ($E[(\gamma, \Gamma)[\Psi^{\cl{x}}]]$) is variational with respect to the true solution, this is lost when the low-scaling `in-cluster' approximation is used for the two-body cumulant ($E[(\gamma, K^*)[\Psi^{\cl{x}}]]$).

In the next section, we will rigorously benchmark these different approaches to computing energies and non-energetic spin--spin correlation functions, and analyze their convergence in realistic molecular and extended systems with systematic expansion of the bath space of each cluster. Surprisingly, we find that energies derived with the `in-cluster' cumulant approximation exhibit an improved convergence to the exact complete bath limit of the embedding theory, due to more favorable cancellation of errors.

\section{Results} \label{sec:results}

We now turn to a larger selection of systems, to consider the convergence of these different energy estimators to their `in-method' exact limit, as the size of the interacting bath expansion of each cluster is enlarged. For this expansion, we consider the `cluster-specific bath natural orbitals' (BNO) introduced in Ref.~\onlinecite{PhysRevX.12.011046}, which can be used as a systematic way to increase the orbital space controlled by a single threshold parameter $\eta$, and we will use CCSD as a cluster solver throughout. As $\eta$ is reduced towards zero, the bath space increases in size towards completeness, capturing longer-ranged and higher-energy coupling of the fragment to its environment. The use of CCSD as a solver allows us to access bath sizes sufficient to converge to `in-method' exactness, and we will move to consider stronger-correlation solvers in future work. We will initially compare the energies from the three introduced energy functionals which are derived from reconstructed wave~function forms. As a recap, these are:
\begin{itemize}
\item $E[\Psi^{\cl{x}}]$: Introduced in Ref.~\onlinecite{PhysRevX.12.011046} and Sec.~\ref{sec:lin_energy_func}, this is the cheapest approach to energy computation of Eq.~\eqref{eq:linear_energy_func_2}, requiring no communication between independent additive cluster contributions.
\item $E[(\gamma, \Gamma)[\Psi^{\cl{x}}]]$: Described in Sec.~\ref{sec:proj_wfn_dms}, this computes the energy from the exact density matrices from the reconstructed wave function partitioned over the clusters. This scales worse than the mean-field for the exact reconstructed $\Gamma$, and rapidly becomes the dominant cost in the calculation. It will primarily be used for reference.
\item $E[(\gamma, K^*)[\Psi^{\cl{x}}]]$: Described in Sec.~\ref{sec:2rdm_part_wf}, this approximates the cumulant contribution to the 2-RDM via the `in-cluster' approximation, which ensures low-scaling with system size for this non-factorizable two-body energy contribution. However, the 2-RDM loses strict $N$-representability. The 1-RDM contributions are still $N$-representable from the partitioned wave function according to the efficient algorithm of Sec.~\ref{sec:1rdm_part_wf}.
\end{itemize}

We do not compare the energies from the density matrix derived expectation values of Sec.~\ref{sec:expect_from_dm}, due to the different constraints on the fragment spaces imposed. For the wave function derived expectation values above, all of these will converge to in-method exactness as the bath space increases as long as the combined fragment spaces span the occupied space, while the density matrix expectation values require the combined fragment spaces to span all degrees of freedom. This allows us to consider simple minimal-size atomic intrinsic atomic orbital fragments, with the only convergence criteria now being the bath size controlled by $\eta$.

\subsection{Molecular results: W4-11 test set} \label{sec:W4}

We first consider the comparison of these energy functionals for the W4-11 test set of 152 small molecules consisting of first and second-row elements, which covers a broad range of bonding character, with static as well as dynamic correlated physics present, as well as a mix of high-spin, low-spin and open-shell systems \cite{KARTON2011165}. In a cc-pVDZ basis, canonical CCSD calculations on the full test set is tractable, to obtain a `ground-truth' total energy for all molecules, and we use ROHF/UHF and UCCSD for open-shell cases. Using the information from the corresponding Hartree--Fock calculation, we then fragment each molecule into a minimal basis set of IAOs on each atom, which we use as the fragments of each clusters, which ensures that the total occupied space of the full system is spanned by the fragments \cite{Knizia2013IAO} (we omit the Beryllium dimer test system, due to a lack of virtual orbitals in the default IAO basis). \toadd{The choice of single atom fragments simplifies their selection, but future work to consider larger (or smaller) disjoint fragmentation causes no conceptual or practical difficulties.} To complete the cluster space of each atom, we augment this with the DMET bath space (which must be at most equal in size to the fragment space), and then with the BNOs defined from a given threshold, $\eta$. Smaller $\eta$ values correspond to larger bath spaces, with $\eta>1$ just returning to the smallest (DMET) bath size. Each cluster is then solved independently at the level of CCSD (restricted or unrestricted), and the total energies reconstructed from the three wave~function-based energy functionals provided above, with no self-consistency or chemical potential optimization.

\begin{figure}
    \centering
    \includegraphics[width=1\linewidth]{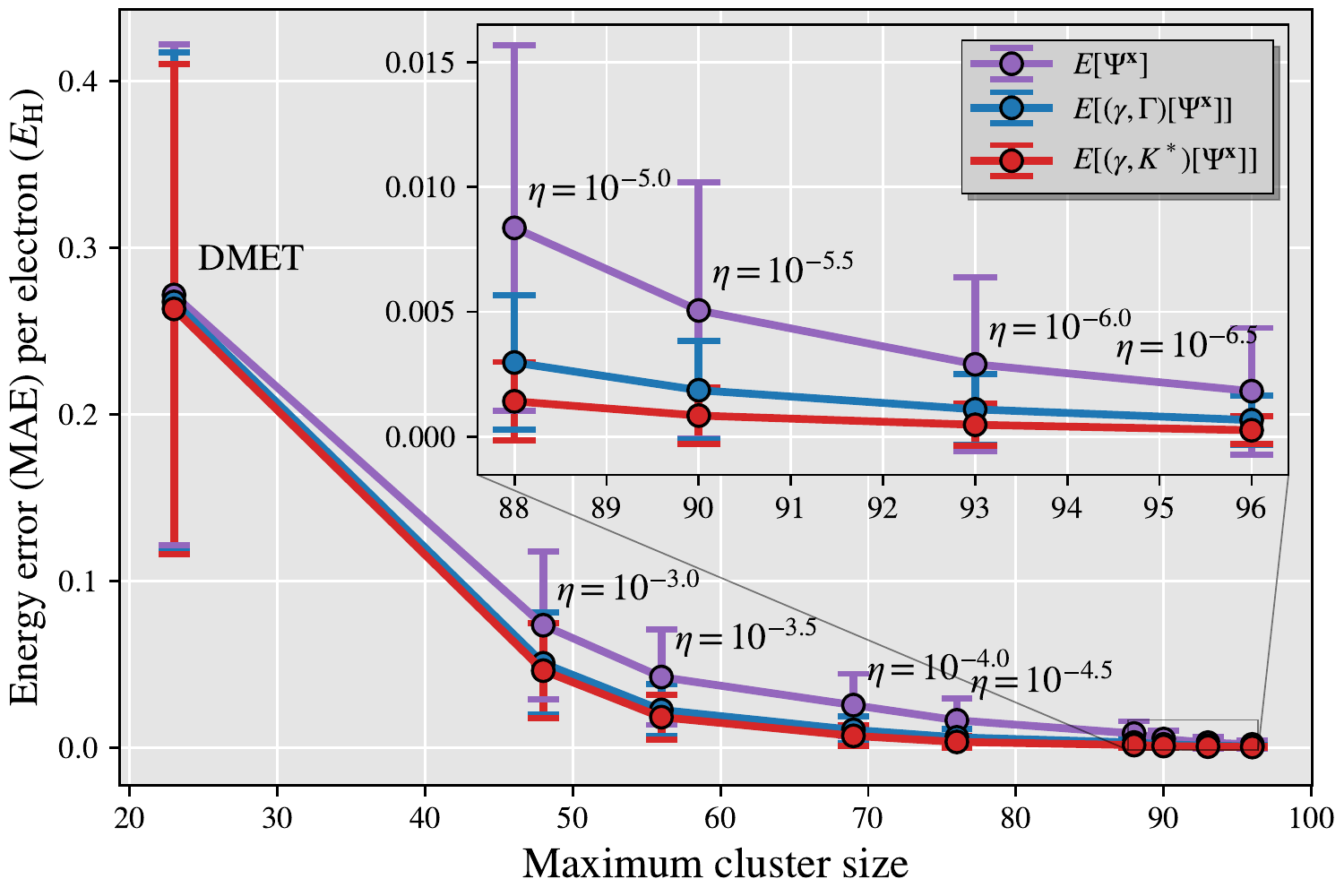}
    \caption{Convergence of the reconstructed system energy from individual cluster solutions to the canonical CCSD with respect to the maximum cluster size for any atomic fragment in any molecule in the W4-11 test set. Cluster sizes are controlled by the bath threshold ($\eta$), starting from just the DMET bath, for the wave~function derived energy functionals. Points indicate the mean absolute error (MAE) per electron in the total energy, while error bars show the standard deviation over the test set. Inset shows a magnified version of the larger bath threshold results.}
    \label{fig:w4_11_conv}
\end{figure}

\begin{figure}
    \centering
    \includegraphics[width=1\linewidth]{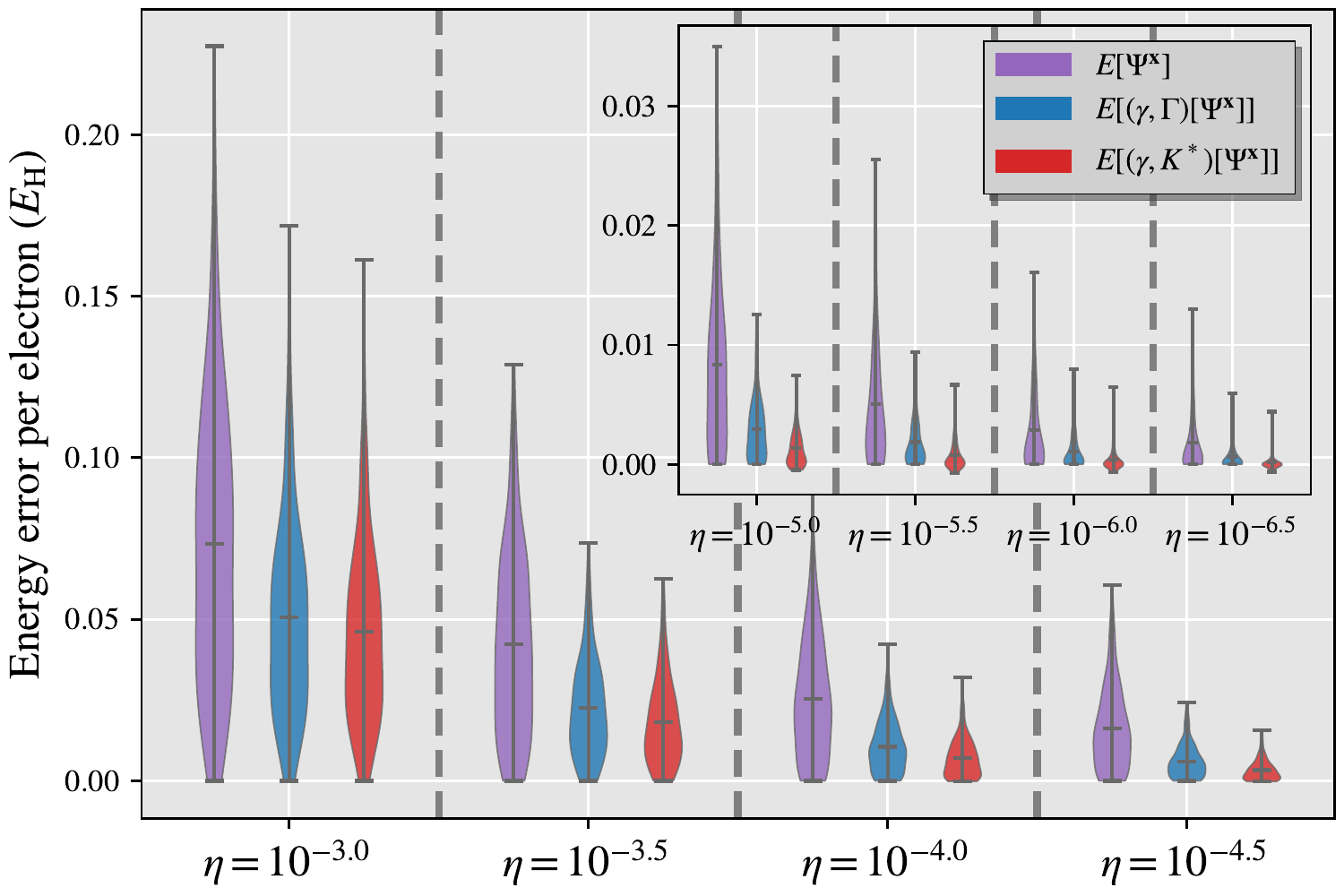}
    \caption{Violin plots showing the distribution of total energy errors per electron over the W4-11 test set of molecules compared to canonical CCSD calculations, for the different partitioned wave~function energy estimators. This is performed with a bath expansion of each IAO atomic fragment for different values of $\eta$, which corresponds to a maximum cluster size as indicated in Fig.~\ref{fig:w4_11_conv}. Total distribution ranges show the minimum and maximum errors over the test set.}
    \label{fig:w4_11_violin}
\end{figure}

These results are shown in Fig.~\ref{fig:w4_11_conv}, with their absolute per electron energy errors aggregated for different values of $\eta$ in Fig.~\ref{fig:w4_11_violin}. It is clear that the bath expansion of these IAO atomic fragments results in a rapid, monotonic and systematic convergence of all energy estimates to in-method exactness, obtaining agreement with canonical CCSD calculations to within tight errors. The largest of these errors, in both MAE and standard deviation over the different molecules in the test set, arises from the $E[\Psi^{\cl{x}}]$ functional, which is to be largely expected, since this functional does not have any `cross-cluster' contributions to the energy, relying on a sum over independent energy contributions from the fragments. A significant improvement in the rate of convergence can be found from the $E[(\gamma, \Gamma)[\Psi^{\cl{x}}]]$ functional, which contains many products of contributions between different cluster solutions in its construction. Furthermore, this energy is rigorously $\Psi$-derivable, and is therefore expected to exhibit an error which is quadratic in the implicitly reconstructed wave function error. Though the energy is not strictly variational due to the non-variationality of coupled-cluster theory, we also find it to be fully variational compared to the canonical CCSD in all cases.

However, somewhat surprisingly, the $E[(\gamma, K^*)[\Psi^{\cl{x}}]]$ energy functional converges to in-method exactness at the fastest rate, obtaining the least energy error across the test set for each bath truncation. This approach was motivated as an efficient approximation to the $E[(\gamma, \Gamma)[\Psi^{\cl{x}}]]$ energy, but due to favourable cancellation of errors between the approximated cumulant energy contribution and the inherent locality approximation in the bath truncation, appears to be the preferred approach (as well as being particularly efficient). For intermediate cluster sizes, the reduction in per-electron MAE can be nearly an order of magnitude compared to the $E[\Psi^{\cl{x}}]$ energy. Due to the approximation in this cumulant non-variational total energies can result (compared to the canonical result), however the appearance of these non-variational results are very rare, and only observed for already tightly converged energies. This is because the bath truncation itself is a variational approximation (since this defines a subspace of the full variational freedom of the cluster), which explains the favourable cancellation of errors, and maintenance of generally variational results, even in this approximation which breaks strict $N$-representability in the 2-RDM.

\subsection{Solid state results: Diamond} \label{sec:SolidState}

A key feature of the embedding approach proposed is the applicability to both molecular and periodic systems, enabled by ensuring all steps scale no worse than the initial Hartree--Fock calculation. We therefore also consider the convergence of structural properties of crystalline fcc diamond in an all-electron cc-pVTZ basis and a 5$\times$5$\times$5 k-point sampling. At 7500 orbitals and 1500 electrons, this is well beyond the capabilities of canonical CCSD for a direct comparison, but we can observe the relative convergence of the different energy estimators, and the resulting equation of state.

The supercell is split into 250 individual embedded cluster problems, comprising of the minimal basis IAOs of a single Carbon atom as the fragment (6 orbitals), the DMET bath space (also 6 orbitals), and the BNOs whose size depends on the $\eta$ parameter. The size of the resulting clusters vary between 50 and $\sim250$ orbitals. Due to the translational and rotational symmetry, all embedded problems are symmetry-equivalent, and therefore only a single cluster needs to be solved, with the linear energy functional trivially exploiting this symmetry, while a slightly more involved exploitation of the symmetry to avoid explicitly solving multiple cluster problems in the accumulation of the long-range RDMs for other expectation values is described in Appendix A. \toadd{We note that exploitation of this symmetry information is not required and only used here for numerical efficiency, with identical results achieved by computing all cluster contributions independently.}

\begin{figure}[t]
    \centering
    \includegraphics[width=1\linewidth]{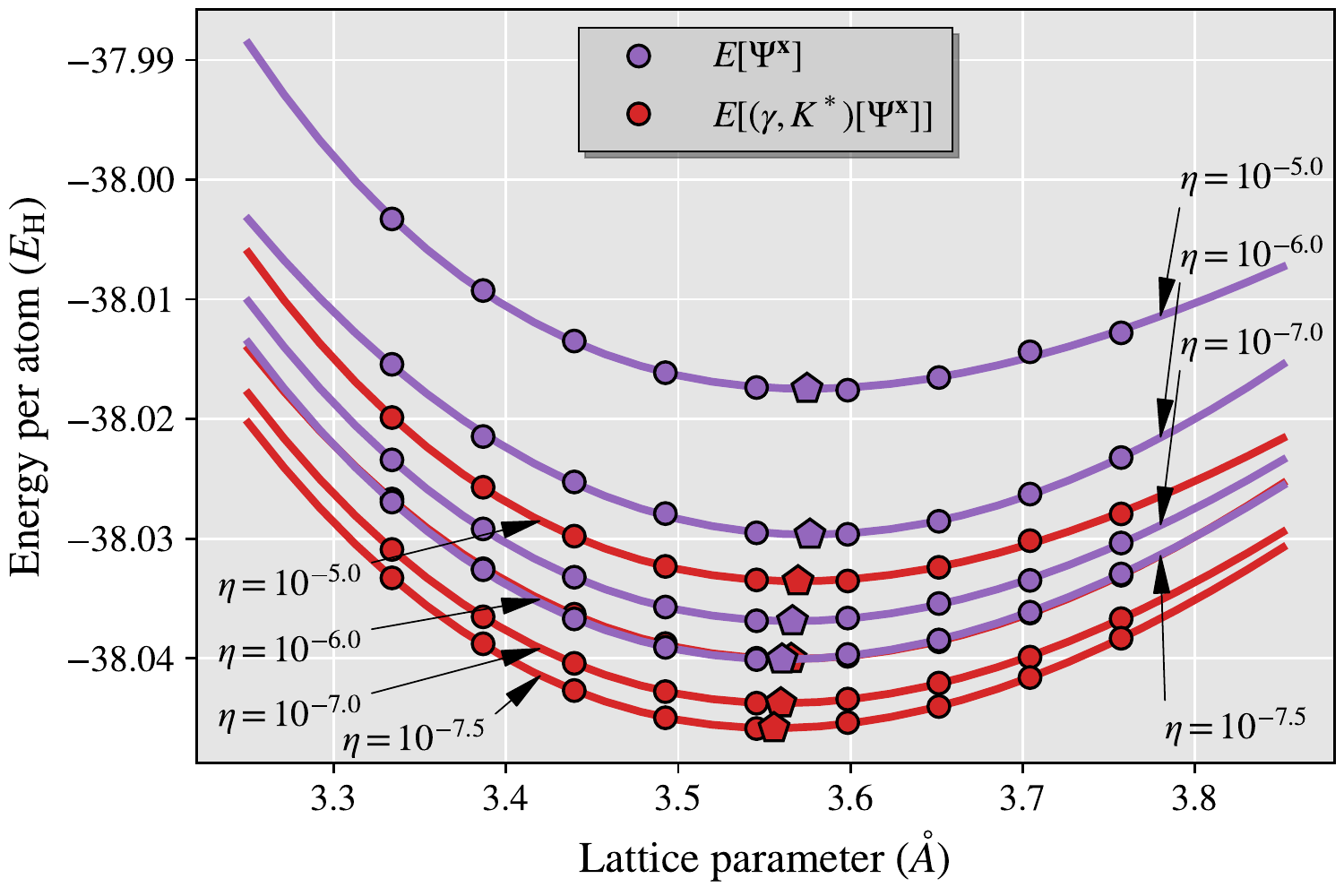}
    \caption{Equation of state for diamond, using both the linear energy functional ($E[\Psi^{\cl{x}}]$, purple) and the RDM energy functional with in-cluster approximated cumulant ($E[(\gamma, K^*)[\Psi^{\cl{x}}]]$, red). Data from different lattice parameters are fit to a Birch-Murnaghan equation of state for different values of $\eta$, systematically expanding the size of the cluster space. $5\times5\times5$ ${\bf k}$-points are used in the supercell, with an all-electron cc-pVTZ basis, while small energy corrections for the basis set superposition error are also included. The calculated energies are shown as circles, the fits are solid lines and the equilibrium lattice parameter from the fit is indicated by pentagons.}
    \label{fig:diamond_eos}
\end{figure}

Results are presented in Fig.~\ref{fig:diamond_eos} for the linear energy functional $E[\Psi^{\cl{x}}]$ and the `in-cluster' cumulant functional, $E[(\gamma, K^*)[\Psi^{\cl{x}}]]$, for a range of lattice parameters of the unit cell. The $E[(\gamma, \Gamma)[\Psi^{\cl{x}}]]$ functional is omitted due to its unfavourable scaling to these system sizes. The input to our {\tt Vayesta} code to generate these results is given in the SI, where periodic integrals for the system are generated from {\tt PySCF}\cite{McClain2017,Sun2017}. 
Changing the lattice vectors leads to an effective change in the basis set coverage at different cell sizes (since the same number of basis functions span different volumes of the supercell, changing the effective completeness of the overall basis). To compensate for this, we correct all energies for this `basis set superposition error' (BSSE) via an adaptation of the counterpoise correction method\cite{Boys1970} as described in Ref.~\onlinecite{PhysRevX.12.011046}. The BSSE per atom is estimated by computing two independent molecular sytems, the first for single carbon atom (30 basis functions) and the second for a single atom in addition to the basis functions of all carbon atoms within a $3\times3\times3$ supercell (1620 basis functions in total), sufficient to describe the changing basis set coverage for the carbon atom under consideration. 
The difference between the two energies is the estimated BSSE error, with this correction changing across the shown lattice range by $\sim 7.2$ m$E_\textrm{H}$, which slightly lowers the energy at large lattice parameters relative to the more compressed cells, leading to a small expansion of the equilibrium volume.

\begin{figure}
    \centering
    \includegraphics[width=1\linewidth]{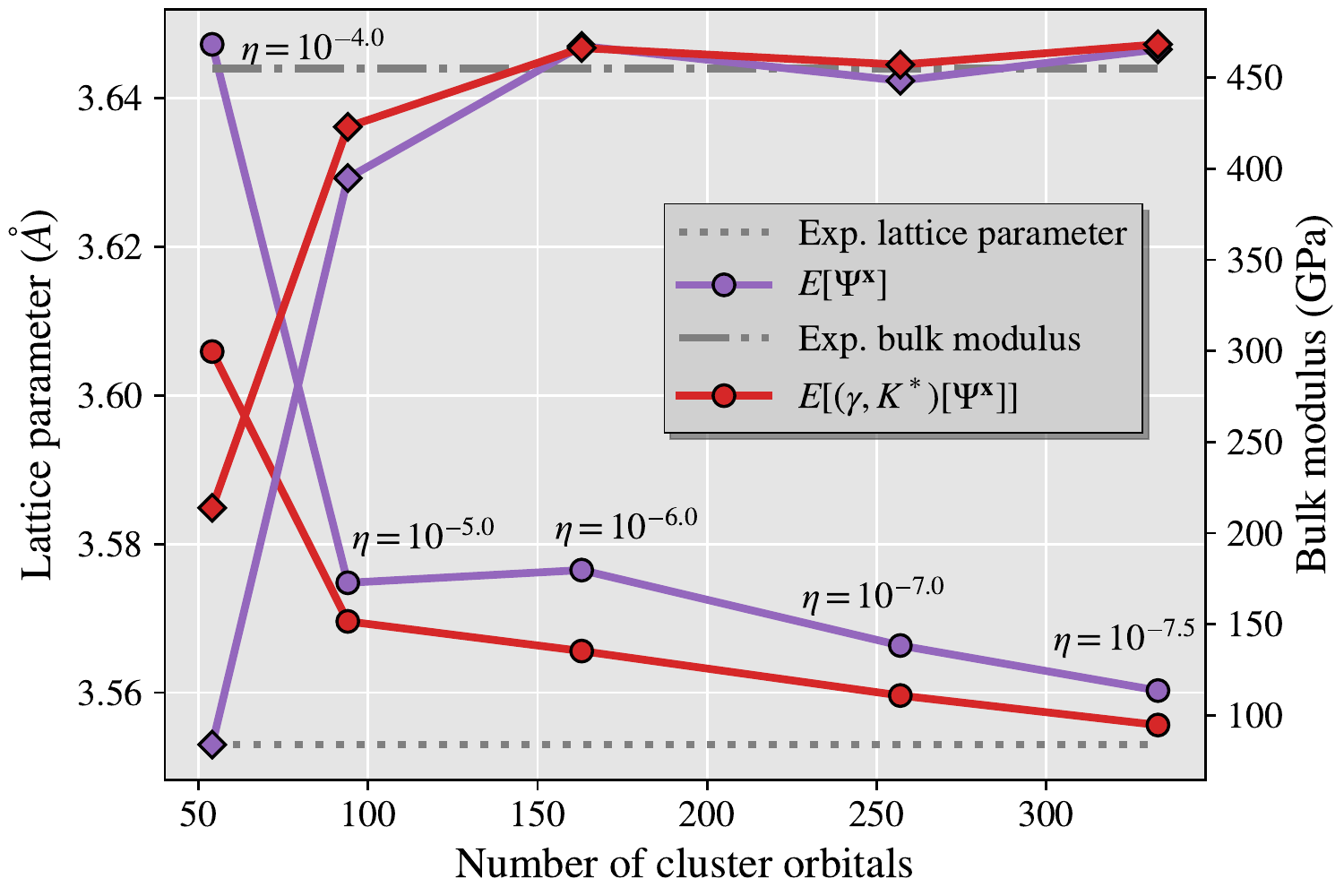}
    \caption{Convergence of equilibrium lattice parameters (circles, left $y$-axis) and bulk modulus (diamonds, right $y$-axis) for a $5\times5\times5$ diamond supercell as the total number of cluster orbitals increases (controlled by the $\eta$ value). Values are estimated from a Birch-Murnaghan fit to the equation of state shown in Fig.~\ref{fig:diamond_eos}. Results are shown for both the linear energy functional ($E[\Psi^{\cl{x}}]$) and the RDM energy functional with in-cluster approximated cumulant ($E[(\gamma, K^*)[\Psi^{\cl{x}}]]$), with experimental values shown as horizontal lines \cite{Schimka2011,McClain2017}.}
    \label{fig:diamond_conv}
\end{figure}

The equation of state for the two different energy functionals and BNO bath threshold ($\eta$) values are shown in Fig.~\ref{fig:diamond_eos}. These demonstrate the systematic improvement in the total energy as $\eta$ decreases. However, while the total energy is not fully converged, we can fit these equations of state to a Birch-Murnaghan form\cite{Birch1947} about approximately $\pm 6\%$ of the equilibrium value, in order to estimate structural properties for each of these curves, and gauge their convergence. We consider the convergence of the equilibrium lattice parameter and bulk modulus in Fig.~\ref{fig:diamond_conv}, showing convergence to experimental values, corrected for zero-point vibrational effects \cite{Schimka2011,McClain2017}. We find a small yet consistently improved convergence for both of these properties from the in-cluster approximated cumulant energy functional at each cluster size compared to the linear energy functional derived properties, in keeping with the conclusions from the convergence of the energy in the molecular test set of Fig.~\ref{fig:w4_11_violin}. Our most accurate result ($\eta=10^{-7.5}$, resulting from a single CCSD calculation on $\sim 330$ orbitals), gives a bulk modulus which agrees with experiment by less than $10$GPa, while the equilibrium lattice parameter is in error by only $\sim0.002$\AA. By way of comparison, literature values from commonly used exchange-correlation functionals of DFT for structural properties give equilibrium lattice constants in error by between $+0.02$\AA (PBE) and $-0.015$\AA~(HSESol) compared to experimental values\cite{Schimka2011}, an order of magnitude greater than these results.


\subsection{Beyond energetics: Non-local spin correlation functions} \label{sec:spin_corr_fns}

While accurate energetics are important, they are far from the only non-local expectation values of interest, and we should also gauge the validity of the cumulant-approximated 2-RDM approach for other two-body expectation values. We therefore evaluate instantaneous spin--spin correlation functions between pairs of atoms ($A$, $B$), given by
\begin{equation}\label{eq:def_ssz}
\begin{split}
    \braket{\hat{S}_z^A \hat{S}_z^B} &= \frac{1}{4}\sum_{ijkl} P_{ij}^A P_{kl}^B 
    \left( \Gamma_{ijkl}^{\alpha\alpha} -\Gamma_{ijkl}^{\alpha\beta} -\Gamma_{ijkl}^{\beta\alpha} + \Gamma_{ijkl}^{\beta\beta}\right) \\
    &+ \frac{1}{4} \sum_{ijk} P_{ik}^A P_{jk}^B 
    \left(\gamma_{ij}^{\alpha} + \gamma_{ij}^{\beta}\right)
    ,
\end{split}
\end{equation}
where $P^A$ is a projector onto an appropriately chosen subspace representing the atom $A$. We define these atomic projectors via symmetrically (L{\"o}wdin) orthogonalized atomic orbitals~(SAO) for each atom, as given in Appendix B.
Note that while the total spin--spin correlations of the full system, $\braket{S_z^2} = \sum_{AB} \braket{\hat{S}_z^A \hat{S}_z^B}$,
are necessarily zero in a spin-restricted formalism (though not in unrestricted), the local contribution from a given atom pair is generally non-zero, indicating the conditional spin-density between different points in the system.

\begin{figure}
    \centering
    \includegraphics{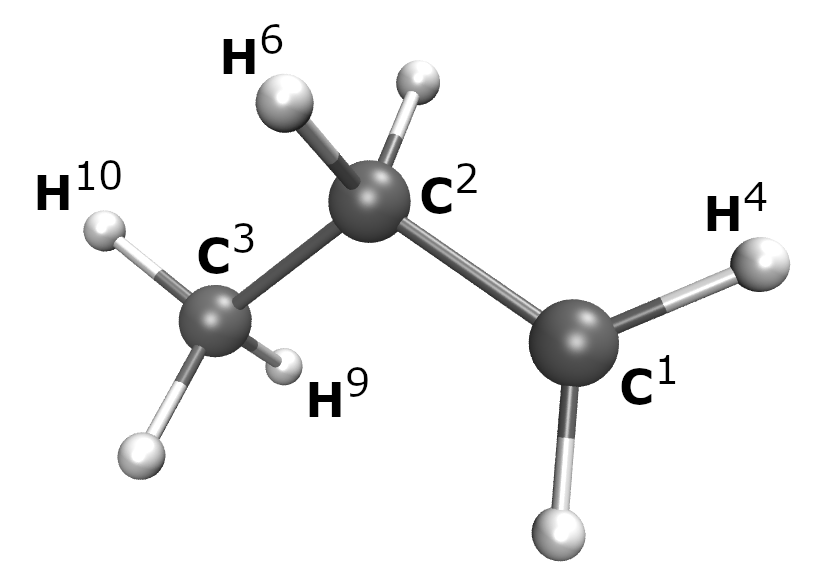}
    \caption{Labels of the atoms in the $n$-propyl radical used in this work.}
    \label{fig:propyl_labels}
\end{figure}

We adapt the fragmentation and BNO bath expansion procedure as detailed in Sec.~\ref{sec:W4} to use a spin-broken UHF reference, spin-dependent bath orbitals, a UCCSD solver, and generalizations of all expectation value accumulation in a unrestricted formalism. We can then consider the convergence of the spin--spin correlation function with bath size for the $n$-propyl radical~(shown in Fig.~\ref{fig:propyl_labels}) in the cc-pVTZ basis between the radical C$^1$ position (atom~$A$) and other atoms in the system (atom~$B$).
In order to compare results from the traditional democratically partitioned 2-RDM of Sec.~\ref{sec:dem_part} ($\Gamma[\Gamma^{\cl{x}}]$), the democratically partitioned cumulant of Sec.~\ref{sec:part_cumulant} ($\Gamma[\gamma^{\cl{x}}, K^{\cl{x}}]$) and the in-cluster cumulant approximated 2-RDM derived from the partitioned wave function of Sec.~\ref{sec:2rdm_part_wf} ($\Gamma[(\gamma, K^*)[\Psi^{\cl{x}}]]$), we choose a complete fragmentation of the system via atomic IAO $\oplus$ PAO fragments, ensuring that the full virtual space of the system is also spanned by the fragmentation. This is a necessary condition for the democratically partitioned approaches to approach exactness as the bath space becomes complete, but is a sufficient but not necessary condition for the $\Gamma[(\gamma, K^*)[\Psi^{\cl{x}}]]$, which only requires fragments to span the occupied space. However, choosing this common fragmentation will allow for results to be compared on the same footing, despite requiring a larger fragment space than would be most efficient for the wave~function partitioned RDMs. In appendix A, we detail how an efficient in-cluster approximated form for $\Gamma[(\gamma, K^*)[\Psi^{\cl{x}}]]$ can be utilized for this spin-spin correlation function, ensuring that all four-index contractions are performed within each cluster, to maintain cubic scaling with system size.


%
%
%
%
\begin{figure}
    \centering
    \includegraphics[width=1\linewidth]{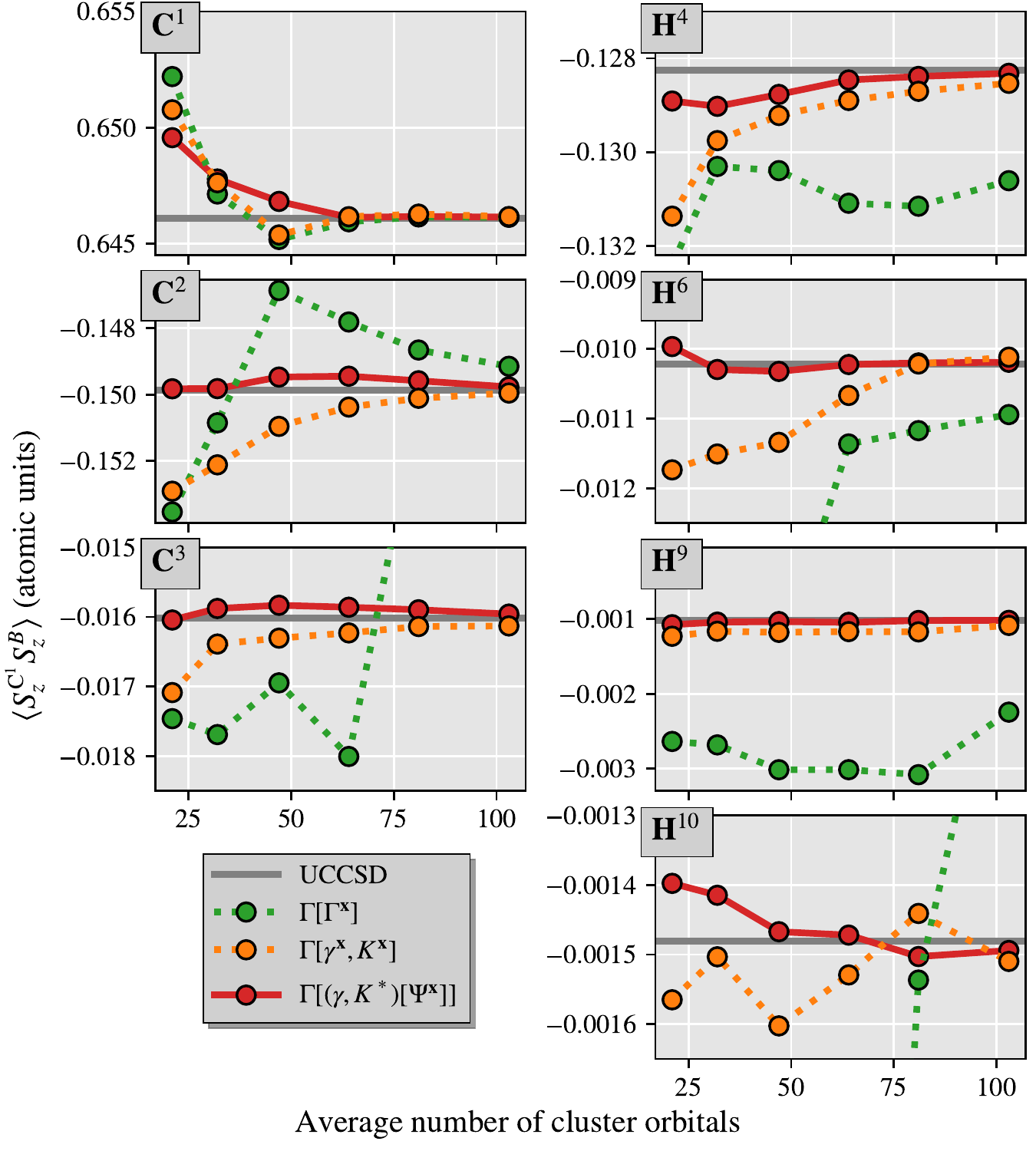}
    \caption{
    $S_z^2$-correlation functions in the $n$-propyl radical
    between the radical position~(C$^1$) and other carbon (first column)
    and hydrogen atoms (second column) in the molecule (see Fig.~\ref{fig:propyl_labels}). Spin correlators are derived from the 2-RDM computed in different ways; traditional democratic partitioning of Sec.~\ref{sec:dem_part} (denoted $\Gamma[\Gamma^{\cl{x}}]$, green), democratic partitioning of the cumulant of Sec.~\ref{sec:part_cumulant} (denoted $\Gamma[\gamma^{\cl{x}}, K^{\cl{x}}]$, yellow) and the wave function derived 2-RDM with in-cluster approximated cumulant of Sec.~\ref{sec:practice} (denoted $\Gamma[(\gamma, K^*)[\Psi^{\cl{x}}]]$, red).
    The calculation was performed in the cc-pVTZ basis set (188 orbitals) with SAO projectors to evaluate the spin correlation functions, and IAO $\oplus$ PAO atomic fragmentation.
    }
    \label{fig:ssz_propyl}
\end{figure}

Figure~\ref{fig:ssz_propyl} plots the convergence of these spin correlation functions from the various functionals as the average cluster (bath) size increases.
Beyond just convergence of any single expectation value, these results also allow us to consider the accuracy of these two-body spin-correlators as a function of their range, indicating the ability of each to compensate for the locality approximation inherent in the embedding for smaller cluster sizes, and converge highly non-local two-point observables. 
While there is only a small difference between the traditional partitioning of the 2-RDM and cumulant partitioning schemes for the on-site correlation function~(top left plot), this changes for the longer-ranged correlations in the lower rows of the figure, where the partitioned cumulant approach of Sec.~\ref{sec:part_cumulant} converges significantly quicker and more smoothly to the full system UCCSD limit, where the direct democratic partitioning of the 2-RDM can become highly erratic.
%
It is particularly noteworthy that the democratically partitioned 2-RDM results can be worse
than then uncorrelated UHF values, even when going to average cluster sizes as large as $\approx$100, at which point the partitioned cumulant results are almost indistinguishable from full system CCSD values. 
%
However, as expected, the wave~function derived expectation values are by far the most reliable across the range of cluster sizes, even with the `in-cluster' approximated cumulant, with an improved and more systematic convergence even for the fully local spin correlator. As such, our results support the use of this approach for both energetics and other one- and two-body properties where possible in this wave~function embedding context.

\section{Conclusions and Outlook}

In this work, we have critically assessed approaches to reconstruct expectation values from wave~function quantum embedding methods. In doing so, we have motivated and compared a number of approaches for calculating total energies across the fragmented system, as well as considering the two-point spin correlation function as an example non-energetic and non-local quantity via the reconstructed reduced density matrices. A fundamental difference exists between two classes of approaches, where expectation values are computed from either partitioned density matrices, or partitioned wave~functions, particularly due to the difference these approaches impose on the choice of fragmentation if a bath expansion of the cluster is to tend to exactness. In the former case, it was found that partitioning the cumulant rather than the 2-RDM directly gave rise to significantly improved two-body properties and energies, and is likely to find applications in solvers where access to wave~function amplitudes is difficult (e.g. the increasing use of quantum computing algorithms as high-level solvers\cite{PhysRevB.105.125117,https://doi.org/10.1002/qua.26975,D2SC01492K,DMET_QC,PhysRevResearch.3.033230}). 

In the latter case where an implicit partitioned wave~function is considered, total energy expressions are motivated and derived from a simple linear energy functional, and from density matrices constructed directly from this wave~function. If obtained via these RDMs, the total energy from the embedding is variational (for a variational solver) and $N$-representable, but incurs significant overhead in the 2-RDM construction. A rigorous and efficient quadratically-scaling algorithm is developed for this $N$-representable 1-RDM, and an `in-cluster' approximation for the 2-RDM is motivated, neglecting many of the cross-cluster contributions in the 2-cumulant. Taken together, these two developments are found to provide the most efficient convergence of energetics and non-local expectation values across a range of systems via a systematic bath expansion of the cluster, while only requiring an atomic IAO fragmentation of the system. We believe that this approach will prove important going forward in the development of robust and accurate wave~function embedding techniques for both chemical and extended systems.

There are a number of further directions we aim to explore as a result of these insights. In systems where the electronic state is required to qualitatively change (e.g. quantum phase transitions), it is possible that a brute-force bath expansion (while necessarily still improvable to exactness) is not going to be the most efficient route in order to capture this changing physics, and feedback from the correlated state to the underlying mean-field reference is required.
The identification of an implicit wave~function, as well as a formally $N$-representable 1-RDM opens new avenues for robust and accurate self-consistency conditions. For instance, the self-consistency could maximize the overlap between this implicit partitioned wave~function and a mean-field state by minimizing the partitioned $T_1$ amplitudes \cite{doi:10.1063/1.463762}. Alternatively, coupling between wave function amplitudes of different cluster are also being formulated. Additionally, the approach of `projected'-DMET could be reconsidered with a full-system 1-RDM which contains more of the non-local physics and is fully $N$-representable compared to the standard democratic partitioning \cite{doi:10.1063/1.5108818}. Finally, the application of a broader range of solvers which can treat significantly strongly correlated systems, as well as excited states, spectra and analytic gradients or forces with these new perspectives of an implicit embedded wave~function are ongoing avenues of investigation.


\ifjcp\else
    \newpage
\fi

\section*{Acknowledgements}
G.H.B. gratefully acknowledges support from the Royal Society via a University Research Fellowship, as well as funding from the European Union's Horizon 2020 research and innovation programme under grant agreement No. 759063. We are grateful to the UK Materials and Molecular Modelling Hub for computational resources, which is partially funded by EPSRC (EP/P020194/1 and EP/T022213/1).

\ifjcp
%
\else
    \bibliographystyle{achemso}
    \bibliography{ref.bib}

\begin{thebibliography}{78}%
\makeatletter
\providecommand \@ifxundefined [1]{%
 \@ifx{#1\undefined}
}%
\providecommand \@ifnum [1]{%
 \ifnum #1\expandafter \@firstoftwo
 \else \expandafter \@secondoftwo
 \fi
}%
\providecommand \@ifx [1]{%
 \ifx #1\expandafter \@firstoftwo
 \else \expandafter \@secondoftwo
 \fi
}%
\providecommand \natexlab [1]{#1}%
\providecommand \enquote  [1]{``#1''}%
\providecommand \bibnamefont  [1]{#1}%
\providecommand \bibfnamefont [1]{#1}%
\providecommand \citenamefont [1]{#1}%
\providecommand \href@noop [0]{\@secondoftwo}%
\providecommand \href [0]{\begingroup \@sanitize@url \@href}%
\providecommand \@href[1]{\@@startlink{#1}\@@href}%
\providecommand \@@href[1]{\endgroup#1\@@endlink}%
\providecommand \@sanitize@url [0]{\catcode `\\12\catcode `\$12\catcode
  `\&12\catcode `\#12\catcode `\^12\catcode `\_12\catcode `\%12\relax}%
\providecommand \@@startlink[1]{}%
\providecommand \@@endlink[0]{}%
\providecommand \url  [0]{\begingroup\@sanitize@url \@url }%
\providecommand \@url [1]{\endgroup\@href {#1}{\urlprefix }}%
\providecommand \urlprefix  [0]{URL }%
\providecommand \Eprint [0]{\href }%
\providecommand \doibase [0]{https://doi.org/}%
\providecommand \selectlanguage [0]{\@gobble}%
\providecommand \bibinfo  [0]{\@secondoftwo}%
\providecommand \bibfield  [0]{\@secondoftwo}%
\providecommand \translation [1]{[#1]}%
\providecommand \BibitemOpen [0]{}%
\providecommand \bibitemStop [0]{}%
\providecommand \bibitemNoStop [0]{.\EOS\space}%
\providecommand \EOS [0]{\spacefactor3000\relax}%
\providecommand \BibitemShut  [1]{\csname bibitem#1\endcsname}%
\let\auto@bib@innerbib\@empty
\bibitem [{\citenamefont {Booth}\ \emph {et~al.}(2013)\citenamefont {Booth},
  \citenamefont {Gr{\"{u}}neis}, \citenamefont {Kresse},\ and\ \citenamefont
  {Alavi}}]{Booth2013}%
  \BibitemOpen
  \bibfield  {author} {\bibinfo {author} {\bibfnamefont {G.~H.}\ \bibnamefont
  {Booth}}, \bibinfo {author} {\bibfnamefont {A.}~\bibnamefont
  {Gr{\"{u}}neis}}, \bibinfo {author} {\bibfnamefont {G.}~\bibnamefont
  {Kresse}},\ and\ \bibinfo {author} {\bibfnamefont {A.}~\bibnamefont
  {Alavi}},\ }\bibfield  {title} {\enquote {\bibinfo {title} {{Towards an exact
  description of electronic wavefunctions in real solids}},}\ }\href
  {https://doi.org/10.1038/nature11770} {\bibfield  {journal} {\bibinfo
  {journal} {Nature}\ }\textbf {\bibinfo {volume} {493}},\ \bibinfo {pages}
  {365--370} (\bibinfo {year} {2013})}\BibitemShut {NoStop}%
\bibitem [{\citenamefont {Pisani}\ \emph {et~al.}(2012)\citenamefont {Pisani},
  \citenamefont {Schütz}, \citenamefont {Casassa}, \citenamefont {Usvyat},
  \citenamefont {Maschio}, \citenamefont {Lorenz},\ and\ \citenamefont
  {Erba}}]{C2CP23927B}%
  \BibitemOpen
  \bibfield  {author} {\bibinfo {author} {\bibfnamefont {C.}~\bibnamefont
  {Pisani}}, \bibinfo {author} {\bibfnamefont {M.}~\bibnamefont {Schütz}},
  \bibinfo {author} {\bibfnamefont {S.}~\bibnamefont {Casassa}}, \bibinfo
  {author} {\bibfnamefont {D.}~\bibnamefont {Usvyat}}, \bibinfo {author}
  {\bibfnamefont {L.}~\bibnamefont {Maschio}}, \bibinfo {author} {\bibfnamefont
  {M.}~\bibnamefont {Lorenz}},\ and\ \bibinfo {author} {\bibfnamefont
  {A.}~\bibnamefont {Erba}},\ }\bibfield  {title} {\enquote {\bibinfo {title}
  {Cryscor: a program for the post-hartree–fock treatment of periodic
  systems},}\ }\href {https://doi.org/10.1039/C2CP23927B} {\bibfield  {journal}
  {\bibinfo  {journal} {Phys. Chem. Chem. Phys.}\ }\textbf {\bibinfo {volume}
  {14}},\ \bibinfo {pages} {7615--7628} (\bibinfo {year} {2012})}\BibitemShut
  {NoStop}%
\bibitem [{\citenamefont {Zhang}\ and\ \citenamefont
  {Gr{\"{u}}neis}(2019)}]{Zhang2019}%
  \BibitemOpen
  \bibfield  {author} {\bibinfo {author} {\bibfnamefont {I.~Y.}\ \bibnamefont
  {Zhang}}\ and\ \bibinfo {author} {\bibfnamefont {A.}~\bibnamefont
  {Gr{\"{u}}neis}},\ }\bibfield  {title} {\enquote {\bibinfo {title} {{Coupled
  Cluster Theory in Materials Science}},}\ }\href
  {https://www.frontiersin.org/article/10.3389/fmats.2019.00123/full}
  {\bibfield  {journal} {\bibinfo  {journal} {Frontiers in Materials}\ }\textbf
  {\bibinfo {volume} {6}} (\bibinfo {year} {2019})}\BibitemShut {NoStop}%
\bibitem [{\citenamefont {Gruber}\ \emph {et~al.}(2018)\citenamefont {Gruber},
  \citenamefont {Liao}, \citenamefont {Tsatsoulis}, \citenamefont {Hummel},\
  and\ \citenamefont {Gr{\"{u}}neis}}]{Gruber2018}%
  \BibitemOpen
  \bibfield  {author} {\bibinfo {author} {\bibfnamefont {T.}~\bibnamefont
  {Gruber}}, \bibinfo {author} {\bibfnamefont {K.}~\bibnamefont {Liao}},
  \bibinfo {author} {\bibfnamefont {T.}~\bibnamefont {Tsatsoulis}}, \bibinfo
  {author} {\bibfnamefont {F.}~\bibnamefont {Hummel}},\ and\ \bibinfo {author}
  {\bibfnamefont {A.}~\bibnamefont {Gr{\"{u}}neis}},\ }\bibfield  {title}
  {\enquote {\bibinfo {title} {{Applying the Coupled-Cluster Ansatz to Solids
  and Surfaces in the Thermodynamic Limit}},}\ }\href
  {https://doi.org/10.1103/PhysRevX.8.021043} {\bibfield  {journal} {\bibinfo
  {journal} {Physical Review X}\ }\textbf {\bibinfo {volume} {8}},\ \bibinfo
  {pages} {021043} (\bibinfo {year} {2018})}\BibitemShut {NoStop}%
\bibitem [{\citenamefont {McClain}\ \emph {et~al.}(2017)\citenamefont
  {McClain}, \citenamefont {Sun}, \citenamefont {Chan},\ and\ \citenamefont
  {Berkelbach}}]{McClain2017}%
  \BibitemOpen
  \bibfield  {author} {\bibinfo {author} {\bibfnamefont {J.}~\bibnamefont
  {McClain}}, \bibinfo {author} {\bibfnamefont {Q.}~\bibnamefont {Sun}},
  \bibinfo {author} {\bibfnamefont {G.~K.-L.}\ \bibnamefont {Chan}},\ and\
  \bibinfo {author} {\bibfnamefont {T.~C.}\ \bibnamefont {Berkelbach}},\
  }\bibfield  {title} {\enquote {\bibinfo {title} {{Gaussian-Based
  Coupled-Cluster Theory for the Ground-State and Band Structure of Solids}},}\
  }\href {https://doi.org/10.1021/acs.jctc.7b00049} {\bibfield  {journal}
  {\bibinfo  {journal} {Journal of Chemical Theory and Computation}\ }\textbf
  {\bibinfo {volume} {13}},\ \bibinfo {pages} {1209--1218} (\bibinfo {year}
  {2017})}\BibitemShut {NoStop}%
\bibitem [{\citenamefont {Xing}\ and\ \citenamefont
  {Lin}(2022)}]{doi:10.1021/acs.jctc.1c00985}%
  \BibitemOpen
  \bibfield  {author} {\bibinfo {author} {\bibfnamefont {X.}~\bibnamefont
  {Xing}}\ and\ \bibinfo {author} {\bibfnamefont {L.}~\bibnamefont {Lin}},\
  }\bibfield  {title} {\enquote {\bibinfo {title} {Staggered mesh method for
  correlation energy calculations of solids: Random phase approximation in
  direct ring coupled cluster doubles and adiabatic connection formalisms},}\
  }\href {https://doi.org/10.1021/acs.jctc.1c00985} {\bibfield  {journal}
  {\bibinfo  {journal} {Journal of Chemical Theory and Computation}\ }\textbf
  {\bibinfo {volume} {18}},\ \bibinfo {pages} {763--775} (\bibinfo {year}
  {2022})},\ \bibinfo {note} {pMID: 34989566},\ \Eprint
  {https://arxiv.org/abs/https://doi.org/10.1021/acs.jctc.1c00985}
  {https://doi.org/10.1021/acs.jctc.1c00985} \BibitemShut {NoStop}%
\bibitem [{\citenamefont {Herbert}(2019)}]{doi:10.1063/1.5126216}%
  \BibitemOpen
  \bibfield  {author} {\bibinfo {author} {\bibfnamefont {J.~M.}\ \bibnamefont
  {Herbert}},\ }\bibfield  {title} {\enquote {\bibinfo {title} {Fantasy versus
  reality in fragment-based quantum chemistry},}\ }\href
  {https://doi.org/10.1063/1.5126216} {\bibfield  {journal} {\bibinfo
  {journal} {The Journal of Chemical Physics}\ }\textbf {\bibinfo {volume}
  {151}},\ \bibinfo {pages} {170901} (\bibinfo {year} {2019})},\ \Eprint
  {https://arxiv.org/abs/https://doi.org/10.1063/1.5126216}
  {https://doi.org/10.1063/1.5126216} \BibitemShut {NoStop}%
\bibitem [{\citenamefont {Gordon}\ \emph {et~al.}(2012)\citenamefont {Gordon},
  \citenamefont {Fedorov}, \citenamefont {Pruitt},\ and\ \citenamefont
  {Slipchenko}}]{doi:10.1021/cr200093j}%
  \BibitemOpen
  \bibfield  {author} {\bibinfo {author} {\bibfnamefont {M.~S.}\ \bibnamefont
  {Gordon}}, \bibinfo {author} {\bibfnamefont {D.~G.}\ \bibnamefont {Fedorov}},
  \bibinfo {author} {\bibfnamefont {S.~R.}\ \bibnamefont {Pruitt}},\ and\
  \bibinfo {author} {\bibfnamefont {L.~V.}\ \bibnamefont {Slipchenko}},\
  }\bibfield  {title} {\enquote {\bibinfo {title} {Fragmentation methods: A
  route to accurate calculations on large systems},}\ }\href
  {https://doi.org/10.1021/cr200093j} {\bibfield  {journal} {\bibinfo
  {journal} {Chemical Reviews}\ }\textbf {\bibinfo {volume} {112}},\ \bibinfo
  {pages} {632--672} (\bibinfo {year} {2012})},\ \bibinfo {note} {pMID:
  21866983},\ \Eprint {https://arxiv.org/abs/https://doi.org/10.1021/cr200093j}
  {https://doi.org/10.1021/cr200093j} \BibitemShut {NoStop}%
\bibitem [{\citenamefont {Saebø}(2002)}]{doi:10.1142/9789812776815_0003}%
  \BibitemOpen
  \bibfield  {author} {\bibinfo {author} {\bibfnamefont {S.}~\bibnamefont
  {Saebø}},\ }\enquote {\bibinfo {title} {Low-scaling methods for electron
  correlation},}\ in\ \href {https://doi.org/10.1142/9789812776815_0003} {\emph
  {\bibinfo {booktitle} {Computational Chemistry: Reviews of Current Trends}}}\
  (\bibinfo  {publisher} {World Scientific},\ \bibinfo {year} {2002})\ pp.\
  \bibinfo {pages} {63--87}\BibitemShut {NoStop}%
\bibitem [{\citenamefont {Usvyat}, \citenamefont {Maschio},\ and\ \citenamefont
  {Sch{\"{u}}tz}(2018)}]{Usvyat2018}%
  \BibitemOpen
  \bibfield  {author} {\bibinfo {author} {\bibfnamefont {D.}~\bibnamefont
  {Usvyat}}, \bibinfo {author} {\bibfnamefont {L.}~\bibnamefont {Maschio}},\
  and\ \bibinfo {author} {\bibfnamefont {M.}~\bibnamefont {Sch{\"{u}}tz}},\
  }\bibfield  {title} {\enquote {\bibinfo {title} {{Periodic and fragment
  models based on the local correlation approach}},}\ }\href
  {https://doi.org/10.1002/wcms.1357} {\bibfield  {journal} {\bibinfo
  {journal} {WIREs Computational Molecular Science}\ }\textbf {\bibinfo
  {volume} {8}},\ \bibinfo {pages} {1--27} (\bibinfo {year}
  {2018})}\BibitemShut {NoStop}%
\bibitem [{\citenamefont {Sch{\"{a}}fer}\ \emph {et~al.}(2021)\citenamefont
  {Sch{\"{a}}fer}, \citenamefont {Libisch}, \citenamefont {Kresse},\ and\
  \citenamefont {Gr{\"{u}}neis}}]{Schafer2021}%
  \BibitemOpen
  \bibfield  {author} {\bibinfo {author} {\bibfnamefont {T.}~\bibnamefont
  {Sch{\"{a}}fer}}, \bibinfo {author} {\bibfnamefont {F.}~\bibnamefont
  {Libisch}}, \bibinfo {author} {\bibfnamefont {G.}~\bibnamefont {Kresse}},\
  and\ \bibinfo {author} {\bibfnamefont {A.}~\bibnamefont {Gr{\"{u}}neis}},\
  }\bibfield  {title} {\enquote {\bibinfo {title} {{Local embedding of coupled
  cluster theory into the random phase approximation using plane waves}},}\
  }\href {https://doi.org/10.1063/5.0036363} {\bibfield  {journal} {\bibinfo
  {journal} {The Journal of Chemical Physics}\ }\textbf {\bibinfo {volume}
  {154}},\ \bibinfo {pages} {011101} (\bibinfo {year} {2021})}\BibitemShut
  {NoStop}%
\bibitem [{\citenamefont {Sun}\ and\ \citenamefont {Chan}(2016)}]{Sun2016}%
  \BibitemOpen
  \bibfield  {author} {\bibinfo {author} {\bibfnamefont {Q.}~\bibnamefont
  {Sun}}\ and\ \bibinfo {author} {\bibfnamefont {G.~K.-l.}\ \bibnamefont
  {Chan}},\ }\bibfield  {title} {\enquote {\bibinfo {title} {{Quantum Embedding
  Theories}},}\ }\href {https://doi.org/10.1021/acs.accounts.6b00356}
  {\bibfield  {journal} {\bibinfo  {journal} {Accounts of Chemical Research}\
  }\textbf {\bibinfo {volume} {49}},\ \bibinfo {pages} {2705--2712} (\bibinfo
  {year} {2016})}\BibitemShut {NoStop}%
\bibitem [{\citenamefont {Fertitta}\ and\ \citenamefont
  {Booth}(2018)}]{Fertitta2018}%
  \BibitemOpen
  \bibfield  {author} {\bibinfo {author} {\bibfnamefont {E.}~\bibnamefont
  {Fertitta}}\ and\ \bibinfo {author} {\bibfnamefont {G.~H.}\ \bibnamefont
  {Booth}},\ }\bibfield  {title} {\enquote {\bibinfo {title} {{Rigorous wave
  function embedding with dynamical fluctuations}},}\ }\href
  {https://doi.org/10.1103/PhysRevB.98.235132} {\bibfield  {journal} {\bibinfo
  {journal} {Physical Review B}\ }\textbf {\bibinfo {volume} {98}},\ \bibinfo
  {pages} {235132} (\bibinfo {year} {2018})}\BibitemShut {NoStop}%
\bibitem [{\citenamefont {Nusspickel}\ and\ \citenamefont
  {Booth}(2020{\natexlab{a}})}]{PhysRevB.102.165107}%
  \BibitemOpen
  \bibfield  {author} {\bibinfo {author} {\bibfnamefont {M.}~\bibnamefont
  {Nusspickel}}\ and\ \bibinfo {author} {\bibfnamefont {G.~H.}\ \bibnamefont
  {Booth}},\ }\bibfield  {title} {\enquote {\bibinfo {title} {Efficient
  compression of the environment of an open quantum system},}\ }\href
  {https://doi.org/10.1103/PhysRevB.102.165107} {\bibfield  {journal} {\bibinfo
   {journal} {Phys. Rev. B}\ }\textbf {\bibinfo {volume} {102}},\ \bibinfo
  {pages} {165107} (\bibinfo {year} {2020}{\natexlab{a}})}\BibitemShut
  {NoStop}%
\bibitem [{\citenamefont {Nusspickel}\ and\ \citenamefont
  {Booth}(2020{\natexlab{b}})}]{PhysRevB.101.045126}%
  \BibitemOpen
  \bibfield  {author} {\bibinfo {author} {\bibfnamefont {M.}~\bibnamefont
  {Nusspickel}}\ and\ \bibinfo {author} {\bibfnamefont {G.~H.}\ \bibnamefont
  {Booth}},\ }\bibfield  {title} {\enquote {\bibinfo {title}
  {Frequency-dependent and algebraic bath states for a dynamical mean-field
  theory with compact support},}\ }\href
  {https://doi.org/10.1103/PhysRevB.101.045126} {\bibfield  {journal} {\bibinfo
   {journal} {Phys. Rev. B}\ }\textbf {\bibinfo {volume} {101}},\ \bibinfo
  {pages} {045126} (\bibinfo {year} {2020}{\natexlab{b}})}\BibitemShut
  {NoStop}%
\bibitem [{\citenamefont {Potthoff}(2001)}]{PhysRevB.64.165114}%
  \BibitemOpen
  \bibfield  {author} {\bibinfo {author} {\bibfnamefont {M.}~\bibnamefont
  {Potthoff}},\ }\bibfield  {title} {\enquote {\bibinfo {title} {Two-site
  dynamical mean-field theory},}\ }\href
  {https://doi.org/10.1103/PhysRevB.64.165114} {\bibfield  {journal} {\bibinfo
  {journal} {Phys. Rev. B}\ }\textbf {\bibinfo {volume} {64}},\ \bibinfo
  {pages} {165114} (\bibinfo {year} {2001})}\BibitemShut {NoStop}%
\bibitem [{\citenamefont {Lu}\ \emph {et~al.}(2019)\citenamefont {Lu},
  \citenamefont {Cao}, \citenamefont {Hansmann},\ and\ \citenamefont
  {Haverkort}}]{PhysRevB.100.115134}%
  \BibitemOpen
  \bibfield  {author} {\bibinfo {author} {\bibfnamefont {Y.}~\bibnamefont
  {Lu}}, \bibinfo {author} {\bibfnamefont {X.}~\bibnamefont {Cao}}, \bibinfo
  {author} {\bibfnamefont {P.}~\bibnamefont {Hansmann}},\ and\ \bibinfo
  {author} {\bibfnamefont {M.~W.}\ \bibnamefont {Haverkort}},\ }\bibfield
  {title} {\enquote {\bibinfo {title} {Natural-orbital impurity solver and
  projection approach for green's functions},}\ }\href
  {https://doi.org/10.1103/PhysRevB.100.115134} {\bibfield  {journal} {\bibinfo
   {journal} {Phys. Rev. B}\ }\textbf {\bibinfo {volume} {100}},\ \bibinfo
  {pages} {115134} (\bibinfo {year} {2019})}\BibitemShut {NoStop}%
\bibitem [{\citenamefont {Ganahl}\ \emph {et~al.}(2015)\citenamefont {Ganahl},
  \citenamefont {Aichhorn}, \citenamefont {Evertz}, \citenamefont
  {Thunstr\"om}, \citenamefont {Held},\ and\ \citenamefont
  {Verstraete}}]{PhysRevB.92.155132}%
  \BibitemOpen
  \bibfield  {author} {\bibinfo {author} {\bibfnamefont {M.}~\bibnamefont
  {Ganahl}}, \bibinfo {author} {\bibfnamefont {M.}~\bibnamefont {Aichhorn}},
  \bibinfo {author} {\bibfnamefont {H.~G.}\ \bibnamefont {Evertz}}, \bibinfo
  {author} {\bibfnamefont {P.}~\bibnamefont {Thunstr\"om}}, \bibinfo {author}
  {\bibfnamefont {K.}~\bibnamefont {Held}},\ and\ \bibinfo {author}
  {\bibfnamefont {F.}~\bibnamefont {Verstraete}},\ }\bibfield  {title}
  {\enquote {\bibinfo {title} {Efficient dmft impurity solver using real-time
  dynamics with matrix product states},}\ }\href
  {https://doi.org/10.1103/PhysRevB.92.155132} {\bibfield  {journal} {\bibinfo
  {journal} {Phys. Rev. B}\ }\textbf {\bibinfo {volume} {92}},\ \bibinfo
  {pages} {155132} (\bibinfo {year} {2015})}\BibitemShut {NoStop}%
\bibitem [{\citenamefont {Scott}\ and\ \citenamefont
  {Booth}(2021)}]{PhysRevB.104.245114}%
  \BibitemOpen
  \bibfield  {author} {\bibinfo {author} {\bibfnamefont {C.~J.~C.}\
  \bibnamefont {Scott}}\ and\ \bibinfo {author} {\bibfnamefont {G.~H.}\
  \bibnamefont {Booth}},\ }\bibfield  {title} {\enquote {\bibinfo {title}
  {Extending density matrix embedding: A static two-particle theory},}\ }\href
  {https://doi.org/10.1103/PhysRevB.104.245114} {\bibfield  {journal} {\bibinfo
   {journal} {Phys. Rev. B}\ }\textbf {\bibinfo {volume} {104}},\ \bibinfo
  {pages} {245114} (\bibinfo {year} {2021})}\BibitemShut {NoStop}%
\bibitem [{\citenamefont {Wu}\ \emph {et~al.}(2019)\citenamefont {Wu},
  \citenamefont {Cui}, \citenamefont {Tong}, \citenamefont {Lindsey},
  \citenamefont {Chan},\ and\ \citenamefont {Lin}}]{doi:10.1063/1.5108818}%
  \BibitemOpen
  \bibfield  {author} {\bibinfo {author} {\bibfnamefont {X.}~\bibnamefont
  {Wu}}, \bibinfo {author} {\bibfnamefont {Z.-H.}\ \bibnamefont {Cui}},
  \bibinfo {author} {\bibfnamefont {Y.}~\bibnamefont {Tong}}, \bibinfo {author}
  {\bibfnamefont {M.}~\bibnamefont {Lindsey}}, \bibinfo {author} {\bibfnamefont
  {G.~K.-L.}\ \bibnamefont {Chan}},\ and\ \bibinfo {author} {\bibfnamefont
  {L.}~\bibnamefont {Lin}},\ }\bibfield  {title} {\enquote {\bibinfo {title}
  {Projected density matrix embedding theory with applications to the
  two-dimensional hubbard model},}\ }\href {https://doi.org/10.1063/1.5108818}
  {\bibfield  {journal} {\bibinfo  {journal} {The Journal of Chemical Physics}\
  }\textbf {\bibinfo {volume} {151}},\ \bibinfo {pages} {064108} (\bibinfo
  {year} {2019})},\ \Eprint
  {https://arxiv.org/abs/https://doi.org/10.1063/1.5108818}
  {https://doi.org/10.1063/1.5108818} \BibitemShut {NoStop}%
\bibitem [{\citenamefont {Wu}\ \emph {et~al.}(2020)\citenamefont {Wu},
  \citenamefont {Lindsey}, \citenamefont {Zhou}, \citenamefont {Tong},\ and\
  \citenamefont {Lin}}]{PhysRevB.102.085123}%
  \BibitemOpen
  \bibfield  {author} {\bibinfo {author} {\bibfnamefont {X.}~\bibnamefont
  {Wu}}, \bibinfo {author} {\bibfnamefont {M.}~\bibnamefont {Lindsey}},
  \bibinfo {author} {\bibfnamefont {T.}~\bibnamefont {Zhou}}, \bibinfo {author}
  {\bibfnamefont {Y.}~\bibnamefont {Tong}},\ and\ \bibinfo {author}
  {\bibfnamefont {L.}~\bibnamefont {Lin}},\ }\bibfield  {title} {\enquote
  {\bibinfo {title} {Enhancing robustness and efficiency of density matrix
  embedding theory via semidefinite programming and local correlation potential
  fitting},}\ }\href {https://doi.org/10.1103/PhysRevB.102.085123} {\bibfield
  {journal} {\bibinfo  {journal} {Phys. Rev. B}\ }\textbf {\bibinfo {volume}
  {102}},\ \bibinfo {pages} {085123} (\bibinfo {year} {2020})}\BibitemShut
  {NoStop}%
\bibitem [{\citenamefont {Faulstich}\ \emph {et~al.}(2022)\citenamefont
  {Faulstich}, \citenamefont {Kim}, \citenamefont {Cui}, \citenamefont {Wen},
  \citenamefont {Kin-Lic~Chan},\ and\ \citenamefont
  {Lin}}]{doi:10.1021/acs.jctc.1c01061}%
  \BibitemOpen
  \bibfield  {author} {\bibinfo {author} {\bibfnamefont {F.~M.}\ \bibnamefont
  {Faulstich}}, \bibinfo {author} {\bibfnamefont {R.}~\bibnamefont {Kim}},
  \bibinfo {author} {\bibfnamefont {Z.-H.}\ \bibnamefont {Cui}}, \bibinfo
  {author} {\bibfnamefont {Z.}~\bibnamefont {Wen}}, \bibinfo {author}
  {\bibfnamefont {G.}~\bibnamefont {Kin-Lic~Chan}},\ and\ \bibinfo {author}
  {\bibfnamefont {L.}~\bibnamefont {Lin}},\ }\bibfield  {title} {\enquote
  {\bibinfo {title} {Pure state v-representability of density matrix embedding
  theory},}\ }\href {https://doi.org/10.1021/acs.jctc.1c01061} {\bibfield
  {journal} {\bibinfo  {journal} {Journal of Chemical Theory and Computation}\
  }\textbf {\bibinfo {volume} {18}},\ \bibinfo {pages} {851--864} (\bibinfo
  {year} {2022})},\ \bibinfo {note} {pMID: 35084855},\ \Eprint
  {https://arxiv.org/abs/https://doi.org/10.1021/acs.jctc.1c01061}
  {https://doi.org/10.1021/acs.jctc.1c01061} \BibitemShut {NoStop}%
\bibitem [{\citenamefont {Metzner}\ and\ \citenamefont
  {Vollhardt}(1989)}]{Metzner1989}%
  \BibitemOpen
  \bibfield  {author} {\bibinfo {author} {\bibfnamefont {W.}~\bibnamefont
  {Metzner}}\ and\ \bibinfo {author} {\bibfnamefont {D.}~\bibnamefont
  {Vollhardt}},\ }\bibfield  {title} {\enquote {\bibinfo {title} {{Correlated
  Lattice Fermions in $\infty$ Dimensions}},}\ }\href
  {https://doi.org/10.1103/PhysRevLett.62.324} {\bibfield  {journal} {\bibinfo
  {journal} {Physical Review Letters}\ }\textbf {\bibinfo {volume} {62}},\
  \bibinfo {pages} {324--327} (\bibinfo {year} {1989})}\BibitemShut {NoStop}%
\bibitem [{\citenamefont {Georges}\ \emph {et~al.}(1996)\citenamefont
  {Georges}, \citenamefont {Kotliar}, \citenamefont {Krauth},\ and\
  \citenamefont {Rozenberg}}]{Georges1996}%
  \BibitemOpen
  \bibfield  {author} {\bibinfo {author} {\bibfnamefont {A.}~\bibnamefont
  {Georges}}, \bibinfo {author} {\bibfnamefont {G.}~\bibnamefont {Kotliar}},
  \bibinfo {author} {\bibfnamefont {W.}~\bibnamefont {Krauth}},\ and\ \bibinfo
  {author} {\bibfnamefont {M.~J.}\ \bibnamefont {Rozenberg}},\ }\bibfield
  {title} {\enquote {\bibinfo {title} {{Dynamical mean-field theory of strongly
  correlated fermion systems and the limit of infinite dimensions}},}\ }\href
  {https://doi.org/10.1103/RevModPhys.68.13} {\bibfield  {journal} {\bibinfo
  {journal} {Reviews of Modern Physics}\ }\textbf {\bibinfo {volume} {68}},\
  \bibinfo {pages} {13--125} (\bibinfo {year} {1996})}\BibitemShut {NoStop}%
\bibitem [{\citenamefont {Kotliar}\ \emph {et~al.}(2006)\citenamefont
  {Kotliar}, \citenamefont {Savrasov}, \citenamefont {Haule}, \citenamefont
  {Oudovenko}, \citenamefont {Parcollet},\ and\ \citenamefont
  {Marianetti}}]{Kotliar2006}%
  \BibitemOpen
  \bibfield  {author} {\bibinfo {author} {\bibfnamefont {G.}~\bibnamefont
  {Kotliar}}, \bibinfo {author} {\bibfnamefont {S.~Y.}\ \bibnamefont
  {Savrasov}}, \bibinfo {author} {\bibfnamefont {K.}~\bibnamefont {Haule}},
  \bibinfo {author} {\bibfnamefont {V.~S.}\ \bibnamefont {Oudovenko}}, \bibinfo
  {author} {\bibfnamefont {O.}~\bibnamefont {Parcollet}},\ and\ \bibinfo
  {author} {\bibfnamefont {C.~A.}\ \bibnamefont {Marianetti}},\ }\bibfield
  {title} {\enquote {\bibinfo {title} {{Electronic structure calculations with
  dynamical mean-field theory}},}\ }\href
  {https://doi.org/10.1103/RevModPhys.78.865} {\bibfield  {journal} {\bibinfo
  {journal} {Reviews of Modern Physics}\ }\textbf {\bibinfo {volume} {78}},\
  \bibinfo {pages} {865--951} (\bibinfo {year} {2006})}\BibitemShut {NoStop}%
\bibitem [{\citenamefont {Karolak}\ \emph {et~al.}(2010)\citenamefont
  {Karolak}, \citenamefont {Ulm}, \citenamefont {Wehling}, \citenamefont
  {Mazurenko}, \citenamefont {Poteryaev},\ and\ \citenamefont
  {Lichtenstein}}]{KAROLAK201011}%
  \BibitemOpen
  \bibfield  {author} {\bibinfo {author} {\bibfnamefont {M.}~\bibnamefont
  {Karolak}}, \bibinfo {author} {\bibfnamefont {G.}~\bibnamefont {Ulm}},
  \bibinfo {author} {\bibfnamefont {T.}~\bibnamefont {Wehling}}, \bibinfo
  {author} {\bibfnamefont {V.}~\bibnamefont {Mazurenko}}, \bibinfo {author}
  {\bibfnamefont {A.}~\bibnamefont {Poteryaev}},\ and\ \bibinfo {author}
  {\bibfnamefont {A.}~\bibnamefont {Lichtenstein}},\ }\bibfield  {title}
  {\enquote {\bibinfo {title} {{Double counting in LDA+DMFT---The example of
  NiO}},}\ }\href
  {https://doi.org/https://doi.org/10.1016/j.elspec.2010.05.021} {\bibfield
  {journal} {\bibinfo  {journal} {Journal of Electron Spectroscopy and Related
  Phenomena}\ }\textbf {\bibinfo {volume} {181}},\ \bibinfo {pages} {11--15}
  (\bibinfo {year} {2010})},\ \bibinfo {note} {proceedings of International
  Workshop on Strong Correlations and Angle-Resolved Photoemission Spectroscopy
  2009}\BibitemShut {NoStop}%
\bibitem [{\citenamefont {Zhu}, \citenamefont {Cui},\ and\ \citenamefont
  {Chan}(2020)}]{doi:10.1021/acs.jctc.9b00934}%
  \BibitemOpen
  \bibfield  {author} {\bibinfo {author} {\bibfnamefont {T.}~\bibnamefont
  {Zhu}}, \bibinfo {author} {\bibfnamefont {Z.-H.}\ \bibnamefont {Cui}},\ and\
  \bibinfo {author} {\bibfnamefont {G.~K.-L.}\ \bibnamefont {Chan}},\
  }\bibfield  {title} {\enquote {\bibinfo {title} {Efficient formulation of ab
  initio quantum embedding in periodic systems: Dynamical mean-field theory},}\
  }\href {https://doi.org/10.1021/acs.jctc.9b00934} {\bibfield  {journal}
  {\bibinfo  {journal} {Journal of Chemical Theory and Computation}\ }\textbf
  {\bibinfo {volume} {16}},\ \bibinfo {pages} {141--153} (\bibinfo {year}
  {2020})}\BibitemShut {NoStop}%
\bibitem [{\citenamefont {Rusakov}\ \emph {et~al.}(2019)\citenamefont
  {Rusakov}, \citenamefont {Iskakov}, \citenamefont {Tran},\ and\ \citenamefont
  {Zgid}}]{doi:10.1021/acs.jctc.8b00927}%
  \BibitemOpen
  \bibfield  {author} {\bibinfo {author} {\bibfnamefont {A.~A.}\ \bibnamefont
  {Rusakov}}, \bibinfo {author} {\bibfnamefont {S.}~\bibnamefont {Iskakov}},
  \bibinfo {author} {\bibfnamefont {L.~N.}\ \bibnamefont {Tran}},\ and\
  \bibinfo {author} {\bibfnamefont {D.}~\bibnamefont {Zgid}},\ }\bibfield
  {title} {\enquote {\bibinfo {title} {Self-energy embedding theory (seet) for
  periodic systems},}\ }\href {https://doi.org/10.1021/acs.jctc.8b00927}
  {\bibfield  {journal} {\bibinfo  {journal} {Journal of Chemical Theory and
  Computation}\ }\textbf {\bibinfo {volume} {15}},\ \bibinfo {pages} {229--240}
  (\bibinfo {year} {2019})}\BibitemShut {NoStop}%
\bibitem [{\citenamefont {Zhu}\ and\ \citenamefont {Chan}(2021)}]{Zhu2021}%
  \BibitemOpen
  \bibfield  {author} {\bibinfo {author} {\bibfnamefont {T.}~\bibnamefont
  {Zhu}}\ and\ \bibinfo {author} {\bibfnamefont {G.~K.-L.}\ \bibnamefont
  {Chan}},\ }\bibfield  {title} {\enquote {\bibinfo {title} {{Ab Initio Full
  Cell $GW$ + DMFT for Correlated Materials}},}\ }\href
  {https://doi.org/10.1103/PhysRevX.11.021006} {\bibfield  {journal} {\bibinfo
  {journal} {Physical Review X}\ }\textbf {\bibinfo {volume} {11}},\ \bibinfo
  {pages} {021006} (\bibinfo {year} {2021})}\BibitemShut {NoStop}%
\bibitem [{\citenamefont {Knizia}\ and\ \citenamefont
  {Chan}(2012)}]{Knizia2012}%
  \BibitemOpen
  \bibfield  {author} {\bibinfo {author} {\bibfnamefont {G.}~\bibnamefont
  {Knizia}}\ and\ \bibinfo {author} {\bibfnamefont {G.~K.-L.}\ \bibnamefont
  {Chan}},\ }\bibfield  {title} {\enquote {\bibinfo {title} {{Density Matrix
  Embedding: A Simple Alternative to Dynamical Mean-Field Theory}},}\ }\href
  {https://doi.org/10.1103/PhysRevLett.109.186404} {\bibfield  {journal}
  {\bibinfo  {journal} {Physical Review Letters}\ }\textbf {\bibinfo {volume}
  {109}},\ \bibinfo {pages} {186404} (\bibinfo {year} {2012})}\BibitemShut
  {NoStop}%
\bibitem [{\citenamefont {Knizia}\ and\ \citenamefont
  {Chan}(2013)}]{Knizia2013}%
  \BibitemOpen
  \bibfield  {author} {\bibinfo {author} {\bibfnamefont {G.}~\bibnamefont
  {Knizia}}\ and\ \bibinfo {author} {\bibfnamefont {G.~K.-l.}\ \bibnamefont
  {Chan}},\ }\bibfield  {title} {\enquote {\bibinfo {title} {{Density Matrix
  Embedding: A Strong-Coupling Quantum Embedding Theory}},}\ }\href
  {https://doi.org/10.1021/ct301044e} {\bibfield  {journal} {\bibinfo
  {journal} {Journal of Chemical Theory and Computation}\ }\textbf {\bibinfo
  {volume} {9}},\ \bibinfo {pages} {1428--1432} (\bibinfo {year}
  {2013})}\BibitemShut {NoStop}%
\bibitem [{\citenamefont {Bulik}, \citenamefont {Scuseria},\ and\ \citenamefont
  {Dukelsky}(2014)}]{Bulik2014}%
  \BibitemOpen
  \bibfield  {author} {\bibinfo {author} {\bibfnamefont {I.~W.}\ \bibnamefont
  {Bulik}}, \bibinfo {author} {\bibfnamefont {G.~E.}\ \bibnamefont
  {Scuseria}},\ and\ \bibinfo {author} {\bibfnamefont {J.}~\bibnamefont
  {Dukelsky}},\ }\bibfield  {title} {\enquote {\bibinfo {title} {Density matrix
  embedding from broken symmetry lattice mean fields},}\ }\href
  {https://doi.org/10.1103/PhysRevB.89.035140} {\bibfield  {journal} {\bibinfo
  {journal} {Phys. Rev. B}\ }\textbf {\bibinfo {volume} {89}},\ \bibinfo
  {pages} {035140} (\bibinfo {year} {2014})}\BibitemShut {NoStop}%
\bibitem [{\citenamefont {Wouters}\ \emph {et~al.}(2016)\citenamefont
  {Wouters}, \citenamefont {Jim{\'{e}}nez-Hoyos}, \citenamefont {Sun},\ and\
  \citenamefont {Chan}}]{Wouters2016}%
  \BibitemOpen
  \bibfield  {author} {\bibinfo {author} {\bibfnamefont {S.}~\bibnamefont
  {Wouters}}, \bibinfo {author} {\bibfnamefont {C.~A.}\ \bibnamefont
  {Jim{\'{e}}nez-Hoyos}}, \bibinfo {author} {\bibfnamefont {Q.}~\bibnamefont
  {Sun}},\ and\ \bibinfo {author} {\bibfnamefont {G.~K.}\ \bibnamefont
  {Chan}},\ }\bibfield  {title} {\enquote {\bibinfo {title} {{A Practical Guide
  to Density Matrix Embedding Theory in Quantum Chemistry}},}\ }\href
  {https://doi.org/10.1021/acs.jctc.6b00316} {\bibfield  {journal} {\bibinfo
  {journal} {Journal of Chemical Theory and Computation}\ }\textbf {\bibinfo
  {volume} {12}},\ \bibinfo {pages} {2706--2719} (\bibinfo {year}
  {2016})}\BibitemShut {NoStop}%
\bibitem [{\citenamefont {Welborn}, \citenamefont {Tsuchimochi},\ and\
  \citenamefont {Van~Voorhis}(2016)}]{Welborn2016}%
  \BibitemOpen
  \bibfield  {author} {\bibinfo {author} {\bibfnamefont {M.}~\bibnamefont
  {Welborn}}, \bibinfo {author} {\bibfnamefont {T.}~\bibnamefont
  {Tsuchimochi}},\ and\ \bibinfo {author} {\bibfnamefont {T.}~\bibnamefont
  {Van~Voorhis}},\ }\bibfield  {title} {\enquote {\bibinfo {title} {Bootstrap
  embedding: An internally consistent fragment-based method},}\ }\href
  {https://doi.org/10.1063/1.4960986} {\bibfield  {journal} {\bibinfo
  {journal} {J. Chem. Phys.}\ }\textbf {\bibinfo {volume} {145}},\ \bibinfo
  {pages} {074102} (\bibinfo {year} {2016})}\BibitemShut {NoStop}%
\bibitem [{\citenamefont {Cui}, \citenamefont {Zhu},\ and\ \citenamefont
  {Chan}(2020)}]{Cui2020}%
  \BibitemOpen
  \bibfield  {author} {\bibinfo {author} {\bibfnamefont {Z.-H.}\ \bibnamefont
  {Cui}}, \bibinfo {author} {\bibfnamefont {T.}~\bibnamefont {Zhu}},\ and\
  \bibinfo {author} {\bibfnamefont {G.~K.-L.}\ \bibnamefont {Chan}},\
  }\bibfield  {title} {\enquote {\bibinfo {title} {{Efficient Implementation of
  Ab Initio Quantum Embedding in Periodic Systems: Density Matrix Embedding
  Theory}},}\ }\href {https://doi.org/10.1021/acs.jctc.9b00933} {\bibfield
  {journal} {\bibinfo  {journal} {Journal of Chemical Theory and Computation}\
  }\textbf {\bibinfo {volume} {16}},\ \bibinfo {pages} {119--129} (\bibinfo
  {year} {2020})}\BibitemShut {NoStop}%
\bibitem [{\citenamefont {Pham}, \citenamefont {Hermes},\ and\ \citenamefont
  {Gagliardi}(2020)}]{Pham2020}%
  \BibitemOpen
  \bibfield  {author} {\bibinfo {author} {\bibfnamefont {H.~Q.}\ \bibnamefont
  {Pham}}, \bibinfo {author} {\bibfnamefont {M.~R.}\ \bibnamefont {Hermes}},\
  and\ \bibinfo {author} {\bibfnamefont {L.}~\bibnamefont {Gagliardi}},\
  }\bibfield  {title} {\enquote {\bibinfo {title} {Periodic electronic
  structure calculations with the density matrix embedding theory},}\ }\href
  {https://doi.org/10.1021/acs.jctc.9b00939} {\bibfield  {journal} {\bibinfo
  {journal} {Journal of Chemical Theory and Computation}\ }\textbf {\bibinfo
  {volume} {16}},\ \bibinfo {pages} {130--140} (\bibinfo {year}
  {2020})}\BibitemShut {NoStop}%
\bibitem [{\citenamefont {Lau}, \citenamefont {Knizia},\ and\ \citenamefont
  {Berkelbach}(2021)}]{Lau2021}%
  \BibitemOpen
  \bibfield  {author} {\bibinfo {author} {\bibfnamefont {B.~T.}\ \bibnamefont
  {Lau}}, \bibinfo {author} {\bibfnamefont {G.}~\bibnamefont {Knizia}},\ and\
  \bibinfo {author} {\bibfnamefont {T.~C.}\ \bibnamefont {Berkelbach}},\
  }\bibfield  {title} {\enquote {\bibinfo {title} {{Regional Embedding Enables
  High-Level Quantum Chemistry for Surface Science}},}\ }\href
  {https://doi.org/10.1021/acs.jpclett.0c03274} {\bibfield  {journal} {\bibinfo
   {journal} {Journal of Physical Chemistry Letters}\ }\textbf {\bibinfo
  {volume} {12}},\ \bibinfo {pages} {1104--1109} (\bibinfo {year}
  {2021})}\BibitemShut {NoStop}%
\bibitem [{\citenamefont {Sekaran}\ \emph {et~al.}(2021)\citenamefont
  {Sekaran}, \citenamefont {Tsuchiizu}, \citenamefont {Sauban\`ere},\ and\
  \citenamefont {Fromager}}]{PhysRevB.104.035121}%
  \BibitemOpen
  \bibfield  {author} {\bibinfo {author} {\bibfnamefont {S.}~\bibnamefont
  {Sekaran}}, \bibinfo {author} {\bibfnamefont {M.}~\bibnamefont {Tsuchiizu}},
  \bibinfo {author} {\bibfnamefont {M.}~\bibnamefont {Sauban\`ere}},\ and\
  \bibinfo {author} {\bibfnamefont {E.}~\bibnamefont {Fromager}},\ }\bibfield
  {title} {\enquote {\bibinfo {title} {Householder-transformed density matrix
  functional embedding theory},}\ }\href
  {https://doi.org/10.1103/PhysRevB.104.035121} {\bibfield  {journal} {\bibinfo
   {journal} {Phys. Rev. B}\ }\textbf {\bibinfo {volume} {104}},\ \bibinfo
  {pages} {035121} (\bibinfo {year} {2021})}\BibitemShut {NoStop}%
\bibitem [{\citenamefont {Hermes}\ and\ \citenamefont
  {Gagliardi}(2019)}]{doi:10.1021/acs.jctc.8b01009}%
  \BibitemOpen
  \bibfield  {author} {\bibinfo {author} {\bibfnamefont {M.~R.}\ \bibnamefont
  {Hermes}}\ and\ \bibinfo {author} {\bibfnamefont {L.}~\bibnamefont
  {Gagliardi}},\ }\bibfield  {title} {\enquote {\bibinfo {title}
  {Multiconfigurational self-consistent field theory with density matrix
  embedding: The localized active space self-consistent field method},}\ }\href
  {https://doi.org/10.1021/acs.jctc.8b01009} {\bibfield  {journal} {\bibinfo
  {journal} {Journal of Chemical Theory and Computation}\ }\textbf {\bibinfo
  {volume} {15}},\ \bibinfo {pages} {972--986} (\bibinfo {year} {2019})},\
  \bibinfo {note} {pMID: 30620876},\ \Eprint
  {https://arxiv.org/abs/https://doi.org/10.1021/acs.jctc.8b01009}
  {https://doi.org/10.1021/acs.jctc.8b01009} \BibitemShut {NoStop}%
\bibitem [{\citenamefont {Lin}\ and\ \citenamefont
  {Lindsey}(2022)}]{https://doi.org/10.1002/cpa.21984}%
  \BibitemOpen
  \bibfield  {author} {\bibinfo {author} {\bibfnamefont {L.}~\bibnamefont
  {Lin}}\ and\ \bibinfo {author} {\bibfnamefont {M.}~\bibnamefont {Lindsey}},\
  }\bibfield  {title} {\enquote {\bibinfo {title} {Variational embedding for
  quantum many-body problems},}\ }\href
  {https://doi.org/https://doi.org/10.1002/cpa.21984} {\bibfield  {journal}
  {\bibinfo  {journal} {Communications on Pure and Applied Mathematics}\
  }\textbf {\bibinfo {volume} {75}},\ \bibinfo {pages} {2033--2068} (\bibinfo
  {year} {2022})},\ \Eprint
  {https://arxiv.org/abs/https://onlinelibrary.wiley.com/doi/pdf/10.1002/cpa.21984}
  {https://onlinelibrary.wiley.com/doi/pdf/10.1002/cpa.21984} \BibitemShut
  {NoStop}%
\bibitem [{\citenamefont {Cui}\ \emph {et~al.}(2022)\citenamefont {Cui},
  \citenamefont {Zhai}, \citenamefont {Zhang},\ and\ \citenamefont
  {Chan}}]{doi:10.1126/science.abm2295}%
  \BibitemOpen
  \bibfield  {author} {\bibinfo {author} {\bibfnamefont {Z.-H.}\ \bibnamefont
  {Cui}}, \bibinfo {author} {\bibfnamefont {H.}~\bibnamefont {Zhai}}, \bibinfo
  {author} {\bibfnamefont {X.}~\bibnamefont {Zhang}},\ and\ \bibinfo {author}
  {\bibfnamefont {G.~K.-L.}\ \bibnamefont {Chan}},\ }\bibfield  {title}
  {\enquote {\bibinfo {title} {Systematic electronic structure in the cuprate
  parent state from quantum many-body simulations},}\ }\href
  {https://doi.org/10.1126/science.abm2295} {\bibfield  {journal} {\bibinfo
  {journal} {Science}\ }\textbf {\bibinfo {volume} {377}},\ \bibinfo {pages}
  {1192--1198} (\bibinfo {year} {2022})},\ \Eprint
  {https://arxiv.org/abs/https://www.science.org/doi/pdf/10.1126/science.abm2295}
  {https://www.science.org/doi/pdf/10.1126/science.abm2295} \BibitemShut
  {NoStop}%
\bibitem [{\citenamefont {Cui}\ \emph {et~al.}(2020)\citenamefont {Cui},
  \citenamefont {Sun}, \citenamefont {Ray}, \citenamefont {Zheng},
  \citenamefont {Sun},\ and\ \citenamefont {Chan}}]{PhysRevResearch.2.043259}%
  \BibitemOpen
  \bibfield  {author} {\bibinfo {author} {\bibfnamefont {Z.-H.}\ \bibnamefont
  {Cui}}, \bibinfo {author} {\bibfnamefont {C.}~\bibnamefont {Sun}}, \bibinfo
  {author} {\bibfnamefont {U.}~\bibnamefont {Ray}}, \bibinfo {author}
  {\bibfnamefont {B.-X.}\ \bibnamefont {Zheng}}, \bibinfo {author}
  {\bibfnamefont {Q.}~\bibnamefont {Sun}},\ and\ \bibinfo {author}
  {\bibfnamefont {G.~K.-L.}\ \bibnamefont {Chan}},\ }\bibfield  {title}
  {\enquote {\bibinfo {title} {Ground-state phase diagram of the three-band
  hubbard model from density matrix embedding theory},}\ }\href
  {https://doi.org/10.1103/PhysRevResearch.2.043259} {\bibfield  {journal}
  {\bibinfo  {journal} {Phys. Rev. Research}\ }\textbf {\bibinfo {volume}
  {2}},\ \bibinfo {pages} {043259} (\bibinfo {year} {2020})}\BibitemShut
  {NoStop}%
\bibitem [{\citenamefont {Mitra}\ \emph {et~al.}(2022)\citenamefont {Mitra},
  \citenamefont {Hermes}, \citenamefont {Cho}, \citenamefont {Agarawal},\ and\
  \citenamefont {Gagliardi}}]{doi:10.1021/acs.jpclett.2c01915}%
  \BibitemOpen
  \bibfield  {author} {\bibinfo {author} {\bibfnamefont {A.}~\bibnamefont
  {Mitra}}, \bibinfo {author} {\bibfnamefont {M.~R.}\ \bibnamefont {Hermes}},
  \bibinfo {author} {\bibfnamefont {M.}~\bibnamefont {Cho}}, \bibinfo {author}
  {\bibfnamefont {V.}~\bibnamefont {Agarawal}},\ and\ \bibinfo {author}
  {\bibfnamefont {L.}~\bibnamefont {Gagliardi}},\ }\bibfield  {title} {\enquote
  {\bibinfo {title} {Periodic density matrix embedding for co adsorption on the
  mgo(001) surface},}\ }\href {https://doi.org/10.1021/acs.jpclett.2c01915}
  {\bibfield  {journal} {\bibinfo  {journal} {The Journal of Physical Chemistry
  Letters}\ }\textbf {\bibinfo {volume} {13}},\ \bibinfo {pages} {7483--7489}
  (\bibinfo {year} {2022})},\ \bibinfo {note} {pMID: 35939641},\ \Eprint
  {https://arxiv.org/abs/https://doi.org/10.1021/acs.jpclett.2c01915}
  {https://doi.org/10.1021/acs.jpclett.2c01915} \BibitemShut {NoStop}%
\bibitem [{\citenamefont {Cao}\ \emph {et~al.}(2022)\citenamefont {Cao},
  \citenamefont {Sun}, \citenamefont {Yuan}, \citenamefont {Hu}, \citenamefont
  {Pham},\ and\ \citenamefont
  {Lv}}]{https://doi.org/10.48550/arxiv.2209.03202}%
  \BibitemOpen
  \bibfield  {author} {\bibinfo {author} {\bibfnamefont {C.}~\bibnamefont
  {Cao}}, \bibinfo {author} {\bibfnamefont {J.}~\bibnamefont {Sun}}, \bibinfo
  {author} {\bibfnamefont {X.}~\bibnamefont {Yuan}}, \bibinfo {author}
  {\bibfnamefont {H.-S.}\ \bibnamefont {Hu}}, \bibinfo {author} {\bibfnamefont
  {H.~Q.}\ \bibnamefont {Pham}},\ and\ \bibinfo {author} {\bibfnamefont
  {D.}~\bibnamefont {Lv}},\ }\href {https://doi.org/10.48550/ARXIV.2209.03202}
  {\enquote {\bibinfo {title} {Ab initio quantum simulation of strongly
  correlated materials with quantum embedding},}\ } (\bibinfo {year}
  {2022})\BibitemShut {NoStop}%
\bibitem [{\citenamefont {Lee}\ \emph {et~al.}(2019)\citenamefont {Lee},
  \citenamefont {Ayral}, \citenamefont {Yao}, \citenamefont {Lanata},\ and\
  \citenamefont {Kotliar}}]{PhysRevB.99.115129}%
  \BibitemOpen
  \bibfield  {author} {\bibinfo {author} {\bibfnamefont {T.-H.}\ \bibnamefont
  {Lee}}, \bibinfo {author} {\bibfnamefont {T.}~\bibnamefont {Ayral}}, \bibinfo
  {author} {\bibfnamefont {Y.-X.}\ \bibnamefont {Yao}}, \bibinfo {author}
  {\bibfnamefont {N.}~\bibnamefont {Lanata}},\ and\ \bibinfo {author}
  {\bibfnamefont {G.}~\bibnamefont {Kotliar}},\ }\bibfield  {title} {\enquote
  {\bibinfo {title} {Rotationally invariant slave-boson and density matrix
  embedding theory: Unified framework and comparative study on the
  one-dimensional and two-dimensional hubbard model},}\ }\href
  {https://doi.org/10.1103/PhysRevB.99.115129} {\bibfield  {journal} {\bibinfo
  {journal} {Phys. Rev. B}\ }\textbf {\bibinfo {volume} {99}},\ \bibinfo
  {pages} {115129} (\bibinfo {year} {2019})}\BibitemShut {NoStop}%
\bibitem [{\citenamefont {Fertitta}\ and\ \citenamefont
  {Booth}(2019)}]{Fertitta2019}%
  \BibitemOpen
  \bibfield  {author} {\bibinfo {author} {\bibfnamefont {E.}~\bibnamefont
  {Fertitta}}\ and\ \bibinfo {author} {\bibfnamefont {G.~H.}\ \bibnamefont
  {Booth}},\ }\bibfield  {title} {\enquote {\bibinfo {title} {{Energy-weighted
  density matrix embedding of open correlated chemical fragments}},}\ }\href
  {https://doi.org/10.1063/1.5100290} {\bibfield  {journal} {\bibinfo
  {journal} {The Journal of Chemical Physics}\ }\textbf {\bibinfo {volume}
  {151}},\ \bibinfo {pages} {014115} (\bibinfo {year} {2019})}\BibitemShut
  {NoStop}%
\bibitem [{\citenamefont {Sriluckshmy}\ \emph {et~al.}(2021)\citenamefont
  {Sriluckshmy}, \citenamefont {Nusspickel}, \citenamefont {Fertitta},\ and\
  \citenamefont {Booth}}]{Sriluckshmy2021}%
  \BibitemOpen
  \bibfield  {author} {\bibinfo {author} {\bibfnamefont {P.~V.}\ \bibnamefont
  {Sriluckshmy}}, \bibinfo {author} {\bibfnamefont {M.}~\bibnamefont
  {Nusspickel}}, \bibinfo {author} {\bibfnamefont {E.}~\bibnamefont
  {Fertitta}},\ and\ \bibinfo {author} {\bibfnamefont {G.~H.}\ \bibnamefont
  {Booth}},\ }\bibfield  {title} {\enquote {\bibinfo {title} {{Fully algebraic
  and self-consistent effective dynamics in a static quantum embedding}},}\
  }\href {https://doi.org/10.1103/PhysRevB.103.085131} {\bibfield  {journal}
  {\bibinfo  {journal} {Physical Review B}\ }\textbf {\bibinfo {volume}
  {103}},\ \bibinfo {pages} {085131} (\bibinfo {year} {2021})}\BibitemShut
  {NoStop}%
\bibitem [{\citenamefont {Mineh}\ and\ \citenamefont
  {Montanaro}(2022)}]{PhysRevB.105.125117}%
  \BibitemOpen
  \bibfield  {author} {\bibinfo {author} {\bibfnamefont {L.}~\bibnamefont
  {Mineh}}\ and\ \bibinfo {author} {\bibfnamefont {A.}~\bibnamefont
  {Montanaro}},\ }\bibfield  {title} {\enquote {\bibinfo {title} {Solving the
  hubbard model using density matrix embedding theory and the variational
  quantum eigensolver},}\ }\href {https://doi.org/10.1103/PhysRevB.105.125117}
  {\bibfield  {journal} {\bibinfo  {journal} {Phys. Rev. B}\ }\textbf {\bibinfo
  {volume} {105}},\ \bibinfo {pages} {125117} (\bibinfo {year}
  {2022})}\BibitemShut {NoStop}%
\bibitem [{\citenamefont {Kirsopp}\ \emph {et~al.}()\citenamefont {Kirsopp},
  \citenamefont {Di~Paola}, \citenamefont {Manrique}, \citenamefont {Krompiec},
  \citenamefont {Greene-Diniz}, \citenamefont {Guba}, \citenamefont {Meyder},
  \citenamefont {Wolf}, \citenamefont {Strahm},\ and\ \citenamefont
  {Muñoz~Ramo}}]{https://doi.org/10.1002/qua.26975}%
  \BibitemOpen
  \bibfield  {author} {\bibinfo {author} {\bibfnamefont {J.~J.~M.}\
  \bibnamefont {Kirsopp}}, \bibinfo {author} {\bibfnamefont {C.}~\bibnamefont
  {Di~Paola}}, \bibinfo {author} {\bibfnamefont {D.~Z.}\ \bibnamefont
  {Manrique}}, \bibinfo {author} {\bibfnamefont {M.}~\bibnamefont {Krompiec}},
  \bibinfo {author} {\bibfnamefont {G.}~\bibnamefont {Greene-Diniz}}, \bibinfo
  {author} {\bibfnamefont {W.}~\bibnamefont {Guba}}, \bibinfo {author}
  {\bibfnamefont {A.}~\bibnamefont {Meyder}}, \bibinfo {author} {\bibfnamefont
  {D.}~\bibnamefont {Wolf}}, \bibinfo {author} {\bibfnamefont {M.}~\bibnamefont
  {Strahm}},\ and\ \bibinfo {author} {\bibfnamefont {D.}~\bibnamefont
  {Muñoz~Ramo}},\ }\bibfield  {title} {\enquote {\bibinfo {title} {Quantum
  computational quantification of protein–ligand interactions},}\ }\href
  {https://doi.org/https://doi.org/10.1002/qua.26975} {\bibfield  {journal}
  {\bibinfo  {journal} {International Journal of Quantum Chemistry}\ }\textbf
  {\bibinfo {volume} {n/a}},\ \bibinfo {pages} {e26975}},\ \Eprint
  {https://arxiv.org/abs/https://onlinelibrary.wiley.com/doi/pdf/10.1002/qua.26975}
  {https://onlinelibrary.wiley.com/doi/pdf/10.1002/qua.26975} \BibitemShut
  {NoStop}%
\bibitem [{\citenamefont {Li}\ \emph {et~al.}(2022)\citenamefont {Li},
  \citenamefont {Huang}, \citenamefont {Cao}, \citenamefont {Huang},
  \citenamefont {Shuai}, \citenamefont {Sun}, \citenamefont {Sun},
  \citenamefont {Yuan},\ and\ \citenamefont {Lv}}]{D2SC01492K}%
  \BibitemOpen
  \bibfield  {author} {\bibinfo {author} {\bibfnamefont {W.}~\bibnamefont
  {Li}}, \bibinfo {author} {\bibfnamefont {Z.}~\bibnamefont {Huang}}, \bibinfo
  {author} {\bibfnamefont {C.}~\bibnamefont {Cao}}, \bibinfo {author}
  {\bibfnamefont {Y.}~\bibnamefont {Huang}}, \bibinfo {author} {\bibfnamefont
  {Z.}~\bibnamefont {Shuai}}, \bibinfo {author} {\bibfnamefont
  {X.}~\bibnamefont {Sun}}, \bibinfo {author} {\bibfnamefont {J.}~\bibnamefont
  {Sun}}, \bibinfo {author} {\bibfnamefont {X.}~\bibnamefont {Yuan}},\ and\
  \bibinfo {author} {\bibfnamefont {D.}~\bibnamefont {Lv}},\ }\bibfield
  {title} {\enquote {\bibinfo {title} {Toward practical quantum embedding
  simulation of realistic chemical systems on near-term quantum computers},}\
  }\href {https://doi.org/10.1039/D2SC01492K} {\bibfield  {journal} {\bibinfo
  {journal} {Chem. Sci.}\ }\textbf {\bibinfo {volume} {13}},\ \bibinfo {pages}
  {8953--8962} (\bibinfo {year} {2022})}\BibitemShut {NoStop}%
\bibitem [{\citenamefont {Kawashima}\ \emph {et~al.}(2021)\citenamefont
  {Kawashima}, \citenamefont {Lloyd}, \citenamefont {Coons}, \citenamefont
  {Nam}, \citenamefont {Matsuura}, \citenamefont {Garza}, \citenamefont
  {Johri}, \citenamefont {Huntington}, \citenamefont {Senicourt}, \citenamefont
  {Maksymov}, \citenamefont {Nguyen}, \citenamefont {Kim}, \citenamefont
  {Alidoust}, \citenamefont {Zaribafiyan},\ and\ \citenamefont
  {Yamazaki}}]{DMET_QC}%
  \BibitemOpen
  \bibfield  {author} {\bibinfo {author} {\bibfnamefont {Y.}~\bibnamefont
  {Kawashima}}, \bibinfo {author} {\bibfnamefont {E.}~\bibnamefont {Lloyd}},
  \bibinfo {author} {\bibfnamefont {M.~P.}\ \bibnamefont {Coons}}, \bibinfo
  {author} {\bibfnamefont {Y.}~\bibnamefont {Nam}}, \bibinfo {author}
  {\bibfnamefont {S.}~\bibnamefont {Matsuura}}, \bibinfo {author}
  {\bibfnamefont {A.~J.}\ \bibnamefont {Garza}}, \bibinfo {author}
  {\bibfnamefont {S.}~\bibnamefont {Johri}}, \bibinfo {author} {\bibfnamefont
  {L.}~\bibnamefont {Huntington}}, \bibinfo {author} {\bibfnamefont
  {V.}~\bibnamefont {Senicourt}}, \bibinfo {author} {\bibfnamefont {A.~O.}\
  \bibnamefont {Maksymov}}, \bibinfo {author} {\bibfnamefont {J.~H.~V.}\
  \bibnamefont {Nguyen}}, \bibinfo {author} {\bibfnamefont {J.}~\bibnamefont
  {Kim}}, \bibinfo {author} {\bibfnamefont {N.}~\bibnamefont {Alidoust}},
  \bibinfo {author} {\bibfnamefont {A.}~\bibnamefont {Zaribafiyan}},\ and\
  \bibinfo {author} {\bibfnamefont {T.}~\bibnamefont {Yamazaki}},\ }\bibfield
  {title} {\enquote {\bibinfo {title} {Optimizing electronic structure
  simulations on a trapped-ion quantum computer using problem decomposition},}\
  }\href {https://doi.org/10.1038/s42005-021-00751-9} {\bibfield  {journal}
  {\bibinfo  {journal} {Communications Physics}\ }\textbf {\bibinfo {volume}
  {4}},\ \bibinfo {pages} {245} (\bibinfo {year} {2021})}\BibitemShut {NoStop}%
\bibitem [{\citenamefont {Tilly}\ \emph {et~al.}(2021)\citenamefont {Tilly},
  \citenamefont {Sriluckshmy}, \citenamefont {Patel}, \citenamefont {Fontana},
  \citenamefont {Rungger}, \citenamefont {Grant}, \citenamefont {Anderson},
  \citenamefont {Tennyson},\ and\ \citenamefont
  {Booth}}]{PhysRevResearch.3.033230}%
  \BibitemOpen
  \bibfield  {author} {\bibinfo {author} {\bibfnamefont {J.}~\bibnamefont
  {Tilly}}, \bibinfo {author} {\bibfnamefont {P.~V.}\ \bibnamefont
  {Sriluckshmy}}, \bibinfo {author} {\bibfnamefont {A.}~\bibnamefont {Patel}},
  \bibinfo {author} {\bibfnamefont {E.}~\bibnamefont {Fontana}}, \bibinfo
  {author} {\bibfnamefont {I.}~\bibnamefont {Rungger}}, \bibinfo {author}
  {\bibfnamefont {E.}~\bibnamefont {Grant}}, \bibinfo {author} {\bibfnamefont
  {R.}~\bibnamefont {Anderson}}, \bibinfo {author} {\bibfnamefont
  {J.}~\bibnamefont {Tennyson}},\ and\ \bibinfo {author} {\bibfnamefont
  {G.~H.}\ \bibnamefont {Booth}},\ }\bibfield  {title} {\enquote {\bibinfo
  {title} {Reduced density matrix sampling: Self-consistent embedding and
  multiscale electronic structure on current generation quantum computers},}\
  }\href {https://doi.org/10.1103/PhysRevResearch.3.033230} {\bibfield
  {journal} {\bibinfo  {journal} {Phys. Rev. Research}\ }\textbf {\bibinfo
  {volume} {3}},\ \bibinfo {pages} {033230} (\bibinfo {year}
  {2021})}\BibitemShut {NoStop}%
\bibitem [{\citenamefont {Nusspickel}\ and\ \citenamefont
  {Booth}(2022)}]{PhysRevX.12.011046}%
  \BibitemOpen
  \bibfield  {author} {\bibinfo {author} {\bibfnamefont {M.}~\bibnamefont
  {Nusspickel}}\ and\ \bibinfo {author} {\bibfnamefont {G.~H.}\ \bibnamefont
  {Booth}},\ }\bibfield  {title} {\enquote {\bibinfo {title} {Systematic
  improvability in quantum embedding for real materials},}\ }\href
  {https://doi.org/10.1103/PhysRevX.12.011046} {\bibfield  {journal} {\bibinfo
  {journal} {Phys. Rev. X}\ }\textbf {\bibinfo {volume} {12}},\ \bibinfo
  {pages} {011046} (\bibinfo {year} {2022})}\BibitemShut {NoStop}%
\bibitem [{\citenamefont {Karton}, \citenamefont {Daon},\ and\ \citenamefont
  {Martin}(2011)}]{KARTON2011165}%
  \BibitemOpen
  \bibfield  {author} {\bibinfo {author} {\bibfnamefont {A.}~\bibnamefont
  {Karton}}, \bibinfo {author} {\bibfnamefont {S.}~\bibnamefont {Daon}},\ and\
  \bibinfo {author} {\bibfnamefont {J.~M.}\ \bibnamefont {Martin}},\ }\bibfield
   {title} {\enquote {\bibinfo {title} {W4-11: A high-confidence benchmark
  dataset for computational thermochemistry derived from first-principles w4
  data},}\ }\href
  {https://doi.org/https://doi.org/10.1016/j.cplett.2011.05.007} {\bibfield
  {journal} {\bibinfo  {journal} {Chemical Physics Letters}\ }\textbf {\bibinfo
  {volume} {510}},\ \bibinfo {pages} {165--178} (\bibinfo {year}
  {2011})}\BibitemShut {NoStop}%
\bibitem [{Note1()}]{Note1}%
  \BibitemOpen
  \bibinfo {note} {Embedding code and documentation can be found at
  https://github.com/BoothGroup/Vayesta}\BibitemShut {NoStop}%
\bibitem [{\citenamefont {Sun}\ \emph {et~al.}(2020)\citenamefont {Sun},
  \citenamefont {Zhang}, \citenamefont {Banerjee}, \citenamefont {Bao},
  \citenamefont {Barbry}, \citenamefont {Blunt}, \citenamefont {Bogdanov},
  \citenamefont {Booth}, \citenamefont {Chen}, \citenamefont {Cui},
  \citenamefont {Eriksen}, \citenamefont {Gao}, \citenamefont {Guo},
  \citenamefont {Hermann}, \citenamefont {Hermes}, \citenamefont {Koh},
  \citenamefont {Koval}, \citenamefont {Lehtola}, \citenamefont {Li},
  \citenamefont {Liu}, \citenamefont {Mardirossian}, \citenamefont {McClain},
  \citenamefont {Motta}, \citenamefont {Mussard}, \citenamefont {Pham},
  \citenamefont {Pulkin}, \citenamefont {Purwanto}, \citenamefont {Robinson},
  \citenamefont {Ronca}, \citenamefont {Sayfutyarova}, \citenamefont
  {Scheurer}, \citenamefont {Schurkus}, \citenamefont {Smith}, \citenamefont
  {Sun}, \citenamefont {Sun}, \citenamefont {Upadhyay}, \citenamefont {Wagner},
  \citenamefont {Wang}, \citenamefont {White}, \citenamefont {Whitfield},
  \citenamefont {Williamson}, \citenamefont {Wouters}, \citenamefont {Yang},
  \citenamefont {Yu}, \citenamefont {Zhu}, \citenamefont {Berkelbach},
  \citenamefont {Sharma}, \citenamefont {Sokolov},\ and\ \citenamefont
  {Chan}}]{Sun2020}%
  \BibitemOpen
  \bibfield  {author} {\bibinfo {author} {\bibfnamefont {Q.}~\bibnamefont
  {Sun}}, \bibinfo {author} {\bibfnamefont {X.}~\bibnamefont {Zhang}}, \bibinfo
  {author} {\bibfnamefont {S.}~\bibnamefont {Banerjee}}, \bibinfo {author}
  {\bibfnamefont {P.}~\bibnamefont {Bao}}, \bibinfo {author} {\bibfnamefont
  {M.}~\bibnamefont {Barbry}}, \bibinfo {author} {\bibfnamefont {N.~S.}\
  \bibnamefont {Blunt}}, \bibinfo {author} {\bibfnamefont {N.~A.}\ \bibnamefont
  {Bogdanov}}, \bibinfo {author} {\bibfnamefont {G.~H.}\ \bibnamefont {Booth}},
  \bibinfo {author} {\bibfnamefont {J.}~\bibnamefont {Chen}}, \bibinfo {author}
  {\bibfnamefont {Z.-H.}\ \bibnamefont {Cui}}, \bibinfo {author} {\bibfnamefont
  {J.~J.}\ \bibnamefont {Eriksen}}, \bibinfo {author} {\bibfnamefont
  {Y.}~\bibnamefont {Gao}}, \bibinfo {author} {\bibfnamefont {S.}~\bibnamefont
  {Guo}}, \bibinfo {author} {\bibfnamefont {J.}~\bibnamefont {Hermann}},
  \bibinfo {author} {\bibfnamefont {M.~R.}\ \bibnamefont {Hermes}}, \bibinfo
  {author} {\bibfnamefont {K.}~\bibnamefont {Koh}}, \bibinfo {author}
  {\bibfnamefont {P.}~\bibnamefont {Koval}}, \bibinfo {author} {\bibfnamefont
  {S.}~\bibnamefont {Lehtola}}, \bibinfo {author} {\bibfnamefont
  {Z.}~\bibnamefont {Li}}, \bibinfo {author} {\bibfnamefont {J.}~\bibnamefont
  {Liu}}, \bibinfo {author} {\bibfnamefont {N.}~\bibnamefont {Mardirossian}},
  \bibinfo {author} {\bibfnamefont {J.~D.}\ \bibnamefont {McClain}}, \bibinfo
  {author} {\bibfnamefont {M.}~\bibnamefont {Motta}}, \bibinfo {author}
  {\bibfnamefont {B.}~\bibnamefont {Mussard}}, \bibinfo {author} {\bibfnamefont
  {H.~Q.}\ \bibnamefont {Pham}}, \bibinfo {author} {\bibfnamefont
  {A.}~\bibnamefont {Pulkin}}, \bibinfo {author} {\bibfnamefont
  {W.}~\bibnamefont {Purwanto}}, \bibinfo {author} {\bibfnamefont {P.~J.}\
  \bibnamefont {Robinson}}, \bibinfo {author} {\bibfnamefont {E.}~\bibnamefont
  {Ronca}}, \bibinfo {author} {\bibfnamefont {E.~R.}\ \bibnamefont
  {Sayfutyarova}}, \bibinfo {author} {\bibfnamefont {M.}~\bibnamefont
  {Scheurer}}, \bibinfo {author} {\bibfnamefont {H.~F.}\ \bibnamefont
  {Schurkus}}, \bibinfo {author} {\bibfnamefont {J.~E.~T.}\ \bibnamefont
  {Smith}}, \bibinfo {author} {\bibfnamefont {C.}~\bibnamefont {Sun}}, \bibinfo
  {author} {\bibfnamefont {S.-N.}\ \bibnamefont {Sun}}, \bibinfo {author}
  {\bibfnamefont {S.}~\bibnamefont {Upadhyay}}, \bibinfo {author}
  {\bibfnamefont {L.~K.}\ \bibnamefont {Wagner}}, \bibinfo {author}
  {\bibfnamefont {X.}~\bibnamefont {Wang}}, \bibinfo {author} {\bibfnamefont
  {A.}~\bibnamefont {White}}, \bibinfo {author} {\bibfnamefont {J.~D.}\
  \bibnamefont {Whitfield}}, \bibinfo {author} {\bibfnamefont {M.~J.}\
  \bibnamefont {Williamson}}, \bibinfo {author} {\bibfnamefont
  {S.}~\bibnamefont {Wouters}}, \bibinfo {author} {\bibfnamefont
  {J.}~\bibnamefont {Yang}}, \bibinfo {author} {\bibfnamefont {J.~M.}\
  \bibnamefont {Yu}}, \bibinfo {author} {\bibfnamefont {T.}~\bibnamefont
  {Zhu}}, \bibinfo {author} {\bibfnamefont {T.~C.}\ \bibnamefont {Berkelbach}},
  \bibinfo {author} {\bibfnamefont {S.}~\bibnamefont {Sharma}}, \bibinfo
  {author} {\bibfnamefont {A.~Y.}\ \bibnamefont {Sokolov}},\ and\ \bibinfo
  {author} {\bibfnamefont {G.~K.-L.}\ \bibnamefont {Chan}},\ }\bibfield
  {title} {\enquote {\bibinfo {title} {{Recent developments in the PySCF
  program package}},}\ }\href {https://doi.org/10.1063/5.0006074} {\bibfield
  {journal} {\bibinfo  {journal} {The Journal of Chemical Physics}\ }\textbf
  {\bibinfo {volume} {153}},\ \bibinfo {pages} {024109} (\bibinfo {year}
  {2020})}\BibitemShut {NoStop}%
\bibitem [{\citenamefont {Sun}\ \emph {et~al.}(2018)\citenamefont {Sun},
  \citenamefont {Berkelbach}, \citenamefont {Blunt}, \citenamefont {Booth},
  \citenamefont {Guo}, \citenamefont {Li}, \citenamefont {Liu}, \citenamefont
  {McClain}, \citenamefont {Sayfutyarova}, \citenamefont {Sharma},
  \citenamefont {Wouters},\ and\ \citenamefont {Chan}}]{Sun2018}%
  \BibitemOpen
  \bibfield  {author} {\bibinfo {author} {\bibfnamefont {Q.}~\bibnamefont
  {Sun}}, \bibinfo {author} {\bibfnamefont {T.~C.}\ \bibnamefont {Berkelbach}},
  \bibinfo {author} {\bibfnamefont {N.~S.}\ \bibnamefont {Blunt}}, \bibinfo
  {author} {\bibfnamefont {G.~H.}\ \bibnamefont {Booth}}, \bibinfo {author}
  {\bibfnamefont {S.}~\bibnamefont {Guo}}, \bibinfo {author} {\bibfnamefont
  {Z.}~\bibnamefont {Li}}, \bibinfo {author} {\bibfnamefont {J.}~\bibnamefont
  {Liu}}, \bibinfo {author} {\bibfnamefont {J.~D.}\ \bibnamefont {McClain}},
  \bibinfo {author} {\bibfnamefont {E.~R.}\ \bibnamefont {Sayfutyarova}},
  \bibinfo {author} {\bibfnamefont {S.}~\bibnamefont {Sharma}}, \bibinfo
  {author} {\bibfnamefont {S.}~\bibnamefont {Wouters}},\ and\ \bibinfo {author}
  {\bibfnamefont {G.~K.}\ \bibnamefont {Chan}},\ }\bibfield  {title} {\enquote
  {\bibinfo {title} {{PySCF: the Python‐based simulations of chemistry
  framework}},}\ }\href {https://doi.org/10.1002/wcms.1340} {\bibfield
  {journal} {\bibinfo  {journal} {WIREs Computational Molecular Science}\
  }\textbf {\bibinfo {volume} {8}} (\bibinfo {year} {2018}),\
  10.1002/wcms.1340}\BibitemShut {NoStop}%
\bibitem [{\citenamefont {Pulay}(1983)}]{PULAY1983151}%
  \BibitemOpen
  \bibfield  {author} {\bibinfo {author} {\bibfnamefont {P.}~\bibnamefont
  {Pulay}},\ }\bibfield  {title} {\enquote {\bibinfo {title} {Localizability of
  dynamic electron correlation},}\ }\href
  {https://doi.org/https://doi.org/10.1016/0009-2614(83)80703-9} {\bibfield
  {journal} {\bibinfo  {journal} {Chemical Physics Letters}\ }\textbf {\bibinfo
  {volume} {100}},\ \bibinfo {pages} {151--154} (\bibinfo {year}
  {1983})}\BibitemShut {NoStop}%
\bibitem [{\citenamefont {Bulik}, \citenamefont {Chen},\ and\ \citenamefont
  {Scuseria}(2014)}]{doi:10.1063/1.4891861}%
  \BibitemOpen
  \bibfield  {author} {\bibinfo {author} {\bibfnamefont {I.~W.}\ \bibnamefont
  {Bulik}}, \bibinfo {author} {\bibfnamefont {W.}~\bibnamefont {Chen}},\ and\
  \bibinfo {author} {\bibfnamefont {G.~E.}\ \bibnamefont {Scuseria}},\
  }\bibfield  {title} {\enquote {\bibinfo {title} {Electron correlation in
  solids via density embedding theory},}\ }\href
  {https://doi.org/10.1063/1.4891861} {\bibfield  {journal} {\bibinfo
  {journal} {The Journal of Chemical Physics}\ }\textbf {\bibinfo {volume}
  {141}},\ \bibinfo {pages} {054113} (\bibinfo {year} {2014})},\ \Eprint
  {https://arxiv.org/abs/https://doi.org/10.1063/1.4891861}
  {https://doi.org/10.1063/1.4891861} \BibitemShut {NoStop}%
\bibitem [{Note2()}]{Note2}%
  \BibitemOpen
  \bibinfo {note} {We note that the two-body cumulant would often be denoted by
  $\lambda $, but define by ${\protect \tilde {K}}$ to avoid potential
  confusions with the lambda-amplitudes of coupled-cluster theory used later in
  the text.}\BibitemShut {Stop}%
\bibitem [{\citenamefont {Knizia}(2013)}]{Knizia2013IAO}%
  \BibitemOpen
  \bibfield  {author} {\bibinfo {author} {\bibfnamefont {G.}~\bibnamefont
  {Knizia}},\ }\bibfield  {title} {\enquote {\bibinfo {title} {{Intrinsic
  Atomic Orbitals: An Unbiased Bridge between Quantum Theory and Chemical
  Concepts}},}\ }\href {https://doi.org/10.1021/ct400687b} {\bibfield
  {journal} {\bibinfo  {journal} {Journal of Chemical Theory and Computation}\
  }\textbf {\bibinfo {volume} {9}},\ \bibinfo {pages} {4834--4843} (\bibinfo
  {year} {2013})}\BibitemShut {NoStop}%
\bibitem [{\citenamefont {Bartlett}\ and\ \citenamefont
  {Musia\l{}}(2007)}]{RevModPhys.79.291}%
  \BibitemOpen
  \bibfield  {author} {\bibinfo {author} {\bibfnamefont {R.~J.}\ \bibnamefont
  {Bartlett}}\ and\ \bibinfo {author} {\bibfnamefont {M.}~\bibnamefont
  {Musia\l{}}},\ }\bibfield  {title} {\enquote {\bibinfo {title}
  {Coupled-cluster theory in quantum chemistry},}\ }\href
  {https://doi.org/10.1103/RevModPhys.79.291} {\bibfield  {journal} {\bibinfo
  {journal} {Rev. Mod. Phys.}\ }\textbf {\bibinfo {volume} {79}},\ \bibinfo
  {pages} {291--352} (\bibinfo {year} {2007})}\BibitemShut {NoStop}%
\bibitem [{\citenamefont {Lehtola}\ \emph {et~al.}(2017)\citenamefont
  {Lehtola}, \citenamefont {Tubman}, \citenamefont {Whaley},\ and\
  \citenamefont {Head-Gordon}}]{Lehtola2017}%
  \BibitemOpen
  \bibfield  {author} {\bibinfo {author} {\bibfnamefont {S.}~\bibnamefont
  {Lehtola}}, \bibinfo {author} {\bibfnamefont {N.~M.}\ \bibnamefont {Tubman}},
  \bibinfo {author} {\bibfnamefont {K.~B.}\ \bibnamefont {Whaley}},\ and\
  \bibinfo {author} {\bibfnamefont {M.}~\bibnamefont {Head-Gordon}},\
  }\bibfield  {title} {\enquote {\bibinfo {title} {Cluster decomposition of
  full configuration interaction wave functions: A tool for chemical
  interpretation of systems with strong correlation},}\ }\href
  {https://doi.org/10.1063/1.4996044} {\bibfield  {journal} {\bibinfo
  {journal} {The Journal of Chemical Physics}\ }\textbf {\bibinfo {volume}
  {147}},\ \bibinfo {pages} {154105} (\bibinfo {year} {2017})},\ \Eprint
  {https://arxiv.org/abs/https://doi.org/10.1063/1.4996044}
  {https://doi.org/10.1063/1.4996044} \BibitemShut {NoStop}%
\bibitem [{\citenamefont {Shee}\ and\ \citenamefont {Zgid}(2019)}]{Shee2019}%
  \BibitemOpen
  \bibfield  {author} {\bibinfo {author} {\bibfnamefont {A.}~\bibnamefont
  {Shee}}\ and\ \bibinfo {author} {\bibfnamefont {D.}~\bibnamefont {Zgid}},\
  }\bibfield  {title} {\enquote {\bibinfo {title} {{Coupled Cluster as an
  Impurity Solver for Green's Function Embedding Methods}},}\ }\href
  {https://doi.org/10.1021/acs.jctc.9b00603} {\bibfield  {journal} {\bibinfo
  {journal} {Journal of Chemical Theory and Computation}\ }\textbf {\bibinfo
  {volume} {15}},\ \bibinfo {pages} {6010--6024} (\bibinfo {year}
  {2019})}\BibitemShut {NoStop}%
\bibitem [{\citenamefont {Zhu}\ \emph {et~al.}(2019)\citenamefont {Zhu},
  \citenamefont {Jim{\'{e}}nez-Hoyos}, \citenamefont {McClain}, \citenamefont
  {Berkelbach},\ and\ \citenamefont {Chan}}]{Zhu2019}%
  \BibitemOpen
  \bibfield  {author} {\bibinfo {author} {\bibfnamefont {T.}~\bibnamefont
  {Zhu}}, \bibinfo {author} {\bibfnamefont {C.~A.}\ \bibnamefont
  {Jim{\'{e}}nez-Hoyos}}, \bibinfo {author} {\bibfnamefont {J.}~\bibnamefont
  {McClain}}, \bibinfo {author} {\bibfnamefont {T.~C.}\ \bibnamefont
  {Berkelbach}},\ and\ \bibinfo {author} {\bibfnamefont {G.~K.-L.}\
  \bibnamefont {Chan}},\ }\bibfield  {title} {\enquote {\bibinfo {title}
  {{Coupled-cluster impurity solvers for dynamical mean-field theory}},}\
  }\href {https://doi.org/10.1103/PhysRevB.100.115154} {\bibfield  {journal}
  {\bibinfo  {journal} {Physical Review B}\ }\textbf {\bibinfo {volume}
  {100}},\ \bibinfo {pages} {115154} (\bibinfo {year} {2019})}\BibitemShut
  {NoStop}%
\bibitem [{\citenamefont {Shee}, \citenamefont {Yeh},\ and\ \citenamefont
  {Zgid}(2022)}]{doi:10.1021/acs.jctc.1c00712}%
  \BibitemOpen
  \bibfield  {author} {\bibinfo {author} {\bibfnamefont {A.}~\bibnamefont
  {Shee}}, \bibinfo {author} {\bibfnamefont {C.-N.}\ \bibnamefont {Yeh}},\ and\
  \bibinfo {author} {\bibfnamefont {D.}~\bibnamefont {Zgid}},\ }\bibfield
  {title} {\enquote {\bibinfo {title} {Exploring coupled cluster green’s
  function as a method for treating system and environment in green’s
  function embedding methods},}\ }\href
  {https://doi.org/10.1021/acs.jctc.1c00712} {\bibfield  {journal} {\bibinfo
  {journal} {Journal of Chemical Theory and Computation}\ }\textbf {\bibinfo
  {volume} {18}},\ \bibinfo {pages} {664--676} (\bibinfo {year} {2022})},\
  \bibinfo {note} {pMID: 34989565},\ \Eprint
  {https://arxiv.org/abs/https://doi.org/10.1021/acs.jctc.1c00712}
  {https://doi.org/10.1021/acs.jctc.1c00712} \BibitemShut {NoStop}%
\bibitem [{\citenamefont {Yeh}\ \emph {et~al.}(2021)\citenamefont {Yeh},
  \citenamefont {Shee}, \citenamefont {Iskakov},\ and\ \citenamefont
  {Zgid}}]{PhysRevB.103.155158}%
  \BibitemOpen
  \bibfield  {author} {\bibinfo {author} {\bibfnamefont {C.-N.}\ \bibnamefont
  {Yeh}}, \bibinfo {author} {\bibfnamefont {A.}~\bibnamefont {Shee}}, \bibinfo
  {author} {\bibfnamefont {S.}~\bibnamefont {Iskakov}},\ and\ \bibinfo {author}
  {\bibfnamefont {D.}~\bibnamefont {Zgid}},\ }\bibfield  {title} {\enquote
  {\bibinfo {title} {Testing the green's function coupled cluster singles and
  doubles impurity solver on real materials within the framework of self-energy
  embedding theory},}\ }\href {https://doi.org/10.1103/PhysRevB.103.155158}
  {\bibfield  {journal} {\bibinfo  {journal} {Phys. Rev. B}\ }\textbf {\bibinfo
  {volume} {103}},\ \bibinfo {pages} {155158} (\bibinfo {year}
  {2021})}\BibitemShut {NoStop}%
\bibitem [{\citenamefont {Riplinger}\ and\ \citenamefont
  {Neese}(2013)}]{Riplinger2013}%
  \BibitemOpen
  \bibfield  {author} {\bibinfo {author} {\bibfnamefont {C.}~\bibnamefont
  {Riplinger}}\ and\ \bibinfo {author} {\bibfnamefont {F.}~\bibnamefont
  {Neese}},\ }\bibfield  {title} {\enquote {\bibinfo {title} {An efficient and
  near linear scaling pair natural orbital based local coupled cluster
  method},}\ }\href {https://doi.org/10.1063/1.4773581} {\bibfield  {journal}
  {\bibinfo  {journal} {J. Chem. Phys.}\ }\textbf {\bibinfo {volume} {138}},\
  \bibinfo {pages} {034106} (\bibinfo {year} {2013})}\BibitemShut {NoStop}%
\bibitem [{\citenamefont {Li}\ and\ \citenamefont
  {Piecuch}(2010)}]{doi:10.1021/jp100782u}%
  \BibitemOpen
  \bibfield  {author} {\bibinfo {author} {\bibfnamefont {W.}~\bibnamefont
  {Li}}\ and\ \bibinfo {author} {\bibfnamefont {P.}~\bibnamefont {Piecuch}},\
  }\bibfield  {title} {\enquote {\bibinfo {title} {Improved design of orbital
  domains within the cluster-in-molecule local correlation framework:
  Single-environment cluster-in-molecule ansatz and its application to local
  coupled-cluster approach with singles and doubles},}\ }\href
  {https://doi.org/10.1021/jp100782u} {\bibfield  {journal} {\bibinfo
  {journal} {The Journal of Physical Chemistry A}\ }\textbf {\bibinfo {volume}
  {114}},\ \bibinfo {pages} {8644--8657} (\bibinfo {year} {2010})},\ \bibinfo
  {note} {pMID: 20373794}\BibitemShut {NoStop}%
\bibitem [{\citenamefont {Knowles}\ and\ \citenamefont
  {Handy}(1989)}]{KNOWLES198975}%
  \BibitemOpen
  \bibfield  {author} {\bibinfo {author} {\bibfnamefont {P.~J.}\ \bibnamefont
  {Knowles}}\ and\ \bibinfo {author} {\bibfnamefont {N.~C.}\ \bibnamefont
  {Handy}},\ }\bibfield  {title} {\enquote {\bibinfo {title} {A determinant
  based full configuration interaction program},}\ }\href
  {https://doi.org/https://doi.org/10.1016/0010-4655(89)90033-7} {\bibfield
  {journal} {\bibinfo  {journal} {Computer Physics Communications}\ }\textbf
  {\bibinfo {volume} {54}},\ \bibinfo {pages} {75--83} (\bibinfo {year}
  {1989})}\BibitemShut {NoStop}%
\bibitem [{\citenamefont {Gauss}, \citenamefont {Stanton},\ and\ \citenamefont
  {Bartlett}(1991)}]{doi:10.1063/1.460915}%
  \BibitemOpen
  \bibfield  {author} {\bibinfo {author} {\bibfnamefont {J.}~\bibnamefont
  {Gauss}}, \bibinfo {author} {\bibfnamefont {J.~F.}\ \bibnamefont {Stanton}},\
  and\ \bibinfo {author} {\bibfnamefont {R.~J.}\ \bibnamefont {Bartlett}},\
  }\bibfield  {title} {\enquote {\bibinfo {title} {Coupled‐cluster
  open‐shell analytic gradients: Implementation of the direct product
  decomposition approach in energy gradient calculations},}\ }\href
  {https://doi.org/10.1063/1.460915} {\bibfield  {journal} {\bibinfo  {journal}
  {The Journal of Chemical Physics}\ }\textbf {\bibinfo {volume} {95}},\
  \bibinfo {pages} {2623--2638} (\bibinfo {year} {1991})},\ \Eprint
  {https://arxiv.org/abs/https://doi.org/10.1063/1.460915}
  {https://doi.org/10.1063/1.460915} \BibitemShut {NoStop}%
\bibitem [{\citenamefont {Edmiston}\ and\ \citenamefont
  {Krauss}(1966)}]{Edmiston1966}%
  \BibitemOpen
  \bibfield  {author} {\bibinfo {author} {\bibfnamefont {C.}~\bibnamefont
  {Edmiston}}\ and\ \bibinfo {author} {\bibfnamefont {M.}~\bibnamefont
  {Krauss}},\ }\bibfield  {title} {\enquote {\bibinfo {title} {{Pseudonatural
  Orbitals as a Basis for the Superposition of Configurations. I. He 2 +}},}\
  }\href {https://doi.org/10.1063/1.1727841} {\bibfield  {journal} {\bibinfo
  {journal} {The Journal of Chemical Physics}\ }\textbf {\bibinfo {volume}
  {45}},\ \bibinfo {pages} {1833--1839} (\bibinfo {year} {1966})}\BibitemShut
  {NoStop}%
\bibitem [{\citenamefont {Meyer}(1973)}]{Meyer1973}%
  \BibitemOpen
  \bibfield  {author} {\bibinfo {author} {\bibfnamefont {W.}~\bibnamefont
  {Meyer}},\ }\bibfield  {title} {\enquote {\bibinfo {title} {{PNO–CI Studies
  of electron correlation effects. I. Configuration expansion by means of
  nonorthogonal orbitals, and application to the ground state and ionized
  states of methane}},}\ }\href {https://doi.org/10.1063/1.1679283} {\bibfield
  {journal} {\bibinfo  {journal} {The Journal of Chemical Physics}\ }\textbf
  {\bibinfo {volume} {58}},\ \bibinfo {pages} {1017--1035} (\bibinfo {year}
  {1973})}\BibitemShut {NoStop}%
\bibitem [{\citenamefont {Sun}\ \emph {et~al.}(2017)\citenamefont {Sun},
  \citenamefont {Berkelbach}, \citenamefont {McClain},\ and\ \citenamefont
  {Chan}}]{Sun2017}%
  \BibitemOpen
  \bibfield  {author} {\bibinfo {author} {\bibfnamefont {Q.}~\bibnamefont
  {Sun}}, \bibinfo {author} {\bibfnamefont {T.~C.}\ \bibnamefont {Berkelbach}},
  \bibinfo {author} {\bibfnamefont {J.~D.}\ \bibnamefont {McClain}},\ and\
  \bibinfo {author} {\bibfnamefont {G.~K.-L.}\ \bibnamefont {Chan}},\
  }\bibfield  {title} {\enquote {\bibinfo {title} {{Gaussian and plane-wave
  mixed density fitting for periodic systems}},}\ }\href
  {https://doi.org/10.1063/1.4998644} {\bibfield  {journal} {\bibinfo
  {journal} {The Journal of Chemical Physics}\ }\textbf {\bibinfo {volume}
  {147}},\ \bibinfo {pages} {164119} (\bibinfo {year} {2017})}\BibitemShut
  {NoStop}%
\bibitem [{\citenamefont {Boys}\ and\ \citenamefont
  {Bernardi}(1970)}]{Boys1970}%
  \BibitemOpen
  \bibfield  {author} {\bibinfo {author} {\bibfnamefont {S.}~\bibnamefont
  {Boys}}\ and\ \bibinfo {author} {\bibfnamefont {F.}~\bibnamefont
  {Bernardi}},\ }\bibfield  {title} {\enquote {\bibinfo {title} {{The
  calculation of small molecular interactions by the differences of separate
  total energies. Some procedures with reduced errors}},}\ }\href
  {https://doi.org/10.1080/00268977000101561} {\bibfield  {journal} {\bibinfo
  {journal} {Molecular Physics}\ }\textbf {\bibinfo {volume} {19}},\ \bibinfo
  {pages} {553--566} (\bibinfo {year} {1970})}\BibitemShut {NoStop}%
\bibitem [{\citenamefont {Birch}(1947)}]{Birch1947}%
  \BibitemOpen
  \bibfield  {author} {\bibinfo {author} {\bibfnamefont {F.}~\bibnamefont
  {Birch}},\ }\bibfield  {title} {\enquote {\bibinfo {title} {{Finite Elastic
  Strain of Cubic Crystals}},}\ }\href {https://doi.org/10.1103/PhysRev.71.809}
  {\bibfield  {journal} {\bibinfo  {journal} {Physical Review}\ }\textbf
  {\bibinfo {volume} {71}},\ \bibinfo {pages} {809--824} (\bibinfo {year}
  {1947})}\BibitemShut {NoStop}%
\bibitem [{\citenamefont {Schimka}, \citenamefont {Harl},\ and\ \citenamefont
  {Kresse}(2011)}]{Schimka2011}%
  \BibitemOpen
  \bibfield  {author} {\bibinfo {author} {\bibfnamefont {L.}~\bibnamefont
  {Schimka}}, \bibinfo {author} {\bibfnamefont {J.}~\bibnamefont {Harl}},\ and\
  \bibinfo {author} {\bibfnamefont {G.}~\bibnamefont {Kresse}},\ }\bibfield
  {title} {\enquote {\bibinfo {title} {{Improved hybrid functional for solids:
  The HSEsol functional}},}\ }\href {https://doi.org/10.1063/1.3524336}
  {\bibfield  {journal} {\bibinfo  {journal} {The Journal of Chemical Physics}\
  }\textbf {\bibinfo {volume} {134}},\ \bibinfo {pages} {024116} (\bibinfo
  {year} {2011})}\BibitemShut {NoStop}%
\bibitem [{\citenamefont {Stanton}, \citenamefont {Gauss},\ and\ \citenamefont
  {Bartlett}(1992)}]{doi:10.1063/1.463762}%
  \BibitemOpen
  \bibfield  {author} {\bibinfo {author} {\bibfnamefont {J.~F.}\ \bibnamefont
  {Stanton}}, \bibinfo {author} {\bibfnamefont {J.}~\bibnamefont {Gauss}},\
  and\ \bibinfo {author} {\bibfnamefont {R.~J.}\ \bibnamefont {Bartlett}},\
  }\bibfield  {title} {\enquote {\bibinfo {title} {On the choice of orbitals
  for symmetry breaking problems with application to no3},}\ }\href
  {https://doi.org/10.1063/1.463762} {\bibfield  {journal} {\bibinfo  {journal}
  {The Journal of Chemical Physics}\ }\textbf {\bibinfo {volume} {97}},\
  \bibinfo {pages} {5554--5559} (\bibinfo {year} {1992})},\ \Eprint
  {https://arxiv.org/abs/https://doi.org/10.1063/1.463762}
  {https://doi.org/10.1063/1.463762} \BibitemShut {NoStop}%
\end{thebibliography}
\fi

\section*{Appendix}

\subsection{Further algorithmic efficiency in one-RDM construction}

In Sec.~\ref{sec:1rdm_part_wf} we detail the specifics of the 1-RDM construction from the implicit global $T$-amplitudes. There are a number of further efficiency gains which we make to further speed up the generation of this 1-RDM, and which can also be used for other non-local quantities which rely on combining different cluster solutions.

1) We perform a compact singular-value decomposition of the overlap matrices $S^{\cl{x} \cl{y}}$ in Eq.~\eqref{eq:overlap_xy}
and remove all left and right singular vectors corresponding to singular values smaller than a cutoff (which we choose to be $10^{-3}$).
For every pair of clusters, the $\Lambda$- and $T$-amplitudes are then rotated into the resulting SVD basis (scaled by the corresponding singular value),
except those open indices which are not contracted with an overlap between the cluster pairs (e.g. $i_x$ and $j_y$ in Eq.~\eqref{eq:dm1_oo_l2t2}).
These singular values characterize the overlap between the occupied or virtual orbitals of clusters $\cl{x}$ and $\cl{y}$.
If no singular value is above the threshold, the clusters are considered to have negligible overlap and
and the contribution~$\Delta\gamma^\cl{xy}$ is assumed zero.
In the asymptotic large system limit, each cluster will only have an appreciable overlap with $\mathcal{O}(1)$ other clusters due to their local nature. This effectively screens the loop over cluster pairs to eventually only loop over $\mathcal{O}(N)$ overlapping clusters.

2) When periodic boundary conditions are used, we always perform the embedding calculations
at the $\Gamma$-point of the $N_1$$\times$$N_2$$\times$$N_3$ Born--von Karman supercell,
corresponding to original Monkhorst--Pack $\mathbf{k}$-point mesh of the same dimensions.
In this case the summation over clusters~$\cl{x}$ is only performed over fragments within the primitive cell at $\mathbf{R}_0$,
whereas the summation over clusters denoted~$\cl{y}$ is unrestricted over all primitive cells within the supercell ($\mathbf{R}_i$).
%
%
Note that even though density matrix contributions between $\cl{x}$ in the primitive cell and ~$\cl{y}$ over all clusters in the supercell are explicitly constructed, individual clusters only in the primitive cell~$\mathbf{R}_0$ need to be solved, as the resulting wave~function amplitudes can be transformed into the AO basis and translated (corresponding to a permutation of the AO indices) to any other, symmetry-equivalent position within the supercell.
The remaining density matrix contributions from clusters~$\cl{x}$ outside of the primitive cell~$\mathbf{R}_0$ can be simply reconstructed via translation, as
\begin{equation}
    \gamma = 
    \sum_{\mathbf{R}_i}^{N_\mathrm{cells}}
    \mathcal{\hat{T}}(\mathbf{R}_i-\mathbf{R}_0)
    \left(
    \sum_\cl{x \in \mathbf{R}_0}^{N_\mathrm{frag}^\mathrm{prim}}
    \sum_{\mathbf{R}_j}^{N_\mathrm{cells}}
    \sum_{\cl{y} \in \mathbf{R}_j}^{N_\mathrm{frag}^\mathrm{prim}}
    \Delta \gamma^{\cl{x}\cl{y}}
    \right)
    ,
\end{equation}
where $\hat{\mathcal{T}}(\mathbf{R})$ is an operator which translates all atomic orbitals in $\Delta \gamma^{\cl{x}\cl{y}}$ to the right by~$\mathbf{R}$, with $\Delta \gamma^{\cl{x}\cl{y}}$ 
denoting the contribution to the correlated density matrix from the cluster pair~$\cl{x}$ and $\cl{y}$ in the AO basis.
If desired, the complete supercell density matrix can be readily Fourier-transformed to $\mathbf{k}$-space according to
\begin{equation}
    \gamma_{\mathbf{k},\tilde{\alpha}\tilde{\beta}} =
    \frac{1}{N_\mathrm{cells}}
    \sum_{\mathbf{R}_i, \mathbf{R}_j}^{N_\mathrm{cells}}
    \mathrm{e}^{\mathrm{i} \mathbf{k} (\mathbf{R_j} - \mathbf{R_i})}
    \gamma_{(\mathbf{R}_i, \tilde{\alpha}),(\mathbf{R}_j, \tilde{\beta})}
    ,
\end{equation}
%
where the AOs of the supercell, $\alpha$, were enumerated in terms of the composite label~$(\mathbf{R}_i, \tilde{\alpha})$,
with~$\tilde{\alpha}$ representing an AO in the primitive cell, $\mathbf{R}_0$.

3) Embedding problems in \texttt{Vayesta} can be solved in parallel using the Message Passing Interface~(MPI) over distributed memory. Symmetry unique fragments are assigned to each MPI rank (generally a node), with their clusters built and solved independently on these MPI processes, and OpenMP threading used within each rank for the dense linear algebra on a shared memory basis. However, when constructing the 1-RDM, the cluster $T$- and $\Lambda$-amplitudes need to be communicated. Classical point-to-point or collective communications do not naturally fit the existing algorithm. For this reason, we employ one-sided communication (also called remote memory access, RMA), which allow each MPI process assigned to a given cluster to access the cluster amplitudes of all other MPI processes, without the need to halt or synchronize with the sending process.

4) When projecting $T_2$- (or $\Lambda_2$-) amplitudes, the projector is applied in a symmetrically averaged fashion between the first and second occupied index, as shown in Eq.~\eqref{eq:part_t2}.
When contracting (for example) a symmetrically projected $T_2$-amplitude with a symmetrically projected $\Lambda_2$-amplitude, the result can be thought of in terms of four contributions (each weighted by $1/4$):
first, the $T_2$-amplitude projected in the first index contracted with the $L_2$-amplitude projected in the first index,
second, the $T_2$-amplitude projected in the first index contracted with the $L_2$-amplitude projected in the second index, etc.
In contrast to the symmetrically projected amplitudes, which require
$(N_\mathrm{cl})^4$ memory to be stored,
the amplitudes projected in a single index can be stored in
$N_\mathrm{f} (N_\mathrm{cl})^3$ memory,
where $N_\mathrm{f}$ is the number of fragment orbitals in the cluster, when the projected index is rotated to be represented in the fragment-bath basis rather than the particle-hole basis of the cluster. 
More crucially, contractions between two such amplitudes, such as the one in Eq.~\eqref{eq:dm1_oo_l2t2}, only scale as $N_\mathrm{f} (N_\mathrm{cl})^4$, instead of $(N_\mathrm{cl})^5$.
Since the number of fragment orbitals is often significantly smaller than the number of cluster orbitals (sometimes by a factor of 100 or more), it is computationally more efficient to perform this contraction four times with the amplitudes projected in different indices and then average the result, instead of first performing the averaging of the projected amplitudes and then performing a single contraction. Furthermore, inter-node MPI communication of these `single-index'-projected $T_2$ and $\Lambda_2$ cluster amplitudes can be significantly reduced by a factor of $N_\mathrm{f}/N_\textrm{cl}$ utilizing this compressed representation.

\subsection{Local spin-spin correlation functions} \label{app:spin-spin}

In Eq.~\ref{eq:def_ssz}, the two-point instantaneous spin--spin correlation function is defined, $\braket{\hat{S}_z^A \hat{S}_z^B}$, between two local subspaces denoted by $A$ and $B$. We choose these atomic spaces to be the set of symmetrically (L{\"o}wdin) orthogonalized atomic orbitals~(SAO) for each atom, resulting in a projector of the form
\begin{equation} \label{eq:sao_projector}
    P^{A}_{ij} = \sum_{s \in A}^{N_\mathrm{SAO}^A}
    \left[ \mathbf{C}^T \mathbf{S} \mathbf{C}_\mathrm{SAO} \right]_{is}
    \left[ \mathbf{C}^T \mathbf{S} \mathbf{C}_\mathrm{SAO} \right]_{js}
    ,
\end{equation}
where $\mathbf{C}_\mathrm{SAO}$ is the coefficient matrix of the SAOs and $s$ runs over the $N_\mathrm{SAO}^A$
orbitals associated with atom~$A$.
However, for an efficient evaluation of this expectation value via the wave~function partitioned approach, we require the two-body contributions via the `in-cluster' approximated cumulant to be constructed in each cluster independently, to avoid the construction and subsequent contraction over rank-4 quantities in the full space which would lead to high computational scaling and memory footprints. This can be achieved in a similar manner to the computation of the $E[(\gamma, K^*)[\Psi^{\cl{x}}]]$ energy in \ref{eq:e_incluster}. The resulting working equations in a restricted basis are
\begin{equation}
\begin{split}
    \braket{\hat{S}_z^A \hat{S}_z^B}&[(\gamma, K^*)[\Psi^\cl{x}]] = \frac{1}{4} \sum_{pqr}^{N_\mathrm{mo}} P^A_{pq} P^B_{rp} \gamma_{qr} \\
    &- \frac{1}{4} \sum_{pq}^{N_\mathrm{mo}} \sum_i^{N_\mathrm{occ}} P^A_{pq} P^B_{ip} \left( 2\gamma_{qi} - \gamma_{qi}^0 \right) \\
    &- \frac{1}{12} \sum_\cl{x}^{N_\mathrm{frag}} \sum_{pqrs}^{N_\mathrm{cl}^\cl{x}}
    P^A_{pq} P^B_{rs}
    \left( K_{pqrs}^* + 2 K_{psrq}^* \right) ,
\end{split}
\end{equation}
while for an unrestricted formalism, they take the form
\begin{equation}
\begin{split}
    & \braket{\hat{S}_z^A \hat{S}_z^B}[(\gamma, K^*)[\Psi^\cl{x}]] = \frac{1}{4}
    \sum_\sigma^{\{\alpha,\beta\}} \sum_{pqr}^{N_\mathrm{mo}} P^A_{pq} P^B_{rp} \gamma_{qr}^\sigma \\
    &\quad - \frac{1}{4} \sum_\sigma^{\{\alpha,\beta\}}  \sum_{pq}^{N_\mathrm{mo}} \sum_i^{N_\mathrm{occ}^\sigma} P^A_{pq} P^B_{ip} \left(2 \gamma_{qi}^\sigma - \gamma_{qi}^{0\sigma} \right) \\
    &\quad + \frac{1}{8}
    \sum_{\sigma \sigma'}^{\{\alpha,\beta\}} 
    \left(2\delta_{\sigma\sigma'} - 1\right)
    \left[
    \sum_i^{N_\mathrm{occ}^\sigma} P^A_{ii}
    \sum_{pq}^{N_\mathrm{mo}} P^B_{pq} \left( 2\gamma_{qp}^{\sigma'} - \gamma_{qp}^{\sigma' 0} \right)
    \right. \\
    &\qquad  \left. + \sum_i^{N_\mathrm{occ}^\sigma} P^B_{ii}
    \sum_{pq}^{N_\mathrm{mo}} P^A_{pq} \left( 2\gamma_{qp}^{\sigma'} - \gamma_{qp}^{\sigma' 0} \right)
    \right] \\
    &\quad+ \frac{1}{4} \sum_\cl{x}^{N_\mathrm{frag}}
    \left[
    \sum_{pqrs}^{N_\mathrm{cl}^{\alpha \cl{x}}}
    P^A_{pq} P^B_{rs}
    K_{pqrs}^{\alpha\alpha*}
    +
    \sum_{pqrs}^{N_\mathrm{cl}^{\beta \cl{x}}}
    P^A_{pq} P^B_{rs}
    K_{pqrs}^{\beta\beta*} \right. \\
    &\qquad - \left.
    \sum_{pq}^{N_\mathrm{cl}^{\alpha \cl{x}}}
    \sum_{rs}^{N_\mathrm{cl}^{\beta \cl{x}}}
    P^A_{pq} P^B_{rs}
    K_{pqrs}^{\alpha\beta*}
    -
    \sum_{pq}^{N_\mathrm{cl}^{\beta \cl{x}}}
    \sum_{rs}^{N_\mathrm{cl}^{\alpha \cl{x}}}
    P^A_{pq} P^B_{rs}
    K_{pqrs}^{\beta\alpha*} \right]
    ,
\end{split}
\end{equation}
where the factors of the type $\left( 2\gamma_{qp}^{\sigma'} - \gamma_{qp}^{\sigma' 0} \right)$ arise due to the specific definition of $\approxcumulant$, as given in Eq.~\eqref{eq:approx_cumulant_1}.
%
%

\newif\ifsi
\sitrue

\ifsi
\clearpage
\onecolumngrid
\section*{Supporting Information}
In this SI, we provide example inputs for the {\tt Vayesta} code v1.0.0 (publicly available via https://github.com/BoothGroup/Vayesta) for the generation of a number of representative results from this work.
\definecolor{LightGray}{gray}{0.9}

\subsection*{Input for Chlorine dimer results of Fig.~\ref{fig:cl2_intro}, providing all energy functionals described}
\pythonexternal{cl2.py}

\subsection*{Input for a single molecule in the W4-11 test set of Sec.~\ref{sec:W4} for a range of bath natural orbital thresholds}
\pythonexternal{nccn.py}

\subsection*{Input for the periodic diamond cell calculations of Sec.~\ref{sec:SolidState}}
\pythonexternal{diamond_2.py}

\subsection*{Input for generation of non-local spin-spin correlation functions of Sec.~\ref{sec:spin_corr_fns}}
\pythonexternal{propyl_szsz.py}
\fi




\end{document}